\documentclass[MSNbibl,nameyear,dvips]{arxstspdf}
\usepackage{graphicx}
\usepackage{flushend}
\usepackage{stfloats}

\volume{28}
\issue{1}
\pubyear{2013}
\firstpage{1}
\lastpage{39}
\doi{10.1214/12-STS402} 

\makeatletter

\newcommand{\rright}{\right}
\newcommand{\lleft}{\left}

\newcommand{\given}{ | }
\newcommand{\Beta}{B}

\newcommand{\sizedmat}[2]{%
\mathord{\mathop{\mathbf{#1}}_{(#2)}}%
}
\newcommand{\Var}{\operatorname{Var}}
\newcommand{\Cov}{\operatorname{Cov}}
\newcommand{\argmin}{\operatorname{argmin}}

\newcommand{\pkg}[1]{\textbf{#1}} 

\newcommand{\MLM}{MvLM}

\setattribute{abstract}    {skip} {20\p@}
\setattribute{keyword}     {skip} {8\p@}

\setattribute{abstract}{width}{360pt}
\setattribute{keyword}{width}{360pt}
\makeatother

\begin{document}
\begin{frontmatter}

\title{Elliptical Insights: Understanding Statistical Methods through
Elliptical Geometry}
\runtitle{Elliptical Insights}

\begin{aug}
\author[a]{\fnms{Michael} \snm{Friendly}\corref{}\ead[label=e1]{friendly@yorku.ca}},
\author[b]{\fnms{Georges} \snm{Monette}\ead[label=e2]{georges@yorku.ca}}
\and
\author[c]{\fnms{John} \snm{Fox}\ead[label=e3]{jfox@mcmaster.ca}}
\runauthor{M. Friendly, G. Monette and J. Fox}

\affiliation{York University, York University and McMaster University}

\address[a]{Michael Friendly is Professor, Psychology Department, York University, 4700 Keele
St, Toronto, Ontario, M3J 1P3, Canada \printead{e1}.}
\address[b]{Georges Monette is Associate Professor, Mathematics and Statistics Department, York University,
4700 Keele St, Toronto, Ontario, M3J 1P3, Canada \printead{e2}.}
\address[c]{John Fox is Senator William McMaster Professor of Social Statistics, Department of Sociology, McMaster University,
1280 Main Street West, Hamilton, Ontario, L8S 4M4, Canada \printead{e3}.}

\end{aug}

%
\begin{abstract}
Visual insights into a wide variety of statistical methods, for both didactic
and data analytic purposes, can often be achieved through geometric
diagrams and
geometrically based statistical graphs. This paper extols and
illustrates the
virtues of the ellipse and her higher-dimensional cousins for both these
purposes in a variety of contexts, including linear models, multivariate
linear models and mixed-effect models.
We emphasize the strong relationships among statistical methods,
matrix-algebraic
solutions and geometry that can often be easily understood in terms of
ellipses.
\end{abstract}

%

\end{frontmatter}

\section{Introduction}\label{sec1}\vspace*{-2pt}

\begin{quote}
Whatever relates to extent and quantity may be represented by
geometrical figures. Statistical projections which speak to the senses without
fatiguing the mind, possess the advantage of fixing the attention on a great
number of important facts.\vspace*{-2pt}
\flushright{Alexander von Humboldt [(\citeyear{Humboldt1811a}),
page~ciii]}\vadjust{\goodbreak}
\end{quote}

In the beginning,
there was an ellipse. As modern statistical
methods progressed from bivariate to multivariate, the ellipse escaped
the plane to a 3D
ellipsoid, and then grew onward to higher dimensions.
This paper extols and illustrates the
virtues of the ellipse and her higher-dimensional cousins for both
didactic and
data analytic purposes.

When
Francis Galton (\citeyear{Galton1886}) first studied the relationship
between heritable traits of
parents and their offspring, he had a remarkable visual
insight---contours of
equal bivariate frequencies in the joint distribution seemed to form concentric
shapes whose outlines were, to Galton, tolerably close to concentric ellipses
differing only in scale.

\begin{figure}

\includegraphics{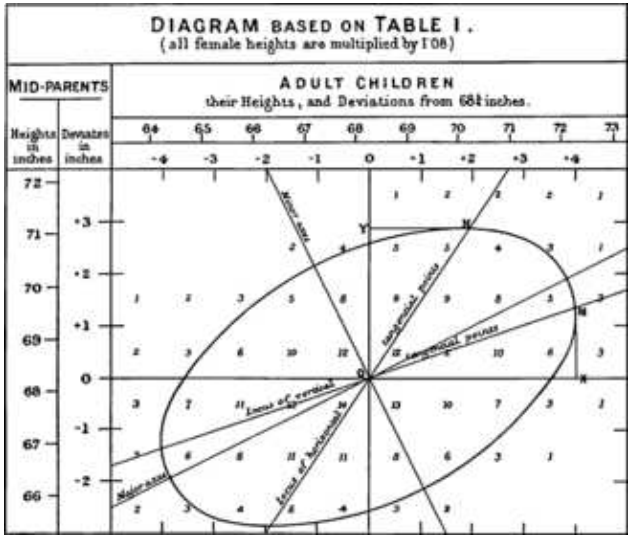}

\caption{Galton's 1886 diagram, showing the relationship of height of children
to the average of their parents' height. The diagram is essentially an overlay
of a geometrical interpretation on a bivariate grouped frequency
distribution, shown
as numbers.}%
\label{figgalton-corr}
\end{figure}

Galton's goal was to to predict (or explain) how a characteristic, $Y$, (e.g.,
height) of children was related to that of their parents, $X$. To this
end, he
calculated summaries, $\operatorname{Ave}(Y\given X)$, and, for symmetry,
$\operatorname{Ave}(X\given Y)$, and plotted
these as lines of means on his diagram. Lo and behold, he had a second visual
insight: the lines of means of ($Y\given X$) and ($X\given Y$)
corresponded approximately to
the locus of horizontal and vertical tangents to the concentric
ellipses. To
complete the picture, he added lines showing the major and minor axes
of the
family of ellipses, with the result shown in Figure~\ref{figgalton-corr}.

It is not stretching the point too far to say that a large part of modern
statistical methods descends from these visual insights:%
\footnote{\citeauthor{Pearson1920} [(\citeyear{Pearson1920}), page
37] later stated, ``that Galton
should have evolved all this from his observations is to my mind one
of the most noteworthy scientific discoveries arising from pure
analysis of observations.''}
correlation and
regression [\citet{Pearson1896}], the bivariate normal distribution,
and principal components [\citet{Pearson1901}, \citet{Hotelling1933}]
all trace
their ancestry to Galton's geometrical
diagram.%
\footnote{Well, not entirely. Auguste Bravais [1811--1863] (\citeyear
{Bravais1846}), an astronomer
and physicist first introduced the mathematical theory of the bivariate
normal distribution
as a model for the joint frequency of errors in the geometric position
of a point.
Bravais derived the formula for level slices as concentric ellipses and
had a rudimentary
notion of correlation but did not appreciate this as a representation
of data.
Nonetheless, \citet{Pearson1920} acknowledged Bravais's contribution,
and the correlation
coefficient is often called the Bravais-Pearson coefficient in France
[\citet{Denis2001}]. }

%
%

Basic geometry goes back at least to Euclid, but the properties of the
ellipse and other
conic sections may be traced to Apollonius of Perga (ca.~262
BC--ca.~190 BC), a
Greek geometer and astronomer who gave the ellipse, parabola and
hyperbola their
modern names. In a work popularly called the Conics [\citet{Boyer91}],
he described
the fundamental properties of ellipses (eccentricity, axes, principles of
tangency, normals as minimum and maximum straight lines to the curve) with
remarkable clarity nearly 2000 years before the development of analytic geometry
by Descartes.

Over time, the ellipse would be called to duty to provide simple
explanations of
phenomena once thought complex. Most notable is Kepler's insight that the
Copernican theory of the orbits of planets as concentric circles (which required
notions of epicycles to account for observations) could be brought into
alignment with the detailed observational data from Tycho Brahe and
others by an exquisitely simple
law: ``The orbit of every planet is an ellipse with the sun at a
focus.'' One
century later, Isaac Newton was able to connect this elliptical
geometry with astrophysics by
deriving all three of Kepler's laws as
simpler consequences of general laws of motion and universal gravitation.

This paper takes up the cause of the ellipse as a geometric form that can
provide similar service to statistical understanding and data analysis. Indeed,
it has been doing that since the time of Galton, but these graphic and geometric
contributions have often been incidental and scattered in the literature
[e.g., \citet{Bryant1984}, \citet{CampbellAtchley81},
\citet{SavilleWood1991},
\citet{Wickens1995}].
We focus here on visual insights through ellipses in the areas of
linear models,
multivariate linear models and mixed-effect models. Our goal is to
provide as
comprehensive a treatment of this topic as possible in a single article
together with online supplements.

\begin{table*}
\caption{Statistical and geometrical measures of ``size'' of an ellipsoid}\label{tab1}
\begin{tabular*}{\textwidth}{@{\extracolsep{\fill}}lccc@{}}
\hline
\textbf{Size} & \textbf{Conceptual formula} & \textbf{Geometry} &
\multicolumn{1}{c@{}}{\textbf{Function}} \\
\hline
(a) Generalized variance: & $\operatorname{det} (\bolds{\Sigma}) =
\prod_i \lambda_i$ & area, (hyper)volume & geometric mean\\
(b) Average variance: & $\operatorname{tr} (\bolds{\Sigma
}) = \sum_i \lambda_i $
& linear sum & arithmetic mean\\ 
(c) Average precision: & $1/ \operatorname{tr} (\bolds
{\Sigma}^{-1}) = 1/\sum_i
(1/\lambda_i) $ & & harmonic mean\\
(d) Maximal variance: & $\lambda_1$ & maximum dimension & supremum\\
\hline
\end{tabular*}
\end{table*}
%


The plan of this paper is as follows: Section~\ref{secnotation} provides
the minimal notation and
properties of ellipsoids%
\footnote{As in this paragraph,
we generally use the term ``ellipsoid'' as to refer to ``ellipse or
ellipsoid'' where dimensionality does not matter or context is clear.}
necessary for the remainder of the paper. Due to length restrictions,
other useful and important properties of geometric and statistical
ellipsoids have been relegated
to the \hyperref[app]{Appendix}. Section~\ref{secdata-ellipse}
describes the use of the
\emph{data ellipsoid}
as a visual summary for multivariate data. In Section~\ref{seclm} we apply
data ellipsoids and
confidence ellipsoids for parameters in linear models to explain a wide
range of phenomena, paradoxes and fallacies
that are clarified by this geometric approach.
This view is extended to multivariate linear models in Section~\ref
{secmlm}, primarily through the use
of ellipsoids to portray hypothesis (H) and error (E) covariation in
what we call \emph{HE plots}.
Finally, in Section~\ref{seckiss} we discuss a diverse collection of
current statistical problems whose solutions
can all be described and visualized in terms of ``kissing ellipsoids.''

\section{Notation and Basic Results}\label{secnotation}

There are various representations of an ellipse (or ellipsoid in three
or more dimensions),
both geometric and statistical. Some basic notation and properties are
described below.

\subsection{Geometrical Ellipsoids}\label{secgeometric}

We refer to the common notion of a bounded ellipsoid (with nonempty
interior) in the $p$-dimensional space $\mathbb{R}^{p}$
as a \emph{proper ellipsoid}.
An origin-centered proper ellipsoid
may be defined by the quadratic form
%
%
\begin{equation}
\label{eqellisoid1} \mathcal{E} := \bigl\{ \mathbf{x}\dvtx\mathbf
{x}^{\mathsf{T}}\mathbf{C} \mathbf{x} \le1 \bigr\} ,
\end{equation}
where equality in equation~(\ref{eqellisoid1}) gives the boundary,
$\mathbf{x} = (x_1, x_2, \ldots, x_p)^{\mathsf{T}}$ is a vector
referring to
the coordinate axes and $\mathbf{C}$ is a symmetric
positive definite $p \times p$ matrix.
If $\mathbf{C}$ is only positive semi-definite, then the ellipsoid will
be \emph{improper}, having the shape of a cylinder with elliptical
cross-sections and unbounded in the direction of the null
space of $\mathbf{C}$.
To extend the definition to \emph{singular} (sometimes known as
``degenerate'') ellipsoids, we turn to a definition that is equivalent
to equation~(\ref{eqellisoid1}) for proper ellipsoids.
Let $\mathcal{S}$ denote the unit sphere in $\mathbb{R}^{p}$,
%
%
\begin{equation}
\mathcal{S} := \bigl\{ \mathbf{x}\dvtx\mathbf{x}^{\mathsf
{T}}\mathbf{x} =1
\bigr\} ,
\end{equation}
and let
%
%
\begin{equation}
\label{eqellisoidsph} \mathcal{E} := \mathbf{A} \mathcal{S} ,
\end{equation}
where $\mathbf{A}$ is a nonsingular $p \times p$ matrix. Then
$\mathcal
{E}$ is a proper ellipsoid that could be defined using
equation~(\ref{eqellisoid1}) with $\mathbf{C} = ( \mathbf{A}
\mathbf
{A}^{\mathsf{T}}
)^{-1}$.
We obtain singular ellipsoids by allowing $\mathbf{A}$ to be any matrix,
not necessarily nonsingular or even square.
A more general representation of ellipsoids based on the singular value
decomposition (SVD) of $\mathbf{C}$ is given in Appendix~\ref{sectaxonomy}.
Some useful properties of geometric ellipsoids are described in
Appendix~\ref{secproperties}.

\subsection{Statistical Ellipsoids}\label{secstatistical}

In statistical applications, $\mathbf{C}$ will often be the inverse of a
covariance
matrix (or a sum of squares and cross-products matrix) and the
ellipsoid will
be centered at the means of variables or at estimates of parameters
under some model.
Hence, we will also use the following notation:

For a positive definite matrix
$\bolds{\Sigma}$ we use $\mathcal{E}(\bolds{\mu},\bolds{\Sigma
})$ to
denote the ellipsoid
%
%
\begin{equation}
\label{eqellipsoid3} \mathcal{E} := \bigl\{ \mathbf{x} \dvtx (\mathbf{x}-\bolds{
\mu})^{\mathsf{T}}\bolds{\Sigma}^{-1} (x-\bolds{\mu}) = 1 \bigr\} .
\end{equation}

When $\bolds{\Sigma}$ is the covariance matrix of a multivariate vector
$\mathbf{x}$ with eigenvalues
$\lambda_1 \ge\lambda_2 \ge\cdots$,
the following
properties represent the ``size'' of the ellipsoid in $\mathbb{R}^{p}$
(see Table~\ref{tab1}).

For testing hypotheses for parameters of multivariate linear
models, these different senses of ``size''
correspond (with suitable transformations) to (a) Wilks's $\Lambda$,
(b) the Hotelling--Lawley trace, (c) the Pillai trace, and~(d) Roy's
maximum root tests, as we describe
below in Section~\ref{secmlm}.

%
%
\begin{figure*}

\includegraphics{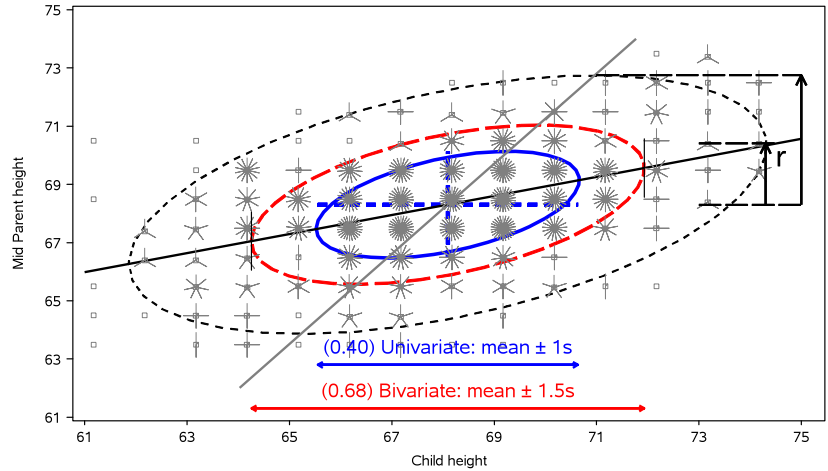}

\caption{Sunflower plot of Galton's data on heights of parents and
their children (in.), with
40\%, 68\% and 95\% data ellipses and the regression lines of $y$ on
$x$ (black) and
$x$ on $y$ (grey). The ratio of the vertical to the regression line
(labeled ``r'') to the vertical
to the top of the ellipse gives a visual estimate of the correlation
($r=0.46$, here).
Shadows (projections) on the coordinate axes give standard intervals,
$\bar{x} \pm k s_x$ and $\bar{y} \pm k s_y$,
with $k=1, 1.5, 2.45$, having
bivariate coverage 40\%, 68\% and 95\% and univariate coverage 68\%,
87\% and 98.6\%, respectively.
Plotting children's height on the abscissa follows Galton.}%
\label{figgalton-reg3}
\end{figure*}

Note that every nonnegative definite matrix $\mathbf{W}$ can be factored
as $\mathbf{W}=\mathbf{A}\mathbf{A}^{\mathsf{T}}$,
and the matrix $\mathbf{A}$ can always be selected so that it is square.
$\mathbf{A}$ will be nonsingular if and only if $\mathbf{W}$ is nonsingular.
A~computational
definition of an ellipsoid that can be used for all nonnegative
definite matrices and that corresponds to the previous definition in
the case of positive-definite matrices is
%
%
\begin{equation}
\label{eqellAS} \mathcal{E}(\bolds{\mu},\mathbf{W}) = \bolds{\mu } + \mathbf{A}
\mathcal{S} ,
\end{equation}
where $\mathcal{S}$ is a unit sphere of conformable dimension and
$\bolds{\mu}$ is the centroid of the ellipsoid.
One convenient choice of $\mathbf{A}$ is the Choleski square root,
$\mathbf{W}^{1/2}$, as we describe in Appendix~\ref{secconjugate}.
Thus, for some results below, a convenient notation in terms of
$\mathbf{W}$ is
%
%
\begin{equation}
\label{eqellipsoidW} \mathcal{E}(\bolds{\mu}, \mathbf{W}) = \bolds {\mu} \oplus
\sqrt{\mathbf{W}} = \bolds{\mu} \oplus\mathbf{W}^{1/2} ,
\end{equation}
where $\oplus$
emphasizes that the ellipsoid is a scaling and rotation of the unit
sphere followed by translation to
a center at $\bolds{\mu}$ and $\sqrt{\mathbf{W}}=\mathbf
{W}^{1/2}=\mathbf{A}$.
This representation is not unique,
however: $\bolds{\mu} \oplus\mathbf{B} = \bolds{\nu} \oplus
\mathbf{C}$
(i.e., they generate the same ellipsoid)
\emph{iff} $\bolds{\mu} = \bolds{\nu}$ and $\mathbf{B}\mathbf
{B}^{\mathsf{T}}=
\mathbf{C}\mathbf{C}^{\mathsf{T}}$.
From this result, it is readily seen that under a linear transformation
given by a matrix
$\mathbf{L}$
the image of the ellipse is
%
%
\begin{eqnarray}
\label{eqLimage} \mathbf{L} \bigl[ \bigl(\mathcal{E}(\bolds{\mu} ,\mathbf{W})
\bigr) \bigr] &= &\mathcal{E}\bigl(\mathbf{L}\bolds{\mu} ,\mathbf {L}\mathbf{W} {
\mathbf{L}}{}^{\mathsf{T}}\bigr)
\nonumber
\\
&=&\mathbf{L}\bolds{\mu} \oplus\sqrt{\mathbf{L}\mathbf{W} {
\mathbf{L}}{}^{\mathsf{T}}}
\\
&=&\mathbf{L}\bolds{\mu} \oplus\mathbf{L}\sqrt{\mathbf{W}} .
\nonumber
\end{eqnarray}

\section{The Data Ellipse and Ellipsoid}\label{secdata-ellipse}

The \emph{data ellipse} [\citet{Monette90}] [or \emph{concentration
ellipse}, \citet{Dempster69}, Chapter 7]
provides a remarkably
simple and effective display for viewing and understanding
bivariate \emph{marginal} relationships in multivariate data.
The data ellipse is typically used to add a visual summary to a scatterplot,
indicating the means, standard deviations, correlation
and slope of the regression line for
two variables. Under classical (Gaussian) assumptions, the data ellipse
provides a statistically sufficient visual summary, as we describe below.

It is historically appropriate to illustrate the data ellipse and
describe its properties using Galton's [(\citeyear{Galton1886}),
Table~I]
data, from which he drew Figure~\ref{figgalton-corr} as a conceptual
diagram,%
\footnote{These data are reproduced in \citeauthor{Stigler1986}
[(\citeyear{Stigler1986}), Table 8.2, page~286].}
shown in Figure~\ref{figgalton-reg3}, where the frequency at
each point is represented by a sunflower symbol. We also overlay the
40\%,
68\% and 95\% data ellipses, as described below.

In Figure~\ref{figgalton-reg3}, the ellipses have the mean vector
$(\bar{x}, \bar{y})$ as their center; the lengths of arms of the
central cross show the standard deviations of the variables, which
correspond to the shadows of the 40\% ellipse. In
addition, the correlation coefficient can be visually represented as
the fraction of a vertical tangent line from $\bar{y}$ to the top of
the ellipse that is below the regression line $\widehat{y} | x$, shown
by the arrow labeled ``r.'' Finally, as Galton noted, the regression line
for $\widehat{y} \given x$ (or $\widehat{x} \given y$)
can be visually estimated as the locus of the points of vertical
(or horizontal) tangents with the family of concentric ellipses.
See \citeauthor{Monette90} [(\citeyear{Monette90}), Figures~5.1--5.2] and
\citeauthor{Friendly91} [(\citeyear{Friendly91}), page~183] for illustrations and further discussion
of the properties of the data ellipse.

%
%

\begin{figure*}
\centering
\begin{tabular}{@{}cc@{}}

\includegraphics{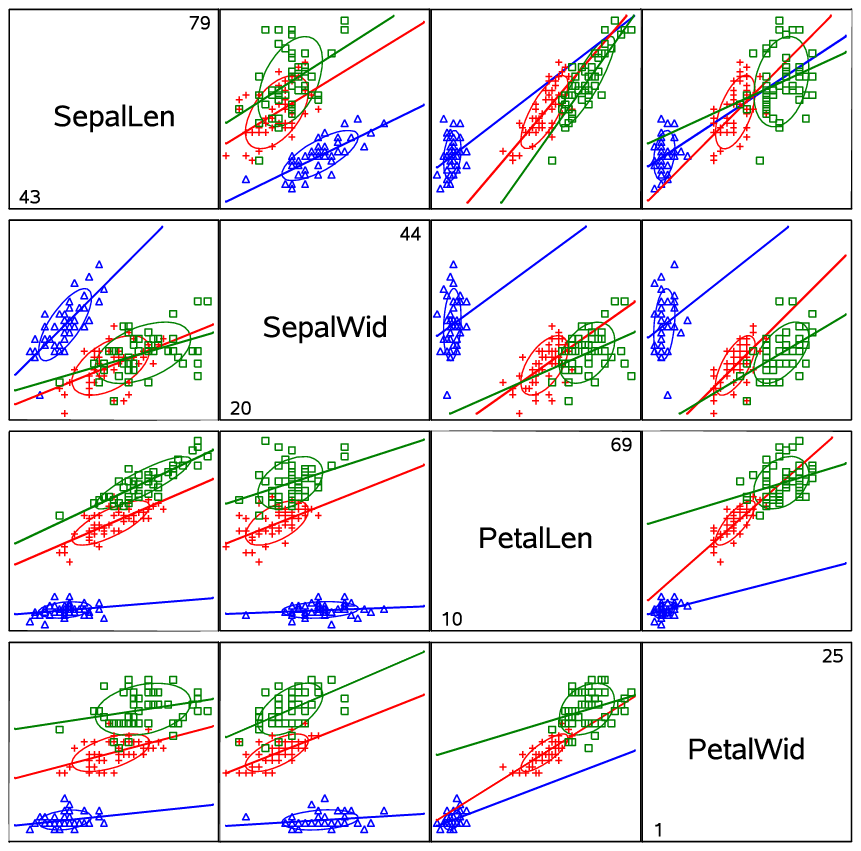}
 & \includegraphics{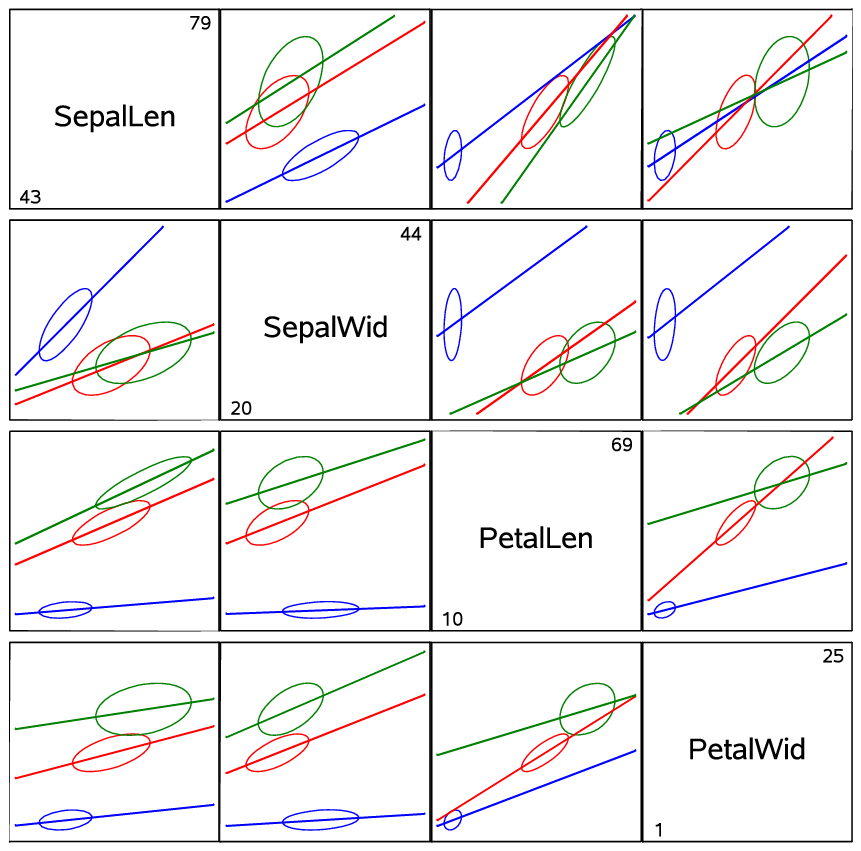}\\
\footnotesize{(a)} & \footnotesize{(b)}
\end{tabular}
\caption{Scatterplot matrices of Anderson's iris data: \textup{(a)} showing
data, separate 68\% data
ellipses, and regression lines for each species; \textup{(b}) showing only
ellipses and regression lines.
Key---\emph{Iris setosa}: blue, $\triangle$; \emph{Iris versicolor}:
red, $+$;
\emph{Iris virginca}: green, $\Box$.}%
\label{figscatirisd1}
\end{figure*}

More formally [\citet{Dempster69}, \citet{Monette90}], for a $p$-dimensional
sample, $\mathbf{Y}_{n \times p}$,
we recognize the quadratic form in equation~(\ref{eqellipsoid3})
as corresponding to the squared Mahalanobis distance,
$D^2_M (\mathbf{y}) = (\mathbf{y} - \bar{\mathbf{y}})^{\mathsf
{T}}\mathbf
{S}^{-1} (\mathbf{y} - \bar{\mathbf{y}})$,
of the point
$\mathbf{y} = (y_1, y_2, \ldots, y_p)^{\mathsf{T}}$
from the centroid of the sample,
$\bar{\mathbf{y}} = (\bar{y}_1, \bar{y}_2, \ldots, \bar
{y}_p)^{\mathsf{T}}$.
Thus, we use a more explicit notation to
define the \emph{data ellipsoid} $\mathcal{E}_c$ of size (``radius'') $c$
as the set of all points $\mathbf{y}$ with $D^2_M (\mathbf{y})$ less
than or
equal to $c^2$,
%
%
\begin{equation}
\label{eqdsq} \mathcal{E}_c ( \bar{\mathbf{y}}, \mathbf{S} ) :=
\bigl\{ \mathbf{y} \dvtx(\mathbf{y} - \bar{\mathbf{y}})^{\mathsf{T}}
\mathbf{S}^{-1} (\mathbf{y} - \bar{\mathbf{y}}) \le c^2
\bigr\} ,
\end{equation}
where
$\mathbf{S} = ({n-1})^{-1} \sum_{i=1}^n (\mathbf{y}_i - \bar
{\mathbf{y}})
(\mathbf{y}_i - \bar{\mathbf{y}}{}^{\mathsf{T}})$
is the sample covariance matrix. In the computational notation of
equation~(\ref{eqellipsoidW}), the boundary of the
data ellipsoid of radius $c$ is thus
%
%
\begin{equation}
\label{eqellipsoidS} \mathcal{E}_c(\bar{\mathbf{y}}, \mathbf{S}) =
\bar{\mathbf{y}} \oplus c \mathbf{S}^{1/2} .
\end{equation}

Many properties of the data ellipsoid hold regardless of the joint
distribution of the
variables; but if the variables are multivariate normal, then the data
ellipsoid approximates
a contour of constant density in their joint distribution. In this case
$D^2_M (x,y)$
has a large-sample $\chi^2_p$ distribution or, in finite samples, approximately
$[p (n-1) / (n-p)] F_{p, n-p}$).

Hence, in the bivariate case, taking $c^2 = \chi^2_2(0.95)= 5.99
\approx6$ encloses approximately
95\% of the data points under normal theory. Other radii also have
useful interpretations:
\begin{itemize}
\item In Figure~\ref{figgalton-reg3} we demonstrate that $c^2 = \chi^2_2(0.40) \approx1$ gives
a data ellipse of 40\% coverage with the property that its projection
on either axis
corresponds to a standard interval, $\bar{x} \pm1 s_x$ and $\bar{y}
\pm1 s_y$. The same property of univariate
coverage pertains to
any linear combination of $x$ and $y$.
\item By analogy with a univariate sample, a 68\% coverage data ellipse with
$c^2 = \chi^2_2(0.68) = 2.28$ gives a bivariate analog of the standard
$\bar{x} \pm1 s_x$ and $\bar{y} \pm1 s_y$ intervals.
The univariate shadows, or those of any linear combination, then
correspond to standard Scheff\'e
intervals taking ``fishing'' (simultaneous interfence) in a
$p=2$-dimensional space into account.
\end{itemize}

As useful as the data ellipse might be for a single, unstructured
sample, its value as a visual summary increases
with the complexity of the data.
For example, Figure~\ref{figscatirisd1} shows scatterplot matrices
of all pairwise plots of the variables from Edgar Anderson's (\citeyear
{Anderson35})
classic
data on three species of iris flowers found in the Gasp\'{e} Peninsula,
later used by \citet{Fisher36} in his development of discriminant analysis.
The data ellipses show clearly that the means, variances, correlations
and regression slopes differ systematically across the three iris species
in all pairwise plots.
We emphasize that the ellipses serve as sufficient visual summaries of
the important
statistical properties (first and second moments)%
\footnote{We recognize that a normal-theory summary (first and second moments),
shown visually or numerically, can be distorted
by multivariate outliers, particularly in smaller samples.
In what follows,
robust covariance estimates can, in principle, be substituted
for the classical, normal-theory estimates in all cases.
To save space, we do not explore these possibilities further here.}
by removing the data points
from the plots in the version at the right.

\section{Linear Models: Data Ellipses and Confidence Ellipses}\label{seclm}

Here we consider how ellipses help to visualize relationships among variables
in connection with linear models (regression, ANOVA).
We begin with views in the space of the variables (data space)
and progress to related views in the space of model parameters
($\bolds{\beta}$ space).

\subsection{Simple Linear Regression}

Various aspects of the standard data ellipse of radius~1 illuminate
many properties
of simple linear regression, as shown in Figure~\ref{figellipses-demo}.
These properties are also useful in more complex contexts:

%
%
\begin{figure}[b]

\includegraphics{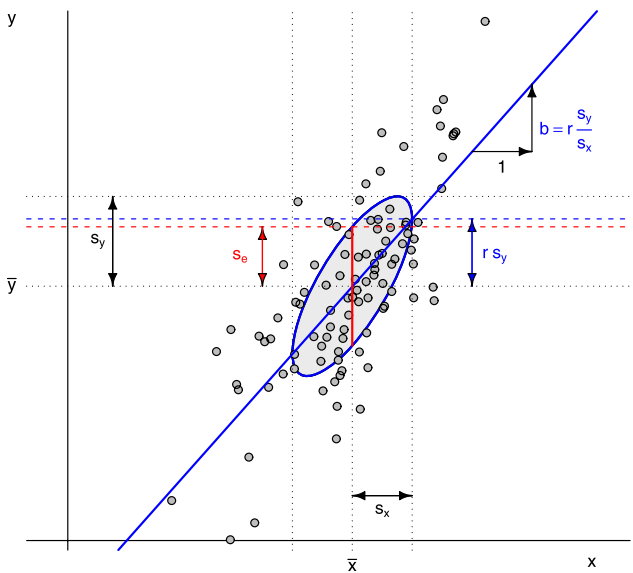}

\caption{Annotated standard data ellipse showing standard deviations
of $x$ and $y$, residual
standard deviation ($s_e$), slope ($b$) and correlation ($r$).}%
\label{figellipses-demo}
\end{figure}

\begin{figure*}

\includegraphics{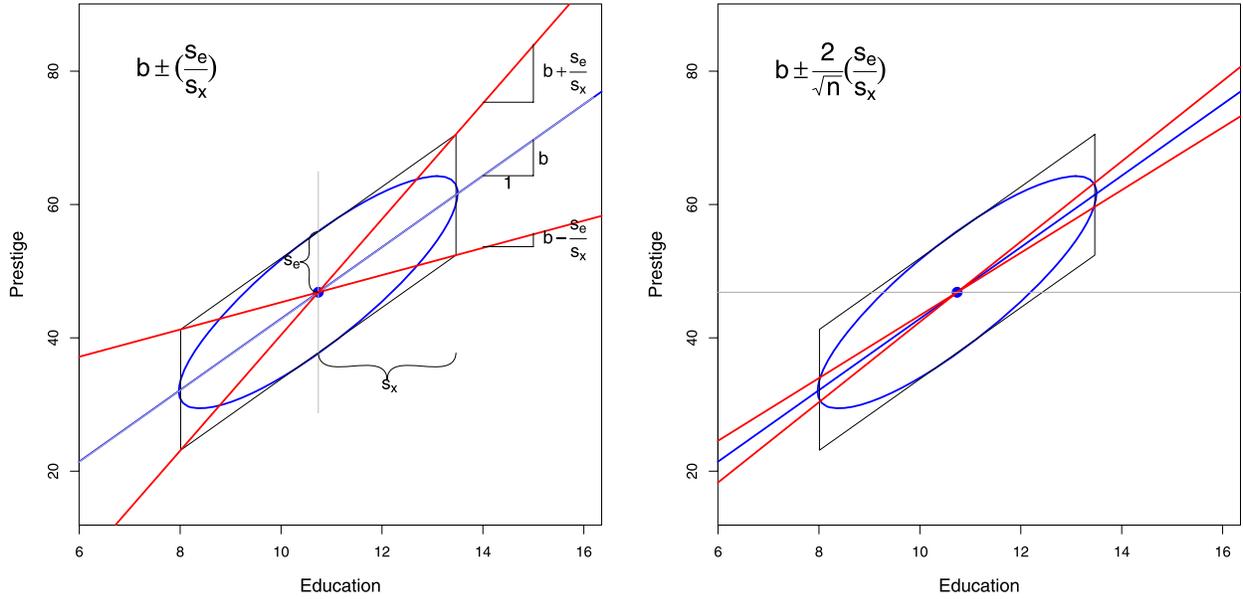}

\caption{Visual 95\% confidence interval for the slope in linear
regression. Left: Standard data ellipse surrounded by the
regression parallelogram. Right: Shrinking the diagonal lines by a
factor of $2/\sqrt{n}$,
gives the approximate 95\% confidence interval for $\beta$.}%
\label{vis-reg-prestige}
\end{figure*}

\begin{itemize}
\item One-half of the widths of the vertical and horizontal projections
(dotted black lines)
give the standard deviations $s_x$ and $s_y$, respectively.
\item Because the perpendicular projection onto any line through the
center of the ellipse, ($\bar{x}, \bar{y}$), corresponds to
some linear combination, $m x + n y$, the half-width of the
corresponding projection of the ellipse
gives the standard deviation of this linear combination.
\item With a multivariate normal distribution the line segment through
the center of the ellipse
shows the mean and standard deviation of the conditional distribution
on that line.
\item The standard deviation of the residuals, $s_e$, can be visualized
as the half-width of the vertical
(red) line at $x=\bar{x}$.
\item The vertical distance between the mean of $y$ and the points
where the ellipse has vertical
tangents is $r s_y$. (As a fraction of $s_y$, this distance is $r =
0.75$ in the figure.)
\item The (blue) regression line of $y$ on $x$ passes through the
points of vertical tangency.
Similarly, the regression of $x$ on $y$ (not shown) passes through the
points of
horizontal tangency.
\end{itemize}

%
%
\begin{figure*}[b]
\centering
\begin{tabular}{@{}c@{\hspace*{2pt}}c@{\hspace*{2pt}}c@{}}

\includegraphics{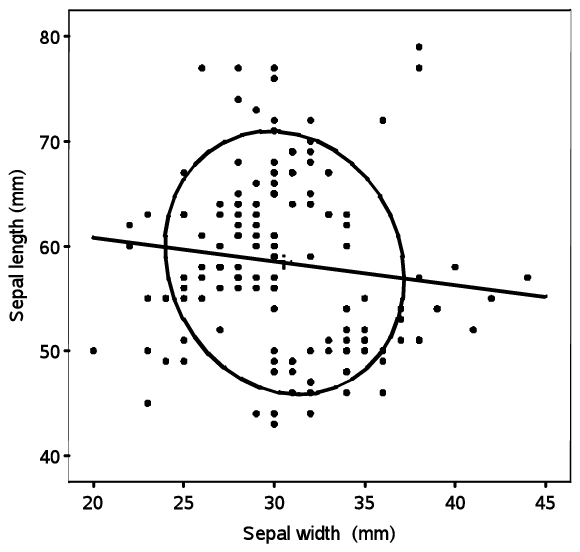}
 & \includegraphics{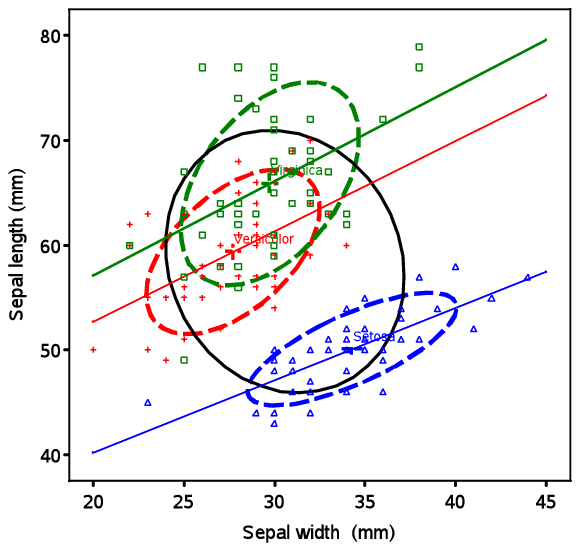} & \includegraphics{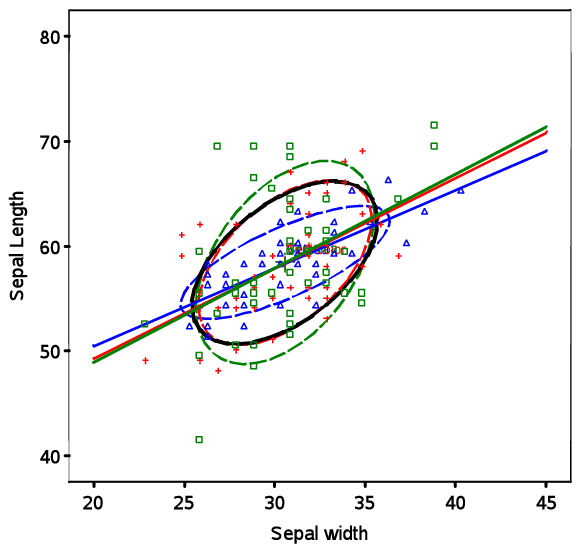}\\
\footnotesize{(a) Total sample, marginal ellipse,} & \footnotesize{(b) Individual sample,} &
\footnotesize{(c) Pooled, within-sample ellipse}\\
\footnotesize{ignoring species} & \footnotesize{conditional ellipses | species} &
\end{tabular}
\caption{Marginal \textup{(a)}, conditional \textup{(b)} and pooled
within-sample \textup{(c)}
relationships
of Sepal length and Sepal width in the iris data. Total-sample data
ellipses are
shown as black, solid curves; individual-group data and ellipses are
shown with
colors and dashed lines.}\label{figcontiris3}
\end{figure*}

\subsection{Visualizing a Confidence Interval for the Slope}

A visual approximation to a 95\% confidence interval for the slope, and
thus a visual test of $H_0\dvtx\beta= 0$,
can be seen in Figure~\ref{vis-reg-prestige}. From the formula for a
95\%
confidence interval,
$\mathrm{CI}_{0.95} (\beta) = b \pm t_{n-2}^{0.975} \times
\operatorname
{SE}(b)$, we can take $t_{n-2}^{0.975} \approx2$
and
$\operatorname{SE}(b) \approx\frac{1}{\sqrt{n}} ( \frac{s_e}{s_x}
)$,
leading to
%
%
\begin{equation}
\label{eqci-approx} \mathrm{CI}_{0.95} (\beta) \approx b \pm
\frac{2}{\sqrt{n}} \times \biggl( \frac{s_e}{s_x} \biggr) .
\end{equation}

To show this visually, the left panel of Figure~\ref{vis-reg-prestige}
displays the standard data ellipse
surrounded by the ``regression parallelogram,'' formed with the
vertical tangent lines and the
tangent lines parallel to the regression line. This corresponds to the
conjugate axes of the
ellipse induced by the Choleski factor of $S_{yx}$ as shown in
Figure~\ref{figconjugate} in Appendix~\ref{secconjugate}.
Simple algebra demonstrates that the diagonal lines through
this parallelogram have slopes of
\[
b \pm\frac{s_{e}}{s_x}.
\]
So, to obtain a visual estimate of the 95\% confidence interval for
$\beta$ (\emph{not}, we note, the 95\% CI for the regression line),
we need only shrink the diagonal lines of the
regression parallelogram toward the regression line by a factor of
$2/\sqrt{n}$, giving the red lines
in the right panel of Figure~\ref{vis-reg-prestige}.
In the data used for this
example, $n=102$, so the factor is approximately 0.2 here.\footnote
{The data are for the rated prestige and average years of education of
102 Canadian occupations circa 1970; see [\citet{FoxSuschnigg89}].}
Now consider the horizontal line through the center of the data
ellipse. If this line is outside the
envelope of the confidence lines, as it is in Figure~\ref
{vis-reg-prestige}, we can reject $H_0\dvtx\beta= 0$ via this simple
visual approximation.

\subsection{Simpson's Paradox, Marginal and Conditional
Relationships}\label{secsimpson-iris}

Because it provides a visual representation of\break means, variances and
correlations,
the data ellipse is ideally suited as a tool for illustrating and
explicating various
phenomena that occur in the analysis of linear models.
One class of simple, but important, examples concerns the difference
between the marginal
relationship between variables, ignoring some important factor or covariate,
and the conditional relationship, adjusting (controlling) for that
factor or covariate.

Simpson's paradox [\citet{Simpson51}] occurs when the marginal and
conditional relationships differ in direction. This may be seen in the plots
of Sepal length against Sepal width for the iris data shown in
Figure~\ref{figcontiris3}. Ignoring
iris species, the marginal, total-sample correlation is slightly negative
as seen in panel (a). The individual-sample ellipses in panel (b) show
that the conditional, within-species correlations are all positive, with
approximately equal regression slopes. The group means have a negative
relationship, accounting for the negative marginal correlation.

A correct analysis of the (conditional) relationship between these
variables, controlling or adjusting for mean
differences among species, is based on the pooled within-sample
covariance matrix,
%
%
\begin{eqnarray}
\label{eqSp} \mathbf{S}_{\mathrm{within}} &= &(N - g)^{-1} \sum
_{i=1}^g \sum_{j=1}^{n_i}
( \mathbf{y}_{ij} - \bar{\mathbf{y}}_{i\cdot} ) (
\mathbf{y}_{ij} - \bar{\mathbf{y}}_{i\cdot} )^{\mathsf{T}}
\nonumber
\\[-8pt]
\\[-8pt]
\nonumber
&=&(N -
g)^{-1} \sum_{i=1}^g
(n_i - 1) \mathbf{S}_i ,
\end{eqnarray}
where $N = \sum n_i$, and the result
is shown in
panel (c) of Figure~\ref{figcontiris3}.
In this graph, the data for \emph{each} species were first
transformed to deviations from the species means on both variables
and then translated back to the grand means.

%
%
\begin{figure*}[t]

\includegraphics{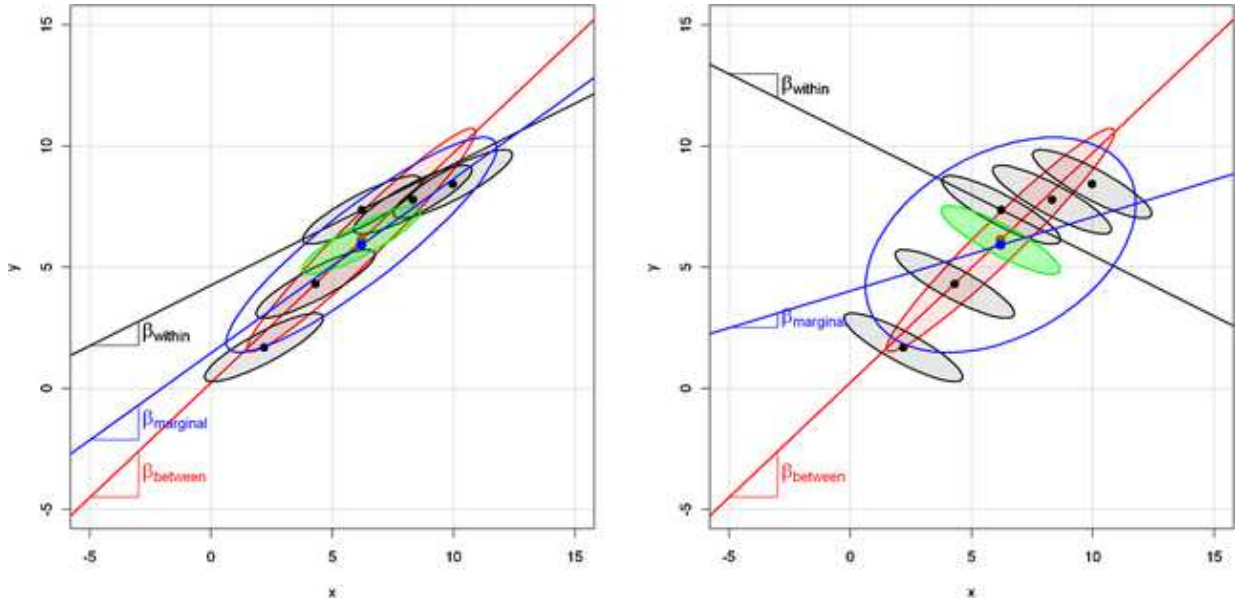}

\caption{Paradoxes and fallacies: between (ecological), within
(conditional) and whole-sample (marginal) associations.
In both panels, the five groups have the same group means, and $\Var
(x)=6$ and $\Var(y)=2$ within each group.
The within-group correlation is $r = +0.87$ in all groups in the left
panel and is $r = -0.87$ in the right panel.
The green ellipse shows the average within-group data ellipse.}
\label{figbetween-within}
\end{figure*}

In a more general context, $ \mathbf{S}_{\mathrm{within}}$
appears as the $\mathbf{E}$ matrix in a multivariate
linear model, adjusting or controlling for all fitted effects (factors
and covariates).
For essentially correlational analyses (principal components,
factor analysis, etc.),
similar displays can be used to show how multi-sample analyses
can be compromised by substantial group mean differences and corrected
by analysis of the pooled within-sample covariance matrix, or by
including important group variables in the model.
Moreover, display of the individual within-group data ellipses can
show visually how well the assumption of
equal covariance matrices,
$\Sigma_1 = \Sigma_2 = \cdots= \Sigma_g$,
is satisfied in the data, for the two variables
displayed.

\subsection{Other Paradoxes and Fallacies}

Data ellipses can also be used to visualize and understand other
paradoxes and
fallacies that occur with linear models. We consider situations in
which there
is a principal relationship between variables $y$ and $x$ of interest,
but (as in the preceding subsection) the data
are stratified in $g$ samples by a factor (``group'') that might correspond
to different subpopulations (e.g., men and women, age groups),
different spatial regions (e.g., states), different points in time or some
combination of the above.

In some cases, group may be unknown or may not have been included in
the model,
so we can only estimate the marginal association between $y$ and $x$,
giving a slope $\beta_{\mathrm{marginal}}$ and correlation
$r_{\mathrm{marginal}}$.
In other cases, we may not have individual data, but only aggregate group
data, $(\bar{y}_i, \bar{x}_i), i=1, \ldots, g$, from which we can estimate
the between-groups (``ecological'') association, with slope
$\beta_{\mathrm{between}}$ and correlation $r_{\mathrm{between}}$.
When all data are available and the model is an ANCOVA model of the form
$y \sim x + \mathrm{group}$,
we can estimate a common conditional, within-group slope,
$\beta_{\mathrm{within}}$, or, with the model $y \sim x + \mathrm
{group} + x \times\mathrm{group}$,
the separate within-group slopes, $\beta_i$.

%
%
\begin{figure*}[b]

\includegraphics{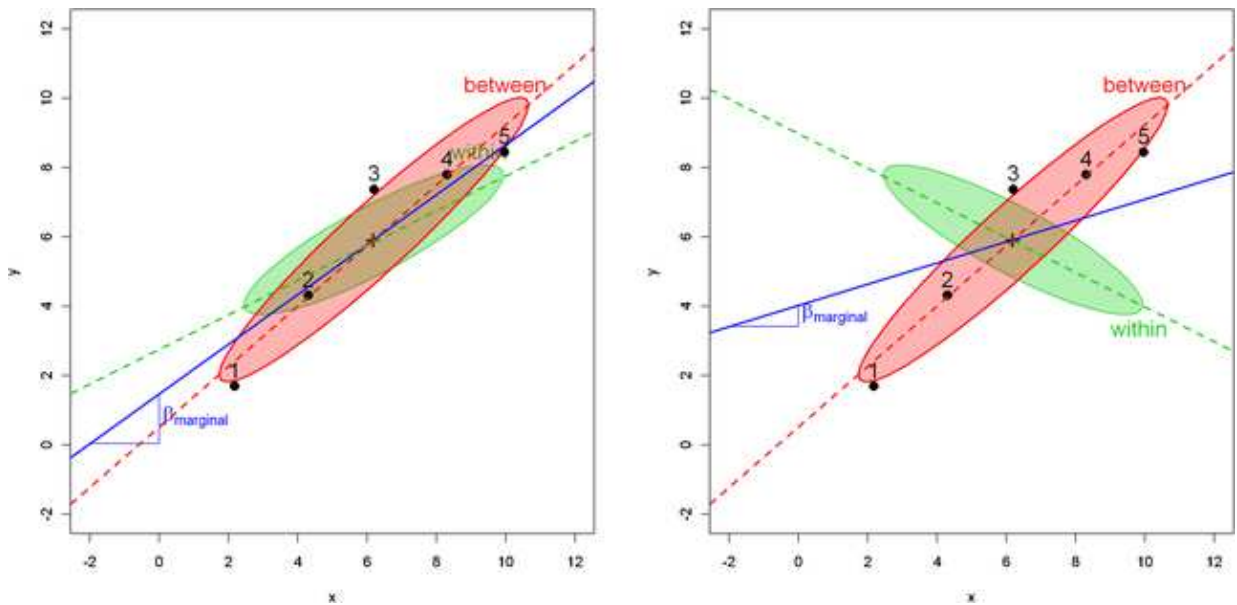}

\caption{Visual demonstration that $\bolds{\beta}_{\mathrm
{marginal}}$ lies between $\bolds{\beta}_{\mathrm{within}}$ and
$\bolds{\beta}_{\mathrm{between}}$.
Each panel shows an HE plot for the MANOVA model $(x, y) \sim\mathrm
{group}$, in which the within and between ellipses are identical to
those in Figure~\protect\ref{figbetween-within}, except for scale.}
\label{figbetween-HE}
\end{figure*}

Figure~\ref{figbetween-within} illustrates these estimates in a
simulation of five groups, with $n_i=10$, means
$\bar{x}_i = 2 i + \mathcal{U}(-0.4, 0.4)$ and
$\bar{y}_i = \bar{x}_i + \mathcal{N}(0, 0.5^2)$,
so that\break $r_{\mathrm{between}} \approx0.95$.
Here $\mathcal{U}(a,b)$ represents the uniform distribution between
$a$ and $b$, and
$\mathcal{N}(\mu,\sigma^2)$ represents the normal distribution with
mean $\mu$ and
variance $\sigma^2$.
For simplicity, we have set the within-group covariance matrices to be
identical in all groups, with
$\Var(x)=6$, $\Var(y)=2$ and $\Cov(x,y)=\pm3$ in the left and right
panels, respectively, giving
$r_{\mathrm{within}} = \pm0.87$.

In the left panel, the conditional, within-group slope is smaller than
the ecological, between-group slope,
reflecting the smaller within-group than between-group correlation.
In general, however, it can be shown that
\[
\bolds{\beta}_{\mathrm{marginal}} \in[\bolds{\beta}_{\mathrm
{within}} , \bolds{
\beta}_{\mathrm{between}} ] ,
\]
which is also evident in the right panel, where the within-group slope
is negative.
This result follows from the fact that the marginal data ellipse for
the total sample
has a shape that is a convex combination (weighted average) of the
average within-group
covariance of $(x, y)$, shown by the green ellipse in Figure~\ref
{figbetween-within},
and the covariance of the means $(\bar{x}_i, \bar{y}_i)$, shown by
the red between-group ellipse.
In fact, the between and within data ellipses in Figure~\ref
{figbetween-within}
are just (a scaling of) the $\mathbf{H}$ and $\mathbf{E}$ ellipses in an
hypothesis-error (HE) plot for the
MANOVA model, $(x, y) \sim\mathrm{group}$, as will be developed in
Section~\ref{secmlm}.
See Figure~\ref{figbetween-HE} for a visual demonstration, using the same
data as in Figure~\ref{figbetween-within}.

The right panels of Figures~\ref{figbetween-within} and~\ref
{figbetween-HE}
provide a prototypical illustration of Simpson's paradox,
where $\beta_{\mathrm{within}}$ and $\beta_{\mathrm{marginal}}$ can
have opposite signs. Underlying this is a
more general \emph{marginal fallacy} (requiring only substantively
different estimates, but not necessarily
different signs)
that can occur when some important factor or covariate is unmeasured
or has been ignored. The fallacy consists of estimating the
unconditional or marginal
relationship ($\beta_{\mathrm{marginal}}$) and believing that it
reflects the conditional relationship, or that
those pesky ``other'' variables will somehow average out. In practice,
the marginal fallacy probably occurs most
often when one views a scatterplot matrix of $(y, x_1, x_2, \ldots)$
and believes that the slopes of
relationships in the separate panels reflect the pairwise conditional
relationships with other variables
controlled. In a regression context, the antidote to the marginal
fallacy is the added-variable
plot (described in Section~\ref{secavp}),
which displays the conditional relationship between the response and a
predictor directly, controlling for all other predictors.

%
%
\begin{figure*}
\centering
\begin{tabular}{@{}cc@{}}

\includegraphics{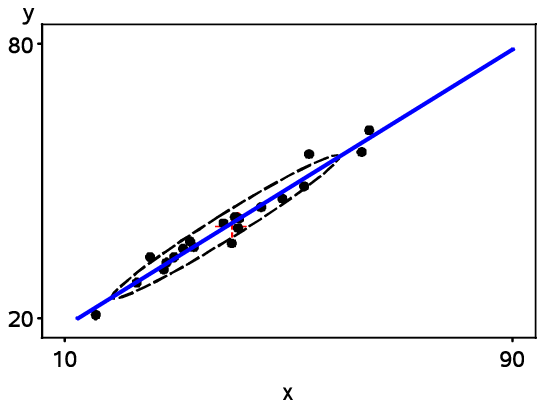}
 & \includegraphics{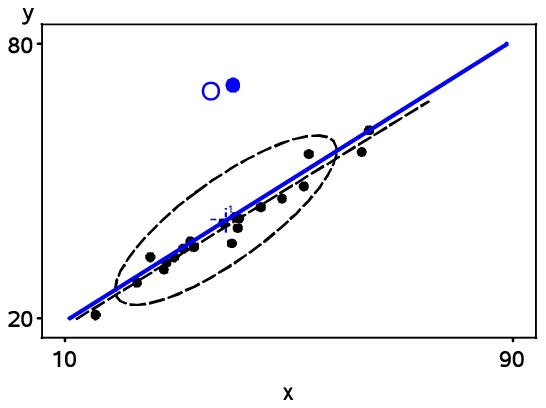}\\
\footnotesize{(a) Original data} & \footnotesize{(b) Low leverage, Outlier}\\[6pt]

\includegraphics{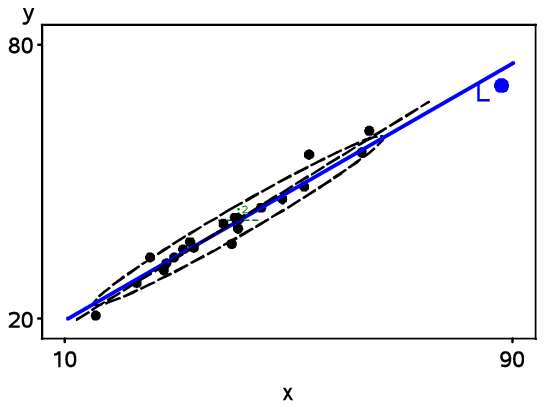}
 & \includegraphics{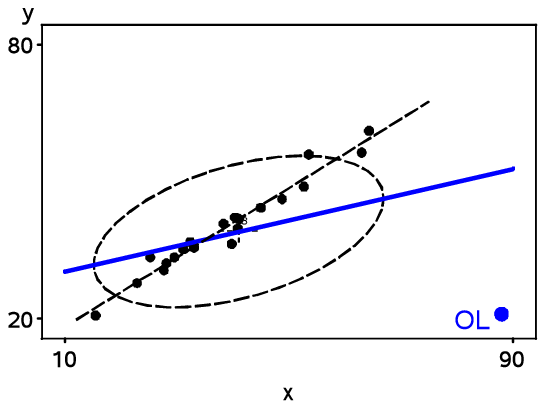}\\
\footnotesize{(c) High leverage, good fit} & \footnotesize{(d) High leverage, Outlier}
\end{tabular}
\caption{Leverage-Influence quartet with data ellipses. \textup{(a)}
Original data;
\textup{(b)} adding one low-leverage outlier (O); \textup{(c)} adding
one ``good''
leverage point (L);
\textup{(d)} adding one ``bad'' leverage point (OL).
In panels \textup{(b)--(d)} the dashed black line is the fitted line
for the original
data, while the thick solid blue line reflects the regression including
the additional point.
The data ellipses show the effect of the additional point on precision.}
\label{figlevdemo21}
\end{figure*}

The right panels of Figures~\ref{figbetween-within} and~\ref
{figbetween-HE} also illustrate Robinson's paradox [\citet{Robinson1950}],
where $\beta_{\mathrm{within}}$ and $\beta_{\mathrm{between}}$ can
have opposite signs.%
\footnote{William Robinson (\citeyear{Robinson1950}) examined the
relationship between literacy rate and percentage
of foreign-born immigrants in the U.S. states from the 1930 Census.
He showed that there was a surprising
positive correlation, $r_{\mathrm{between}}= 0.526$ at the state level,
suggesting that foreign birth was associated with greater literacy;
at the individual level, the correlation $r_{\mathrm{within}}$ was
$-0.118$, suggesting the opposite.
An explanation for the paradox was that immigrants tended to settle in
regions of greater than
average literacy.}
The more general \emph{ecological fallacy} [e.g., \citet
{Lichtman1974}, \citet{Kramer1983}]
is to draw conclusions from aggregated data, estimating
$\beta_{\mathrm{between}}$ or $r_{\mathrm{between}}$, believing that
they reflect relationships
at the individual level, estimating $\beta_{\mathrm{within}}$ or
$r_{\mathrm{within}}$.
Perhaps the earliest instance of this was Andr\'e-Michel Guerry's
(\citeyear{Guerry1833}) use of thematic maps of
France depicting rates of literacy, crime, suicide and other ``moral
statistics'' by department to argue
about the relationships of these moral variables as if they reflected
individual behavior.%
\footnote{Guerry was certainly aware of the logical problem of
ecological inference, at least in general terms
[\citet{Friendly07guerry}], and carried out several side analyses to
examine potential confounding
variables.}
As can be seen in Figure~\ref{figbetween-within}, the ecological fallacy
can often be resolved
by accounting for some confounding variable(s) that vary between groups.

Finally, there are situations where only a subset of the relevant data
are available (e.g.,
one group in Figure~\ref{figbetween-within}) or when the relevant data
are available only
at the individual level,
so that only the conditional relationship,
$\beta_{\mathrm{within}}$, can be estimated. The \emph{atomistic fallacy}
(also called the \emph{fallacy of composition} or the
\emph{individualistic fallacy}), for example, \citet{Alker1969},
\citet{Riley1963},
is the inverse to the
ecological fallacy and consists of believing that one can draw conclusions
about the ecological relationship, $\beta_{\mathrm{between}}$, from
the conditional one.

The atomistic fallacy occurs most often in the context of multilevel
models [\citet{Diez-Roux1998}]
where it is desired to draw inferences regarding variability of
higher-level units
(states, countries) from data collected from lower-level units.
For example, imagine that the right panel of Figure~\ref
{figbetween-within} depicts the negative
relationship of mortality from heart disease ($y$) with individual
income ($x$) for
individuals within countries. It would be fallacious to infer that the
same slope
(or even its sign) applies to a between-country analysis of heart
disease mortality vs.
GNP per capita. A positive value of $\beta_{\mathrm{between}}$ in
this context might
result from the fact that, across countries, higher GNP per capita is
associated with
less healthy diet (more fast food, red meat, larger portions), leading
to increased heart disease.

\subsection{Leverage, Influence \emph{and} Precision}

The topic of leverage and influence in regression is often introduced
with graphs
similar to Figure~\ref{figlevdemo21}, what we call
the ``leverage-influence quartet.''
In these graphs, a bivariate sample of $n=20$ points was first generated
with $x \sim\mathcal{N}(40, 10^2)$ and $y \sim10 + 0.75 x + \mathcal
{N}(0, 2.5^2)$.
Then, in each of
panels (b)--(d) a single point was added at the locations shown, to
represent, respectively,
a low-leverage point with a large residual,%
\footnote{In this context, a residual is ``large'' when the point in
question deviates substantially from the
regression line for the rest of the data---what is sometimes termed a
``deleted residual;'' see below.}
a high-leverage point with small residual
(a~``good'' leverage point) and a high-leverage point with large residual
(a ``bad'' leverage point). The goal is to visualize how leverage
[$\propto(x-\bar{x})^2$] and
residual ($y - \hat{y}^\star_i$) (where $\hat{y}^\star_{i}$ is the
fitted value for observation $i$, computed on the basis of an auxiliary
regression in which observation $i$ is deleted) combine to produce
influential points---those that affect
the estimates of $\bolds{\beta} = (\beta_0 , \beta_1)^{\mathsf{T}}$.

The ``standard'' version of this graph shows \emph{only} the fitted
regression lines for each
panel. So, for the moment, ignore the data ellipses in the plots.
The canonical, first-moment-only, story behind the standard version is
that the points added in panels
(b) and (c) are not harmful---the fitted line does not change very much
when these
additional points are included. Only the bad leverage point, ``OL,'' in
panel (d) is harmful.

%
%
\begin{figure}

\includegraphics{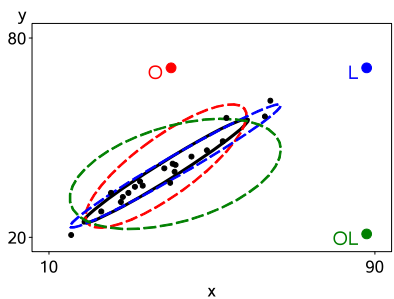}

\caption{Data ellipses in the Leverage-Influence quartet. This graph
overlays the data ellipses
and additional points from the four panels of Figure~\protect\ref
{figlevdemo21}. It can be seen that only the
OL point affects the slope, while the O and L points affect precision
of the estimates in opposite
directions.}%
\label{figlevdemo22}
\end{figure}

Adding the data ellipses to each panel immediately makes it clear that
there is a second-moment
part to the story---the effect of unusual points on the \emph
{precision} of our estimates
of $\bolds{\beta}$. Now,
we see \emph{directly} that there is a big difference in impact between
the low-leverage outlier [panel (b)] and the high-leverage,
small-residual case [panel (c)],
even though their effect on coefficient estimates is negligible.
In panel (b), the single outlier inflates the estimate of residual
variance (the size of the
vertical slice of the data ellipse at $\bar{x}$).

To make the added value of the data ellipse more apparent, we overlay
the data ellipses from
Figure~\ref{figlevdemo21} in a single graph, shown in
Figure~\ref{figlevdemo22}, to allow direct comparison. Because you now
know that regression lines
can be visually estimated as the locus of vertical tangents, we
suppress these lines in the
plot to focus on precision. Here, we can also see why the high-leverage
point ``L'' [added in panel (c) of Figure~\ref{figlevdemo21}] is
called a
``good leverage point.''
By increasing the standard deviation of $x$, it makes the data ellipse
somewhat more elongated,
giving increased precision of our estimates of $\bolds{\beta}$.

Whether a ``good'' leverage point is \emph{really} good depends upon
our faith in the regression model (and in the point),
and may be regarded either as increasing the precision of $\hat
{\bolds{\beta}}$ or providing an illusion of precision.
In either case, the data ellipse for the modified data shows the effect
on precision directly.

\subsection{\texorpdfstring{Ellipsoids in Data Space and $\beta$ Space}
{Ellipsoids in Data Space and beta Space}}\label{secbetaspace}

It is most common to look at data and fitted models in ``data space,''
where axes correspond to
variables, points represent observations, and fitted models are plotted
as lines (or planes) in this space.
As we've suggested, data ellipsoids provide informative summaries of
relationships in data space.
For linear models, particularly regression models with quantitative
predictors, there is another space---``$\bolds{\beta}$ space''---that
provides deeper views of models and the relationships among them.
In $\bolds{\beta}$ space, the axes pertain to coefficients and points
are models (true, hypothesized, fitted) whose coordinates
represent values of parameters.

%
\begin{figure*}

\includegraphics{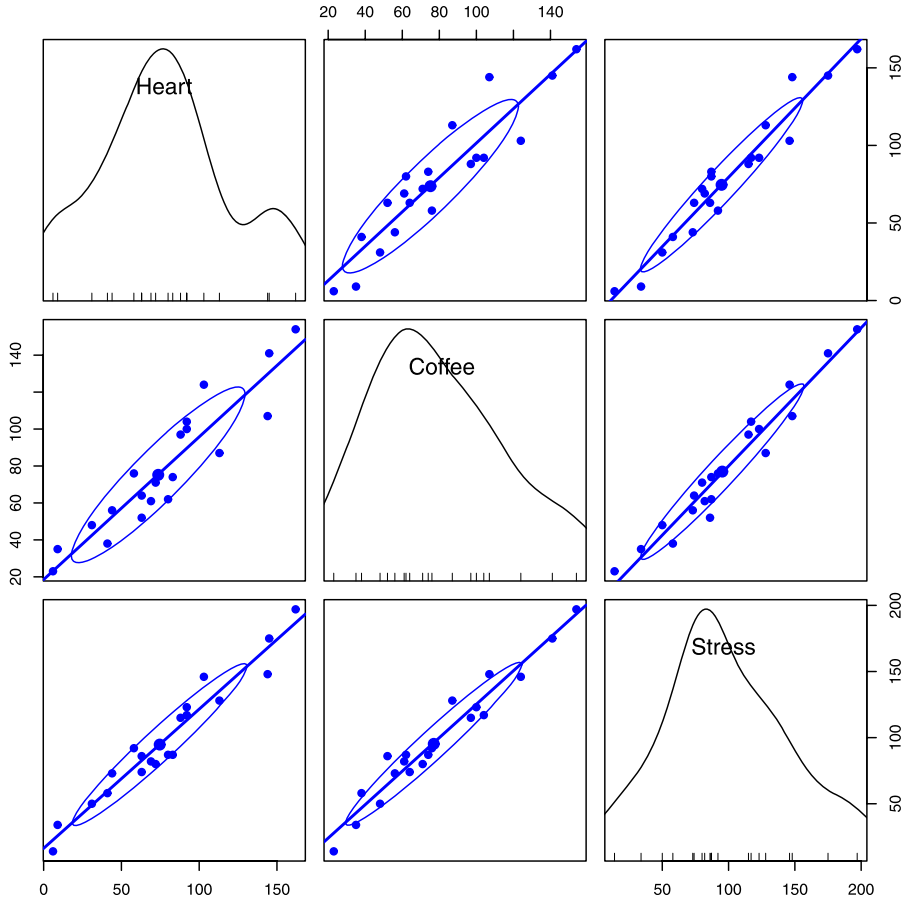}

\caption{Scatterplot matrix, showing the pairwise relationships among
Heart ($y$), Coffee ($x_1$) and Stress ($x_2$),
with linear regression lines and 68\% data ellipses for the marginal
bivariate relationships.
}%
\label{figvis-reg-coffee11}\vspace*{-3pt}
\end{figure*}

In the sense described below, data space and $\bolds{\beta}$ space are
\emph{dual} to each other.
In simple linear regression, for example, each line in data space
corresponds to a point in $\bolds{\beta}$ space,
the set of points on any line in $\bolds{\beta}$ space corresponds to a
pencil of lines through a given point
in data space, and the proposition that every pair of points defines a
line in one space corresponds to
the proposition that every two lines intersect in a point in the other space.

Moreover, ellipsoids in these spaces are dual and inversely related to
each other.
In data space, joint confidence intervals for the mean vector or joint
prediction
regions for the data are given by the ellipsoids $(\bar{x}_1, \bar
{x}_2)^{\mathsf{T}}\oplus c \sqrt{\mathbf{S}}$.
In the dual $\bolds{\beta}$ space, joint confidence regions for the
coefficients of
a response variable $y$ on $(x_1, x_2)$
are given by ellipsoids of the form $\widehat{\bolds{\beta}} \oplus c
\sqrt{\mathbf{S}^{-1}}$.
We illustrate these relationships in the example below.

Figure~\ref{figvis-reg-coffee11} shows a scatterplot matrix among the
variables
Heart ($y$), an index of cardiac damage, Coffee ($x_1$), a measure of daily
coffee consumption, and Stress ($x_2$), a measure of occupational
stress, in a contrived
sample of $n=20$. For the sake of the example we assume that the main
goal is
to determine whether or not coffee is good or bad for your heart, and stress
represents one potential confounding variable among others (age,
smoking, etc.)
that might be useful to control statistically.

The plot in Figure~\ref{figvis-reg-coffee11} shows only the marginal
relationship
between each pair of variables. The mar\-ginal message seems to be that
coffee is
bad for your heart, stress is bad for your heart and coffee consumption is
also related to occupational stress.
Yet, when we fit both variables together, we obtain the following results,
suggesting that coffee is good for you (the coefficient for coffee is now
negative, though nonsignificant). How can this be? (See Table
\ref{tab2}).

%
\begin{table}
\caption{Coefficients and tests for the joint model predicting heart disease from coffee and stress}\label{tab2}
\tabcolsep=0pt
\begin{tabular*}{\columnwidth}{@{\extracolsep{\fill}}lcccc@{}}
 \hline
& \textbf{Estimate} ($\bolds{\widehat{\beta}}$) & \textbf{Std. error} & $\bolds{t}$ \textbf{value} & $\bolds{\operatorname{Pr}(>|t|)}$
\\
\hline
Intercept & $-7.7943$ & 5.7927 & $-1.35$ & 0.1961 \\
Coffee & $-0.4091$ & 0.2918 & $-1.40$ & 0.1789 \\
Stress & \phantom{$-$}1.1993 & 0.2244 & \phantom{$-$}5.34 & 0.0001 \\
\hline
\end{tabular*}
\end{table}
%

%
%
\begin{figure*}[b]

\includegraphics{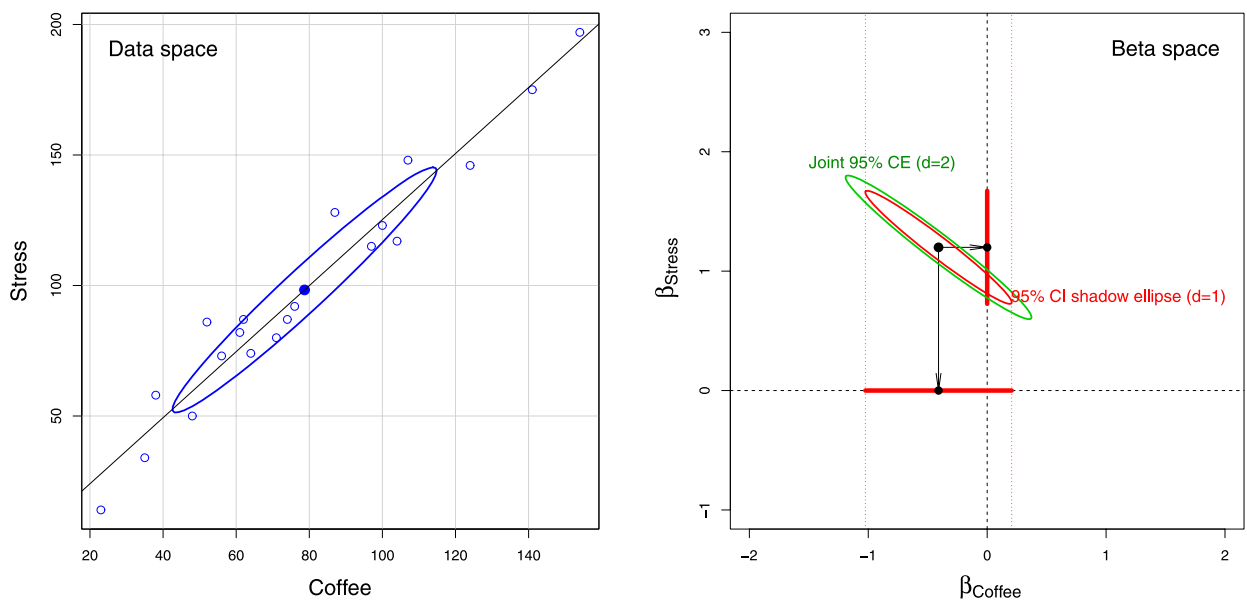}

\caption{Data space and $\bolds{\beta}$ space representations of
Coffee and Stress.
Left: Standard (40\%) data ellipse. Right: Joint 95\% confidence
ellipse (green) for
($\beta_{\mathrm{Coffee}}, \beta_{\mathrm{Stress}}$), CI ellipse
(red) with 95\% univariate shadows.
}%
\label{figvis-reg-coffee12}
\end{figure*}

Figure~\ref{figvis-reg-coffee12} shows the relationship between the
predictors in data space and how this translates into joint and
individual confidence intervals for the coefficients in
$\bolds{\beta}$ space. The left panel is the same as the corresponding
(Coffee, Stress) panel in Figure~\ref{figvis-reg-coffee11}, but with
a standard (40\%) data ellipse. The right panel shows the joint 95\% confidence
region and the individual 95\% confidence intervals in $\bolds{\beta}$
space, determined as
\[
\widehat{\bolds{\beta}} \oplus\sqrt{d F^{0.95}_{d, \nu}}
\times s_e \times\mathbf{S}_X^{-1/2},
\]
where $d$ is the number of dimensions for which we want coverage,
$\nu$ is the residual degrees of freedom for $s_e$, and $\mathbf{S}_X$
is the covariance matrix of the predictors.

Thus, the green ellipse in Figure~\ref{figvis-reg-coffee12} is the
ellipse of joint 95\% coverage, using the factor $\sqrt{2 F^{0.95}_{2,
\nu}}$
and covering the true values of ($\beta_{\mathrm{Stress}}, \beta_{\mathrm{Coffee}}$)
in 95\% of samples. Moreover:
\begin{itemize}
\item Any \emph{joint} hypothesis (e.g., $H_0\dvtx\beta_{\mathrm
{Stress}}=1, \beta_{\mathrm{Coffee}}=1$)
can be tested visually, simply by observing\break whether the
hypothesized point, $(1, 1)$ here, lies inside or outside the joint
confidence ellipse.
\item The shadows of this ellipse on the horizontal and vertical axes
give the Scheff\'e joint 95\% confidence intervals for the parameters,
with protection for
simultaneous inference (``fishing'')
in a 2-dimen\-sional space.
\item Similarly, using the factor
$\sqrt{F^{1-\alpha/d}_{1, \nu}} = t^{1-\alpha/2d}_\nu$\break would give an
ellipse whose 1D shadows are $1-\alpha$ Bonferroni confidence intervals
for $d$ posterior hypotheses.
\end{itemize}

Visual hypothesis tests and $d=1$ confidence intervals for the
parameters \emph{separately}
are obtained from the red ellipse in Figure~\ref{figvis-reg-coffee12},
which is scaled by $\sqrt{F^{0.95}_{1, \nu}} = t^{0.975}_\nu$. We call
this the ``confidence-interval generating ellipse'' (or, more
compactly, the ``confidence-interval ellipse'').
The shadows of the confidence-interval ellipse on the axes (thick red
lines) give the
corresponding individual 95\% confidence intervals, which are
equivalent to the (partial,\break Type~III) $t$-tests for each coefficient
given in the
standard multiple regression output shown above.
Thus, controlling for Stress, the confidence interval for the slope for
Coffee includes 0,
so we cannot reject the hypothesis that $\beta_{\mathrm{Coffee}}=0$
in the multiple regression model, as we saw above in the numerical output.
On the other hand, the interval for the slope for Stress excludes the origin,
so we reject the null hypothesis that $\beta_{\mathrm{Stress}}=0$,
controlling for Coffee consumption.\looseness=1

Finally, consider the relationship between the data ellipse and the
confidence ellipse. These have exactly the same shape, but
the confidence ellipse
is exactly a $90^o$ rotation and rescaling of the data ellipse. In
directions in
data space where the slice of the data ellipse is wide---where we have
more information
about the relationship between Coffee and Stress---the projection of
the confidence ellipse is
narrow, reflecting greater precision of the estimates of coefficients.
Conversely, where slice of the the data ellipse is narrow (less
information), the projection of the
confidence ellipse is wide (less precision). See Figure~\ref{figinverse}
for the underlying geometry.

%
%
\begin{figure}

\includegraphics{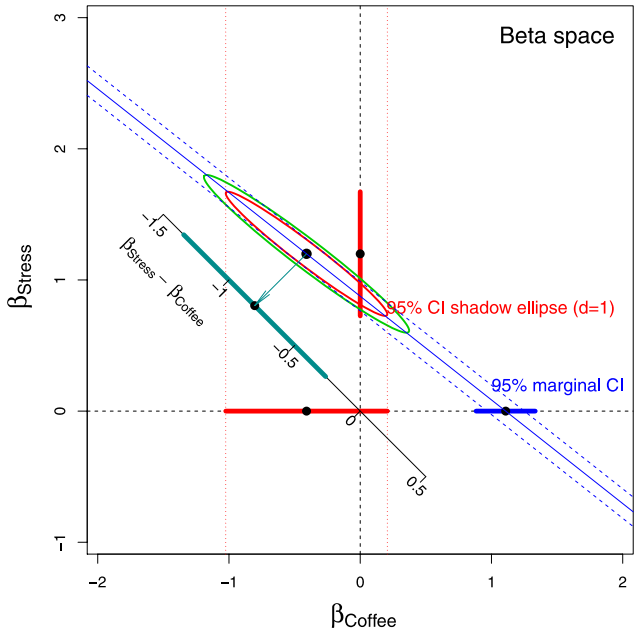}

\caption{Joint 95\% confidence ellipse for ($\beta_{\mathrm
{Coffee}}, \beta_{\mathrm{Stress}}$),
together with the 1D marginal confidence interval for $\beta_{\mathrm
{Coffee}}$
ignoring Stress (thick blue line), and a visual confidence interval for
$\beta_{\mathrm{Stress}} - \beta_{\mathrm{Coffee}}=0$
(dark cyan).
}%
\label{figvis-reg-coffee13}
\end{figure}

The virtues of the confidence ellipse for visualizing hypothesis tests
and interval estimates
do not end here. Say we wanted to test the hypothesis that Coffee was
unrelated to Heart damage
in the \emph{simple} regression ignoring Stress. The (Heart, Coffee)
panel in Figure~\ref{figvis-reg-coffee11}
showed the strong marginal relationship between the variables. This can
be seen in Figure~\ref{figvis-reg-coffee13} as
the oblique projection of the confidence ellipse to the horizontal axis
where $\beta_{\mathrm{Stress}}=0$.
The estimated slope for Coffee in the simple regression is exactly the
oblique shadow of
the center of the ellipse $(\widehat{\beta}_{\mathrm{Coffee}},
\widehat{\beta}_{\mathrm{Stress}})$
through the point where the ellipse has a horizontal tangent onto the
horizontal axis at
$\beta_{\mathrm{Stress}}=0$. The thick blue line in this figure shows
the confidence interval
for the slope for Coffee in the simple regression model. The confidence
interval does not cover the origin, so
we reject $H_0\dvtx\beta_{\mathrm{Coffee}} = 0$ in the simple
regression model.
The oblique shadow of the red 95\% confidence-interval ellipse onto the
horizontal axis
is slightly smaller. How much smaller is a function of the $t$-value of
the coefficient for Stress?


We can go further. As we noted earlier, all linear combinations of
variables or parameters
in data or models correspond graphically to projections (shadows) onto
certain subspaces.
Let's assume that Coffee and Stress were measured on the same scales so
it makes sense
to ask if they have equal impacts on Heart disease in the joint model
that includes them both.
Figure~\ref{figvis-reg-coffee13} also shows an auxiliary axis through the
origin with slope $=-1$
corresponding to values of $\beta_{\mathrm{Stress}} - \beta_{\mathrm
{Coffee}}$. The orthogonal
projection of the coefficient vector on this axis is the point estimate of
$\widehat{\beta}_{\mathrm{Stress}} - \widehat{\beta}_{\mathrm{Coffee}}$
and the shadow of the red ellipse along this axis is the 95\%
confidence interval
for the difference in slopes. This interval excludes 0, so we would
reject the hypothesis
that Coffee and Stress have equal coefficients.

\subsection{Measurement Error}

In classical linear models, the predictors are often considered to be fixed
variables or, if random, to be measured without error and independent
of the regression errors;
either condition, along with the assumption of linearity, guarantees
unbiasedness of the standard OLS estimators.
In practice, of course, predictor variables are often also observed
indicators, subject to error, a fact that is recognized in errors-in-variables
regression models and in more general structural equation models
but often ignored otherwise. Ellipsoids in data space and $\beta$ space
are well suited to showing the effect of measurement error in
predictors on OLS estimates.

%
%
\begin{figure*}

\includegraphics{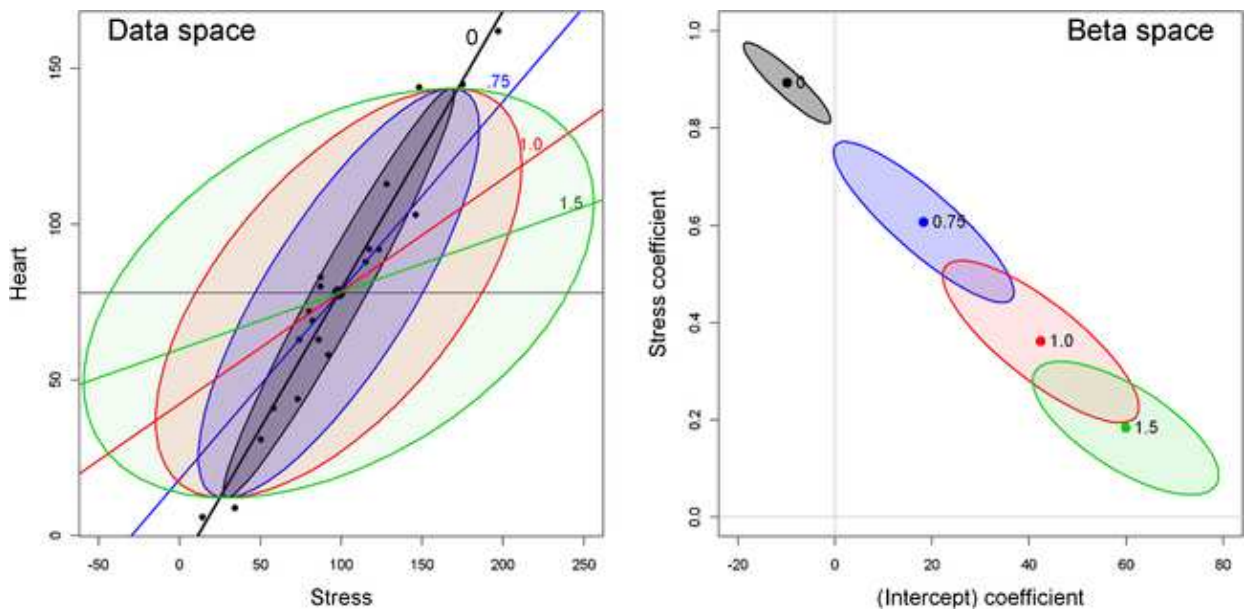}

\caption{Effects of measurement error in Stress on the marginal
relationship between Heart disease and Stress.
Each panel starts with the observed data ($\delta=0$), then adds
random normal error,
$\mathcal{N}(0, \delta\times\mathrm{SD}_{\mathrm{Stress}})$, with $\delta=
\{0.75, 1.0, 1.5\}$, to the value of Stress.
Increasing measurement error biases the slope for Stress toward 0.
Left: 50\% data ellipses; right: 50\% confidence ellipses for $(\beta_0, \beta_{\mathrm{Stress}})$. }
\label{figcoffee-stress}\vspace*{-3pt}
\end{figure*}

The statistical facts are well known, though perhaps counter-intuitive
in certain details:
measurement error in a predictor biases regression coefficients, while
error in the measurement in $y$
increases the standard errors of the regression coefficients but does
not introduce
bias.

In the top row of
Figure~\ref{figvis-reg-coffee11}, adding measurement error to the Heart
disease variable
would expand the data ellipses vertically, but
(apart from random variation)
leaves the slopes of the regression lines unchanged.
Measurement error in a predictor variable, however, biases the corresponding
estimated coefficient toward
zero (sometimes called \emph{regression attenuation}) as well as
increasing standard errors.

Figure~\ref{figcoffee-stress} demonstrates this effect for the marginal
relation between Heart disease and Stress,
with data ellipses in data space and the corresponding confidence
ellipses in $\beta$ space.
Each panel starts with the observed data (the darkest ellipse, marked
$0$), then adds random normal error,
$\mathcal{N}(0, \delta\times\mathrm{SD}_{\mathrm{Stress}})$, with $\delta=
\{0.75, 1.0, 1.5\}$, to the value of Stress,
while keeping the mean of Stress the same.
All of the data ellipses have the same vertical shadows ($\mathrm
{SD}_{\mathrm{Heart}}$), while the horizontal shadows
increase with $\delta$, driving the slope for Stress toward 0.
In $\beta$ space, it can be seen that the estimated coefficients,
$(\beta_0, \beta_{\mathrm{Stress}})$,
vary along a line and approach $\beta_{\mathrm{Stress}}=0$ for
$\delta$ sufficiently large.
The vertical shadows of
ellipses for $(\beta_0, \beta_{\mathrm{Stress}})$ along the $\beta_{\mathrm{Stress}}$ axis
also demonstrate the effects of measurement error
on the standard error of $\beta_{\mathrm{Stress}}$.

\begin{figure}[t]

\includegraphics{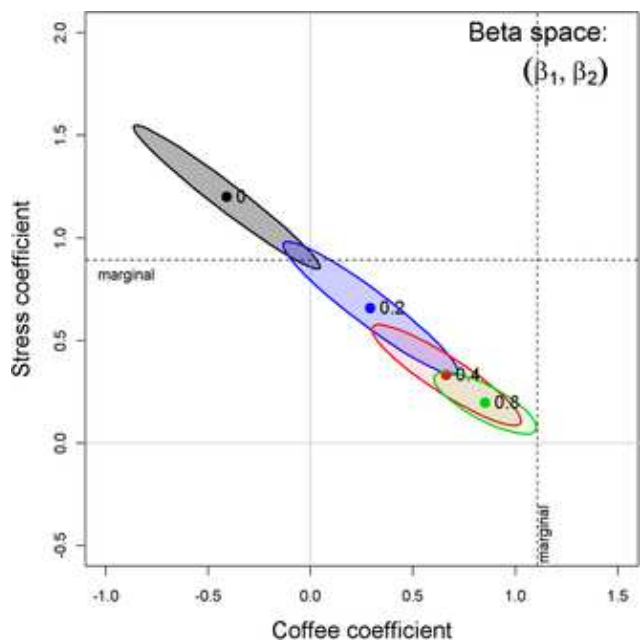}

\caption{Biasing effect of measurement error in one variable (Stress)
on the coefficient of another variable
(Coffee) in a multiple regression. The coefficient for Coffee is driven
toward its value in the marginal
model using Coffee alone, as measurement error in Stress makes it less
informative in the joint model.}\vspace*{-3pt}
\label{figcoffee-measerr}
\end{figure}

Perhaps less well-known, but both more surprising and interesting, is
the effect that measurement error in one variable,
$x_1$,
has on the estimate of the coefficient for an \emph{other} variable,
$x_2$, in a multiple regression model.
Figure~\ref{figcoffee-measerr}
shows the confidence ellipses for $(\beta_{\mathrm{Coffee}}, \beta_{\mathrm{Stress}})$ in the multiple regression
predicting Heart disease, adding random normal error
$\mathcal{N}(0, \delta\times\mathrm{SD}_{\mathrm{Stress}})$, with $\delta=
\{0, 0.2, 0.4, 0.8\}$, to the value of Stress
alone.
As can be plainly seen, while this measurement error in Stress
attenuates its coefficient,
it also has the effect of biasing the coefficient for Coffee toward
that in the \emph{marginal}
regression of Heart disease on Coffee alone.

%
%
%
%

\begin{figure*}

\includegraphics{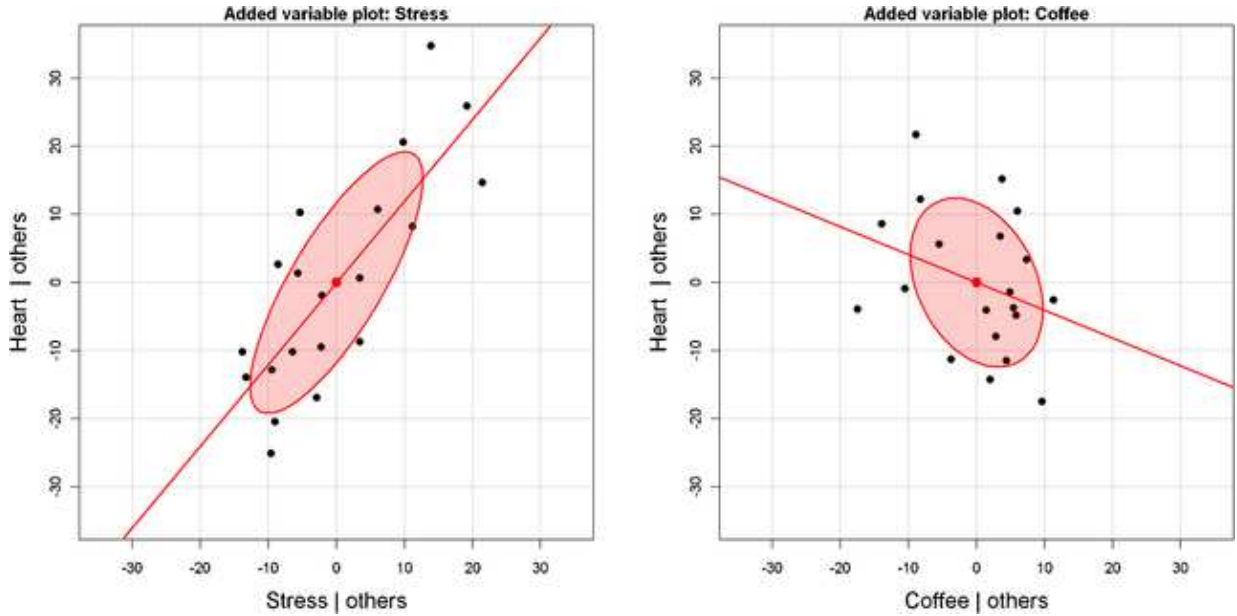}

\caption{Added variable plots for Stress and Coffee in the multiple
regression predicting Heart disease.
Each panel also shows the 50\% conditional data ellipse for residuals
$(\mathbf{x}_k^\star, \mathbf{y}^\star)$, shaded red.}
\label{figcoffee-avplot-A}
\end{figure*}



\subsection{Ellipsoids in Added-Variable Plots}\label{secavp}

In contrast to the marginal, bivariate views of the relationships of a
response to several predictors (e.g., such as shown in the top row of
the scatterplot matrix in Figure~\ref{figvis-reg-coffee11}), \emph
{added-variable plots}
(aka \emph{partial regression plots}) show the partial relationship
between the response and
each predictor, where the effects of all other predictors have been
controlled or adjusted for.
Again we find that such plots have remarkable geometric properties, particularly
when supplemented by ellipsoids.

Formally, we express the fitted standard linear model in vector form as
$\widehat{\mathbf{y}} \equiv\widehat{\mathbf{y}} \given\mathbf
{X} =
\widehat{\beta}_0 \mathbf{1} + \widehat{\beta}_1 \mathbf{x}_1 +
\widehat{\beta}_2 \mathbf{x}_2 + \cdots+ \widehat{\beta}_p \mathbf{x}_p$,
with model matrix $\mathbf{X} = [ \mathbf{1}, \mathbf{x}_1,
\ldots,\break
\mathbf{x}_p ]$.
Let $\mathbf{X}_{[-k]}$ be the model matrix omitting the column for
variable $k$.
Then, algebraically, the added variable plot for variable $k$ is the
scatterplot of the residuals $(\mathbf{x}^\star_k, \mathbf{y}^\star
)$ from
two auxillary regressions,%
\footnote{These quantities can all be computed [\citet
{VellemanWelsh81}] from the results of a
single regression for the full model.}
fitting $\mathbf{y}$ and $\mathbf{x}_k$ from $\mathbf{X}_{[-k]}$,
\begin{eqnarray*}
\mathbf{y}^\star& \equiv& \mathbf{y} \given\mathrm{others} = \mathbf{y}
- \widehat{\mathbf{y}} \given\mathbf{X}_{[-k]},
\\
\mathbf{x}^\star_k & \equiv& \mathbf{x}_k
\given\mathrm{others} = \mathbf{x}_k - \widehat{\mathbf{x}}_k
\given\mathbf{X}_{[-k]} .
\end{eqnarray*}

Geometrically, in the space of the observations,\footnote{The ``space
of the observations'' is yet a third, $n$-dimensional, space, in which
the observations are the axes and each variable is represented as a
point (or vector). See, for example, \citeauthor{Fox2008} [(\citeyear
{Fox2008}), Chapter~10].}
the fitted vector $\widehat{\mathbf{y}}$ is the orthogonal projection of
$\mathbf{y}$
onto the subspace spanned by $\mathbf{X}$. Then $\mathbf{y}^\star$ and
$\mathbf{x}^\star_k$ are the projections onto
the orthogonal complement of the subspace spanned by $\mathbf
{X}_{[-k]}$, so
the simple regression of $\mathbf{y}^\star$ on $\mathbf{x}^\star_k$ has
slope $\hat{\beta}_k$ in the full model,
and the residuals from the line
$\widehat{\mathbf{y}}{}^\star= \hat{\beta}_k \mathbf{x}^\star_k$
in this
plot are identically
the residuals from the overall regression of $\mathbf{y}$ on $\mathbf{X}$.

\begin{figure*}[t]

\includegraphics{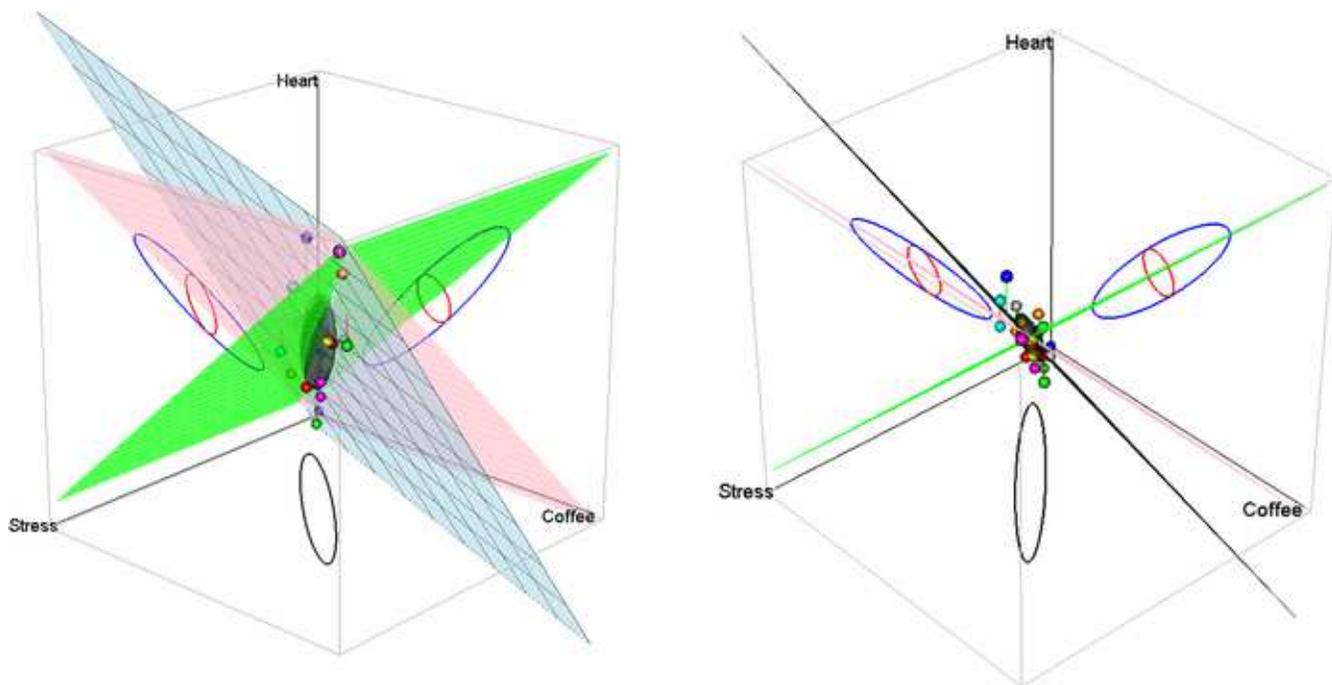}

\caption{3D views of the relationship between Heart, Coffee and
Stress, showing the three regression planes
for the marginal models, Heart $\sim$ Coffee (green), Heart $\sim$
Stress (pink), and the joint model, Heart $\sim$ Coffee $+$ Stress
(light blue).
Left: a standard view; right: a view showing all three regression
planes on edge.
The ellipses in the side panels are 2D projections of the standard
conditional (red) and marginal (blue) ellipsoids, as shown in
Figure~\protect\ref{figcoffee-avplot-B}.}\vspace*{-5pt}
\label{figcoffee-av3D}
\end{figure*}

Another way to describe the added-variable plot (AVP) for $x_k$ is as a
2D projection of the space of
$(\mathbf{y}, \mathbf{X})$, viewed in a plane projecting the data along
the intersection of two hyperplanes:
the plane of the regression of $\mathbf{y}$ on all of $\mathbf{X}$,
and the
plane of regression of
$\mathbf{y}$ on $\mathbf{X}_{[-k]}$. A third plane, that of the regression
of $x_k$ on $\mathbf{X}_{[-k]}$,
also intersects in this space and defines the horizontal axis in the AVP.
This is illustrated in Figure~\ref{figcoffee-av3D}, showing one view
defined by the intersection of
the three planes in the right panel.%
\footnote{Animated 3D movies of this plot are included among the
supplementary materials for this paper.}

%
%
%


Figure~\ref{figcoffee-avplot-A} shows added-variable plots for Stress and
Coffee in the multiple regression predicting Heart disease,
supplemented by data ellipses for the residuals $(\mathbf{x}_k^\star,
\mathbf{y}^\star)$. With reference to the properties
of data ellipses in marginal scatterplots (see Figure~\ref
{figellipses-demo}), the following visual properties (among others)
are useful in this discussion. These results follow simply from
translating ``marginal'' into ``conditional'' (or ``partial'')
in the present context.
The essential idea is that the data ellipse of the AVP for $(x_k^\star
, y^\star)$ is to the estimate of a
coefficient in a multiple regression as
the data ellipse of $(x, y)$ is to simple regression. Thus:

%
\begin{figure*}

\includegraphics{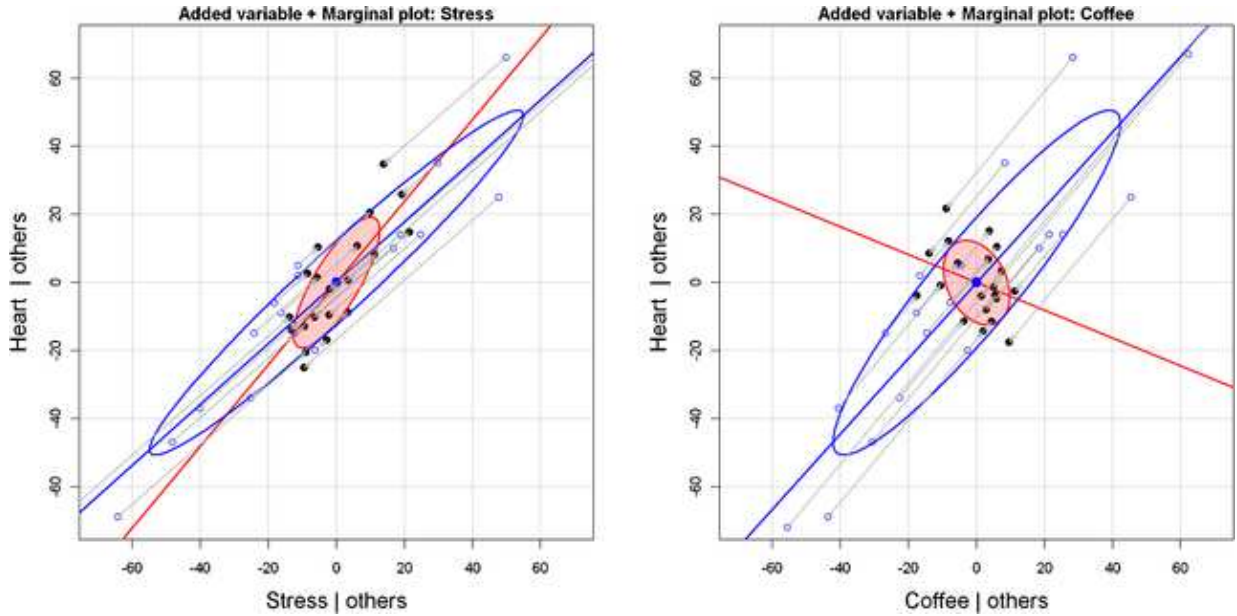}

\caption{Added-variable $+$ marginal plots for Stress and Coffee in
the multiple regression predicting Heart disease.
Each panel shows the 50\% conditional data ellipse for $x_k^\star,
y^\star$ residuals (shaded, red) as well as the marginal 50\%
data ellipse for the $(x_k, y)$ variables, shifted to the origin.
Arrows connect the mean-centered marginal points (open circles) to the
residual points (filled circles).}
\label{figcoffee-avplot-B}
\end{figure*}

\begin{longlist}[(1)]

\item[(1)] The simple regression least squares fit of $\mathbf{y}^\star$ on
$\mathbf{x}_k^\star$ has slope $\hat{\beta}_k$,
the partial slope for $x_k$ in the full model (and intercept = 0).

\item[(2)] The residuals, $(\mathbf{y}^\star- \widehat{\mathbf{y}}{}^\star)$,
shown in this plot are the residuals for $\mathbf{y}$ in the full model.

\item[(3] The correlation between $\mathbf{x}_k^\star$ and $\mathbf
{y}^\star
$, seen in the shape of the
data ellipse for these variables,
is the partial correlation between $y$ and $x_k$ with the other
predictors in $\mathbf{X}_{[-k]}$ partialled out.

\item[(4)] The horizontal half-width of the AVP data ellipse is proportional
to the conditional standard deviation of
$x_k$ remaining after all other predictors have been accounted for,
providing a visual interpretation
of variance inflation due to collinear predictors, as we describe below.

\item[(5)] The vertical half-width of the data ellipse is proportional to
the residual standard deviation $s_e$ in the multiple regression.

\item[(6)] The squared horizontal positions, $(\mathbf{x}_k^\star)^2$, in the
plot give the partial contributions
to leverage on the coefficient $\hat{\beta}_k$ of $x_k$.

\item[(7)] Items (3) and (7) imply that
the AVP for $x_k$ shows the \emph{partial} influence of individual
observations on the coefficient $\hat{\beta}_k$,
in the same way as in Figure~\ref{figlevdemo21} for marginal models.
These influence statistics are
often shown numerically
as DFBETA statistics [\citet{Belsley-etal80}].

\item[(8)] The last three items imply that the collection of added-variable
plots for $\mathbf{y}$ and
$\mathbf{X}$ provide an easy way to visualize the leverage and
influence that individual observations---and indeed the joint influence
of subsets of observations---have on
the estimation of \emph{each} coefficient in a given model.
\end{longlist}

Elliptical insight also permits us to go further, to depict the
relationship between conditional and marginal views
directly.
Figure~\ref{figcoffee-avplot-B} shows the same added-variable plots for
Heart disease on Stress and Coffee
as in Figure~\ref{figcoffee-avplot-A} (with a zoomed-out scaling), but
here we also overlay the
marginal data ellipses for $(x_k, y)$, and marginal regression lines
for Stress and Coffee separately. In 3D data space,
these are the shadows (projections) of the data ellipsoid onto the
planes defined by the
partial variables. In 2D AVP space, they are just the marginal data
ellipses translated to
the origin.

The most obvious feature of Figure~\ref{figcoffee-avplot-B} is that the
AVP for Coffee has a negative slope in the conditional
plot (suggesting that controlling for Stress, coffee consumption is
good for your heart), while
in the marginal plot increasing coffee seems to be bad for your heart.
This serves as a
regression example of Simpson's paradox, which we considered earlier.

Less obvious is the fact that
the marginal and AVP ellipses are easily visualized as a shadow versus
a slice of the full data ellipsoid.
Thus, the AVP ellipse must be contained in the marginal ellipse, as we
can see in Figure~\ref{figcoffee-avplot-B}.
If there are only two $x$'s, then the AVP ellipse must touch the
marginal ellipse at two points.
The shrinkage of the intersection of the AVP ellipse with the $y$ axis
represents improvement in fit due to other $x$'s.

More importantly, the shrinkage of the width (projected onto a
horizontal axis) represents the
square root of the variance inflation factor (VIF), which can be shown
to be the ratio of the horizontal
width of the marginal ellipse of $(x_k, y)$, with standard deviation
$s(x_k)$ to the width of the conditional
ellipse of $(x_k^\star, y^\star)$, with standard deviation $s(x_k
\given\mathrm{others})$.
This geometry implies interesting constraints among the three
quantities: improvement in fit, VIF, and change from the marginal to
conditional slope.

Finally, Figure~\ref{figcoffee-avplot-B} also shows how conditioning on
other predictors works for individual
observations, where each point of $(\mathbf{x}_k^\star, \mathbf
{y}^\star
)$ is the image of $(\mathbf{x}_k, \mathbf{y})$
along the path of the marginal regression. This reminds us that the AVP
is a 2D projection of the full space,
where the regression plane of $\mathbf{y}$ on $\mathbf{X}_{[-k]}$ becomes
the vertical axis and
the regression plane of $\mathbf{x}_k$ on $\mathbf{X}_{[-k]}$ becomes the
horizontal axis.

%
%
\begin{table*}
\caption{Multivariate test statistics as functions of the eigenvalues
$\lambda_i$ solving $\operatorname{det} (\mathbf{H} - \lambda
\mathbf{E})=0$
or\break eigenvalues $\rho_i$ solving $\operatorname{det} [\mathbf{H} -
\rho(\mathbf{H}+\mathbf{E})]=0$}\label{tabcriteria}
\begin{tabular*}{\textwidth}{@{\extracolsep{\fill}}lccc@{}}
\hline
\textbf{Criterion} & \textbf{Formula} & \textbf{``mean'' of} $\bolds{\rho}$ & \textbf{Partial} $\bolds{\eta^2}$ \\
\hline
Wilks's $\Lambda$ & $\Lambda= \prod^s_i \frac{1}{1+\lambda_i} =
\prod^s_i (1-\rho_i)$ & geometric & $\eta^2 = 1-\Lambda^{1/s}$ \\
Pillai trace & $V = \sum^s_i \frac{\lambda_i}{1+\lambda_i} = \sum^s_i \rho_i$ & arithmetic & $\eta^2 = \frac{V}{s} $ \\
Hotelling--Lawley trace & $H = \sum^s_i \lambda_i = \sum^s_i \frac
{\rho_i}{1-\rho_i} $ & harmonic & $\eta^2 = \frac{H}{H+s}$ \\
Roy maximum root & $R = \lambda_1 = \frac{\rho_1}{1-\rho_1}$ &
supremum & $ \eta^2 = \frac{\lambda_1}{1+\lambda_1} = \rho_1$ \\
\hline
\end{tabular*}
\end{table*}

\section{Multivariate Linear Models: HE~Plots}\label{secmlm}


Multivariate linear models (\MLM{}s) have a special affinity with
ellipsoids and elliptical geometry,
as described in this section. To set the stage and establish notation,
we consider
the \MLM\ [e.g., \citet{Timm75}] given by
the equation $\mathbf{Y}=\mathbf{XB}+\mathbf{U}$, where $\mathbf
{Y}$ is an $%
n\times p$ matrix of responses in which each column represents a distinct
response variable; $\mathbf{X}$ is the $n\times q$ model matrix of full
column rank for the regressors; $\bolds{\Beta}$ is the $q \times p$ matrix
of regression coefficients or model parameters; and $\mathbf{U}$ is the
$n \times p$
matrix of errors,
with $\mathrm{vec}(\mathbf{U}) \sim\mathcal{N}_p ( \mathbf{0},
\mathbf{I}_n \otimes\bolds{\Sigma} )$,
where $\otimes$ is the Kronecker product.

A convenient feature of the \MLM\ for general multivariate responses
is that
\emph{all} tests of linear hypotheses (for null effects) can be
represented in the form of a general
linear test,
%
%
\begin{equation}
\label{eqmglt} H_0\dvtx\sizedmat{L} {h \times q} \sizedmat{\Beta} {q
\times p} = \sizedmat{0} {h \times p} ,
\end{equation}
where $\mathbf{L}$ is a rank $h \leq q$ matrix of constants whose rows specify
$h$ linear combinations or contrasts
of the parameters to be tested simultaneously
by a multivariate test.

For \emph{any} such hypothesis of the form given in equation~(\ref{eqmglt}),
the analogs of the univariate
sums of\break squares for hypothesis ($\mathrm{SS}_H$) and error ($\mathrm{SS}_E$)
are the $p \times p$ sum of squares and cross-products (SSP) matrices
given by
%
%
\begin{equation}\qquad
\label{eqhmat} \mathbf{H} \equiv\mathbf{SSP}_H = (\mathbf{L}
\widehat{\mathbf{B}})^{\mathsf{T}} \bigl[\mathbf{L} \bigl(\mathbf
{X}^{\mathsf{T}}\mathbf{X} \bigr)^{-} \mathbf{L}^{\mathsf{T}}
\bigr]^{-1} (\mathbf{L} \widehat{ \mathbf{B}})
\end{equation}
and
%
%
\begin{equation}\qquad
\label{eqemat} \mathbf{E} \equiv\mathbf{SSP}_E = \mathbf
{Y}^{\mathsf{T}} \mathbf{Y} - \widehat{\mathbf{B}}{}^{\mathsf{T}}\bigl(
\mathbf{X}^{\mathsf
{T}}\mathbf{X}\bigr) \widehat{ \mathbf{B}} = \widehat{
\mathbf{U}}{}^{\mathsf{T}}\widehat{\mathbf {U}} ,
\end{equation}
where $\widehat{\mathbf{U}} = \mathbf{Y} - \mathbf{X} \widehat
{\mathbf{B}}$
is the matrix of residuals.
Multivariate test statistics (Wilks's $\Lambda$, Pillai trace,
Hotel\-ling--Lawley trace, Roy's maximum root)
for testing equation~(\ref{eqmglt}) are based on the $s = \min(p, h)$ nonzero
\mbox{latent} roots
$\lambda_{1}>\lambda_{2}>\cdots>\lambda_{s}$ of
the matrix $\mathbf{H}$ relative to the matrix $\mathbf{E}$, that is,
the values of $\lambda$ for which $
\operatorname{det} (\mathbf{H}-\lambda\mathbf{E})=0$ or, equivalently,
the latent roots $\rho_i$ for which $\operatorname{det} [\mathbf{H}
- \rho(\mathbf{H}+\mathbf{E})]=0$.
The details are shown in Table~\ref{tabcriteria}.
These measures
attempt to capture how ``large'' $\mathbf{H}$ is, relative to
$\mathbf{E}$ in $s$ dimensions, and correspond to various ``means'' as we
described earlier.
All of these statistics have transformations to $F$ statistics
giving either exact or approximate null-hypothesis $F$ distributions.
The corresponding latent vectors provide a
set of $s$ orthogonal linear combinations of the responses that produce
maximal univariate $F$ statistics for the hypothesis in equation~(\ref
{eqmglt});
we refer to these as the \emph{canonical discriminant} dimensions.

\begin{figure}

\includegraphics{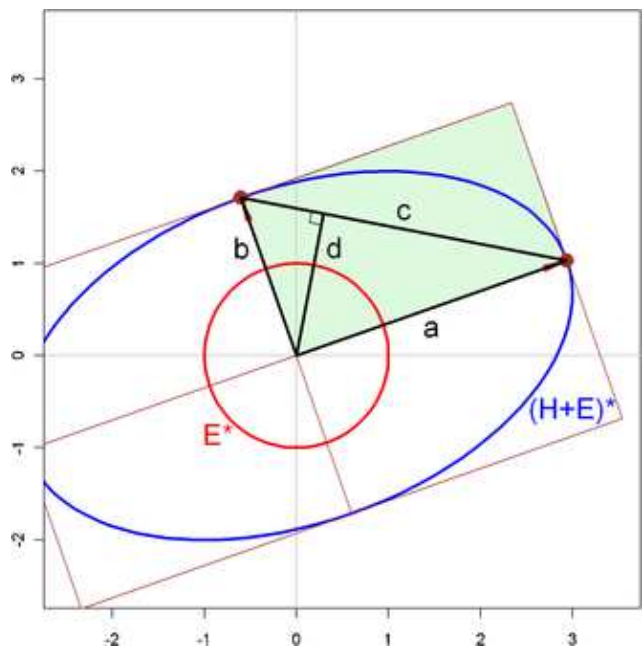}

\caption{Geometry of the classical test statistics used in tests of
hypotheses in multivariate linear models.
The figure shows the representation of the ellipsoid generated by
$(\mathbf{H} + \mathbf{E})$ relative to $\mathbf{E}$
in canonical space where $\mathbf{E}^\star= \mathbf{I}$ and
$(\mathbf{H}+\mathbf{E})^\star$ is the corresponding transformation
of $(\mathbf{H} + \mathbf{E}).$}%
\label{figmtests}
\end{figure}

Beyond the informal characterization of the four classical tests of
hypotheses for multivariate linear
models given in
Table~\ref{tabcriteria}, there is an interesting geometrical representation
that helps one to appreciate their relative power for various alternatives.
This can be illustrated most simply in terms of
the canonical representation, $(\mathbf{H}+\mathbf{E})^\star$,
of the ellipsoid generated by $(\mathbf{H} + \mathbf{E})$ relative to
$\mathbf{E}$,
as shown in Figure~\ref{figmtests} for $p=2$.

%
%

%
%
\begin{figure*}
\centering
\begin{tabular}{@{}cc@{}}

\includegraphics{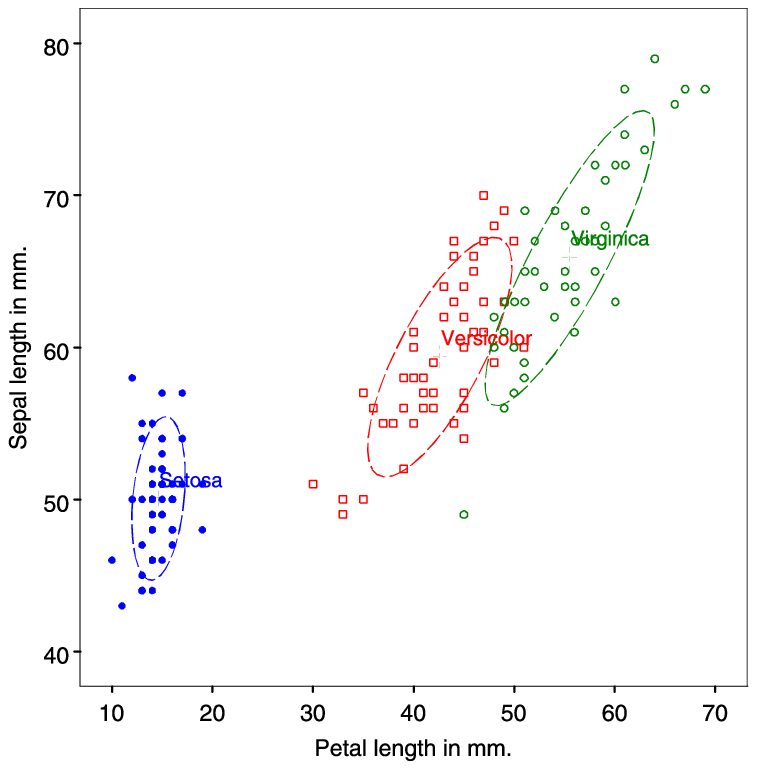}
 & \includegraphics{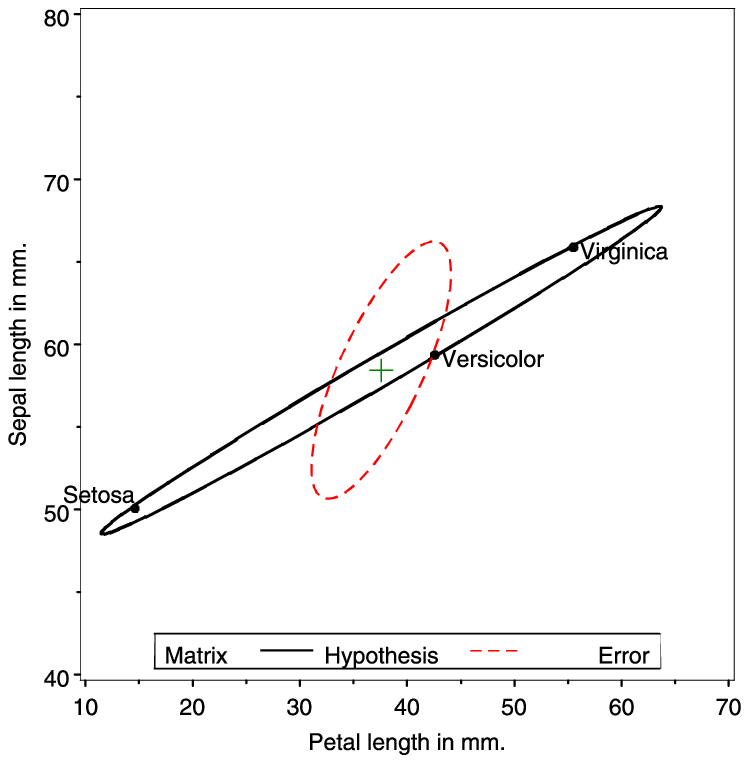}\\
\footnotesize{(a) Data ellipses} & \footnotesize{(b) H and E matrices}
\end{tabular}
\caption{\textup{(a}) Data ellipses and \textup{(b)} corresponding
HE plot for sepal
length and petal length in the iris data set.
The $\mathbf{H}$ ellipse is the data ellipse of the fitted values defined
by the group means, $\bar{\mathbf{y}}_{i \cdot}$
The $\mathbf{E}$ ellipse is the data ellipse of the residuals,
$(\mathbf{y}_{ij} - \bar{\mathbf{y}}_{i \cdot})$.
Using evidence (``significance'') scaling of the $\mathbf{H}$ ellipse, the
plot has the property that
the multivariate test for a given hypothesis is significant by Roy's
largest root test \emph{iff}
the $\mathbf{H}$ ellipse protrudes anywhere outside the $\mathbf{E}$
ellipse.}%
\label{figheplot3a}
\end{figure*}

With $\lambda_i$ as described above,
the eigenvalues and squared radii of $(\mathbf{H}+\mathbf{E})^\star$ are
$\lambda_i + 1$,
so the lengths of the major and minor axes are $a=\sqrt{\lambda_1 +
1}$ and $b=\sqrt{\lambda_2 + 1}$, respectively.
The diagonal of the triangle comprising the segments $a, b$ (labeled
$c$) has length
$c = \sqrt{a^2 + b^2}$.
Finally, a line segment from the origin dropped perpendicularly to the
diagonal joining the two ellipsoid axes is labeled $d$.

In these terms,
Wilks's test, based on $\prod{(1+\lambda_i)^{-1}}$, is equivalent to
a test based on
$a \times b$ which is proportional to the area of the framing
rectangle, shown shaded in Figure~\ref{figmtests}.
The Hotelling--Lawley trace test, based on
$\sum{\lambda_i}$,
is equivalent to a test based on
$c=\sqrt{\sum{\lambda_i} + p}$.
Finally, the Pillai Trace test, based on $\sum{\lambda_i (1+\lambda_i)^{-1}}$, can be shown to be equal to
$2-d^{-2}$ for $p=2$. Thus, it is strictly monotone in $d$ and
equivalent to a test based directly on $d$.

The geometry makes it easy to see that if there is a large discrepancy
between $\lambda_1$ and $\lambda_2$, Roy's test depends
only on $\lambda_1$
while the Pillai test depends more on $\lambda_2$.
Wilks's $\Lambda$ and the Hotelling--Lawley trace criterion are also
functional averages of $\lambda_1$ and $\lambda_2$, with the
former being penalized when $\lambda_2$ is small. In practice, when $s
\le2$, all four test criteria are equivalent, in that
their standard transformations to $F$ statistics are exact and give
rise to identical $p$-values.

\subsection{Hypothesis-Error (HE) Plots}

The essential idea behind HE plots is that any multivariate hypothesis
test, equation~(\ref{eqmglt}), can be represented visually by ellipses (or
ellipsoids beyond 2D) that express
the size of covariation against a multivariate null hypothesis
($\mathbf{H}$) relative to error covariation~($\mathbf{E}$).
The multivariate tests, based on the latent roots of $\mathbf{H}
\mathbf{E}^{-1}$,
are thus translated directly to the sizes of the $\mathbf{H}$ ellipses for
various hypotheses, relative to the size of the $\mathbf{E}$ ellipse.
Moreover, the shape and orientation of these ellipses show something more---the
directions (linear combinations of the responses) that lead to
various effect sizes and significance.

Figure~\ref{figheplot3a} illustrates this idea for two variables from the
iris data set.
Panel (a) shows the data ellipses for sepal length and petal length,
equivalent to
the corresponding plot in Figure~\ref{figscatirisd1}. Panel (b) shows the
HE plot for these
variables from the one-way MANOVA model $\mathbf{y}_{ij} = \bolds{\mu}_i
+ \mathbf{u}_{ij}$
testing equal mean vectors across species, $H_0\dvtx\bolds{\mu}_1 =
\bolds{\mu}_2 = \bolds{\mu}_3$.
Let $\widehat{\mathbf{Y}}$ be the $n \times p$ matrix of fitted values
for this model,
that is, $\widehat{\mathbf{Y}} = \{ \bar{\mathbf{y}}_{i \cdot} \}$.
Then $\mathbf{H}= \widehat{\mathbf{Y}}{}^{\mathsf{T}}\widehat
{\mathbf{Y}}
- n\bar
{\mathbf{y}} \bar{\mathbf{y}}{}^{\mathsf{T}}$ (where $\bar{\mathbf
{y}}$ is the
grand-mean vector), and the $\mathbf{H}$ ellipse in the figure is then
just the
2D projection of the data ellipsoid
of the fitted values, scaled as described below.
Similarly, $\widehat{\mathbf{U}} = \mathbf{Y} - \widehat{\mathbf
{Y}}$, and
$\mathbf{E} = \widehat{\mathbf{U}}{}^{\mathsf{T}}\widehat{\mathbf
{U}} =
(N-g) \mathbf{S}_{\mathrm{pooled}}$, so the $\mathbf{E}$ ellipse is
the 2D projection of the data ellipsoid of the residuals.
Visually, the $\mathbf{E}$ ellipsoid corresponds to shifting the separate
within-group data ellipsoids to the centroid,
as illustrated above in Figure~\ref{figcontiris3}(c).

In HE plots, the $\mathbf{E}$ matrix is first scaled to a covariance matrix
$\mathbf{E}/df_e$, dividing by the error degrees of freedom, $df_e$.
The ellipsoid drawn is
translated to the centroid $\overline{\mathbf{y}}$ of the variables,
giving $\overline{\mathbf{y}} \oplus c \mathbf{E}^{1/2}/df_e$.
This scaling and translation
also allows the means for levels of the factors
to be displayed in the same space,
facilitating interpretation.
In what follows, we show these as
``standard'' bivariate ellipses of 68\% coverage,
using $c=\sqrt{2 F_{2, df_e}^{0.68}}$, except where noted otherwise.

The ellipse for $\mathbf{H}$ reflects the size and orientation of covariation
against the null hypothesis.
In relation to the $\mathbf{E}$ ellipse, the $\mathbf{H}$ ellipse
can be scaled to show either the \emph{effect size} or strength of
\emph{evidence} against $H_0$ (significance).

For effect-size scaling, each $\mathbf{H}$ is divided by $df_e$ to conform
to $\mathbf{E}$. The resulting ellipse is then exactly the data ellipse
of the fitted values, and corresponds visually to a multivariate analog of
univariate effect-size measures [e.g., $(\bar{y}_1 - \bar{y}_2)/s_e$
where $s_e$ is the within-group standard deviation].

For significance scaling, it turns out to be most visually convenient to
use Roy's largest root statistic as the test criterion.
In this case,
the $\mathbf{H}$ ellipse is scaled to $\mathbf{H}/(\lambda_\alpha df_e)$,
where $\lambda_\alpha$ is the critical value of Roy's statistic.%
\footnote{The $F$ test based on Roy's largest root uses the approximation
$ F = (df_2 / df_1) \lambda_1$ with degrees of freedom $df_1, df_2$,
where $df_1 = \max(df_h, df_e)$ and $df_2 = df_e - df_1 + df_h$.
Inverting the $F$ statistic gives the critical value for an $\alpha
$-level test:
$\lambda_\alpha= (df_1/df_2) F^{1-\alpha}_{df_1,df_2}$.}
Using this scaling gives a simple visual test of
$H_0$: Roy's test rejects $H_0$ at a given $\alpha$ level \emph{iff}
the corresponding $\alpha$-level $\mathbf{H}$ ellipse protrudes \emph
{anywhere} outside the $\mathbf{E}$
ellipse.%
\footnote{Other multivariate tests (Wilks's $\Lambda$,
Hotelling--Lawley trace,
Pillai trace) also have geometric interpretations
in HE plots [e.g., Wilks's $\Lambda$ is the ratio of areas (volumes)
of the $\mathbf{H}$ and $\mathbf{E}$ ellipses (ellipsoids);
Hotelling--Lawley trace
is based on the sum of the $\lambda_i$], but these statistics do not provide
such simple visual comparisons. All HE plots shown in this paper use
significance scaling, based on Roy's test.}
Moreover, the directions in which the hypothesis ellipse exceed the
error ellipse
are informative about the responses and their linear combinations that
depart significantly
from $H_0$. Thus, in Figure~\ref{figheplot3a}(b), the variation of the
means of the iris species
shown for these two variables
appears to be largely one-dimensional, corresponding to a weighted sum
(or average) of petal length and
sepal length, perhaps a measure of overall size.

\subsection{Linear Hypotheses: Geometries of Contrasts and Sums of Effects}

Just as in univariate ANOVA designs, important overall effects
($\mathrm{df}_h>1$) in MANOVA may be usefully
explored and interpreted by the use of contrasts among the levels of
the factors involved.
In the general linear hypothesis test of equation~(\ref{eqmglt}), contrasts
are easily specified as one or more $(h_i \times q)$ $\mathbf{L}$
matrices, $\mathbf{L}_1, \mathbf{L}_2, \ldots,$ each of whose rows
sums to zero.

As an important special case,
for an overall effect with
$\mathrm{df}_h$ degrees of freedom (and balanced sample sizes), a set
of $\mathrm{df}_h$ pairwise orthogonal $(1 \times q)$
$\mathbf{L}$ matrices ($\mathbf{L}_i^{\mathsf{T}}\mathbf{L}_j =0$ for
$i\ne j$)
gives rise to a set of $\mathrm{df}_h$ rank-one $\mathbf{H}_i$
matrices that additively decompose the overall hypothesis SSCP matrix
(by a multivariate analog of Pythagoras' Theorem),
\[
\mathbf{H} = \mathbf{H}_1 + \mathbf{H}_2 + \cdots+
\mathbf{H}_{\mathrm{df}_h} ,
\]
exactly as the univariate $SS_H$ may be decomposed in an ANOVA. Each of
these rank-one $\mathbf{H}_i$ matrices
will plot as a vector in an HE plot, and their collection provides a
visual summary of the overall
test, as partitioned by these orthogonal contrasts.
Even more generally, where the subhypothesis matrices may be of rank
$>$ 1, the subhypotheses will have hypothesis ellipses of dimension
rank($\mathbf{H}_i$)
that are conjugate with respect to the hypothesis ellipse for the joint
hypothesis, provided that the
estimators for the subhypotheses are statistically independent.

%
%
\begin{figure}

\includegraphics{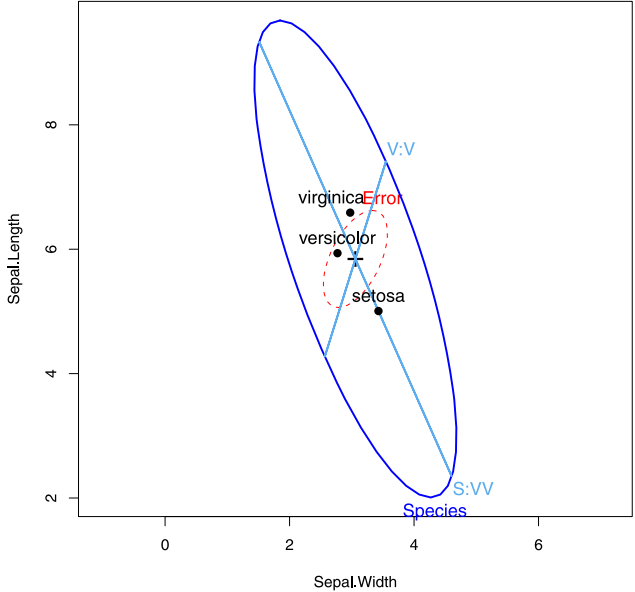}

\caption{$\mathbf{H}$ and $\mathbf{E}$ matrices for sepal width and sepal
length in the iris
data, together with $\mathbf{H}$ matrices for testing two orthogonal
contrasts in the species effect.}\label{figHE-contrasts-iris}
\end{figure}

To illustrate, we show in Figure~\ref{figHE-contrasts-iris} an HE plot
for the sepal width and sepal length variables in the iris data,
corresponding to panel (1:2) in Figure~\ref{figscatirisd1}. Overlayed on
this plot are the
one-df $\mathbf{H}$ matrices obtained from testing two orthogonal contrasts
among the iris species:
\emph{setosa} vs. the average of \emph{versicolor} and \emph
{virginica} (labeled ``S:VV''), and \emph{versicolor} vs. \emph
{virginica} (``V:V''), for which the contrast matrices are
\begin{eqnarray*}
\mathbf{L}_1 & = & \pmatrix{ -2
& 1 & 1},
\\
\mathbf{L}_2 & = & \pmatrix{ 0 &
1 & -1},
\end{eqnarray*}
where the species (columns) are taken in alphabetical order. In this
view, the joint hypothesis testing
equality of the species means has its major axis in data space largely
in the direction of sepal length.
The 1D degenerate ``ellipse'' for $\mathbf{H}_1$, representing the
contrast of setosa with the average of the other two species,
is closely aligned with this axis. The ``ellipse'' for $\mathbf{H}_2$ has
a relatively larger component aligned
with sepal width.

\subsection{Canonical Projections: Ellipses in Data Space and
Canonical Space}

HE plots show the covariation leading toward rejection of a hypothesis
relative to
error covariation for two variables in data space. To visualize these
relationships for more than two
response variables, we can use the obvious generalization of a
scatterplot matrix showing the 2D
projections of the $\mathbf{H}$ and $\mathbf{E}$ ellipsoids for all
pairs of variables.
Alternatively, a transformation to canonical space permits
visualization of all response variables
in the reduced-rank 2D (or 3D) space in which $\mathbf{H}$ covariation
is maximal.


In the MANOVA context, the analysis is called canonical discriminant
analysis (CDA), where the emphasis
is on dimension reduction rather than hypothesis testing.
For a one-way design with $g$ groups and $p$-variate
observations $i$ in group $j$, $\mathbf{y}_{ij}$, CDA finds a set of
$s =
\min(p, g-1)$
linear combinations, $z_1 = \mathbf{c}_1^{\mathsf{T}}\mathbf{y},
z_2 = \mathbf{c}_2^{\mathsf{T}}\mathbf{y}, \ldots,
z_s = \mathbf{c}_s^{\mathsf{T}}\mathbf{y}$,
so that: (a) all $z_k$ are mutually uncorrelated; (b) the vector of
weights $\mathbf{c}_1$ maximizes the univariate $F$ statistic for the
linear combination $z_1$; (c) each successive vector of weights,
$\mathbf{c}_k, k=2, \ldots, s$, maximizes the univariate $F$-statistic
for $z_k$, subject to being uncorrelated with all other linear
combinations.

The canonical projection of $\mathbf{Y}$ to canonical scores $\mathbf
{Z}$ is
given by
%
%
\begin{equation}
\mathbf{Y}_{n \times p} \mapsto\mathbf{Z}_{n \times s} = \mathbf{Y}
\mathbf{E}^{-1} \mathbf{V} / df_e ,
\end{equation}
where $\mathbf{V}$ is the matrix whose columns are the eigenvectors of
$\mathbf{H} {\mathbf{E}}^{-1}$
associated with the ordered nonzero eigenvalues, $\lambda_i,
i=1,\ldots, s$.
A MANOVA of all $s$ linear combinations is statistically
equivalent to that of the raw data.
The $\lambda_i$
are proportional to the fractions of between-group variation
expressed by these linear combinations.
Hence, to the extent that the first one or two
eigenvalues are relatively large, a two-dimensional display will
capture the bulk of between-group differences. The 2D canonical
discriminant HE plot is then simply an HE plot of the scores
$\mathbf{z}_1$ and $\mathbf{z}_2$ on the first two canonical dimensions.
(If $s\ge3$, an analogous 3D version may also be obtained.)

Because the $\mathbf{z}$ scores are all mutually uncorrelated, the
$\mathbf{H}$ and
$\mathbf{E}$ matrices will always have their axes aligned with the
canonical dimensions. When, as here, the $\mathbf{z}$ scores are
standardized, the $\mathbf{E}$ ellipse will be circular, assuming that
the axes in the plot are equated so that a unit data length has the same
physical length on both axes.

Moreover, we can show the contributions of the original variables to
discrimination
as follows: Let $\mathbf{P}$ be the $p \times s$ matrix of the
correlations of
each column of $\mathbf{Y}$ with each column of $\mathbf{Z}$, often called
\emph{canonical structure} coefficients.
Then, for variable $j$, a
vector from the origin
to the point whose coordinates $\mathbf{p}_{\cdot j}$ are given in row
$j$ of $\mathbf{P}$
has projections on the canonical axes equal to these structure coefficients
and squared length equal to the sum squares of these correlations.

\begin{figure}

\includegraphics{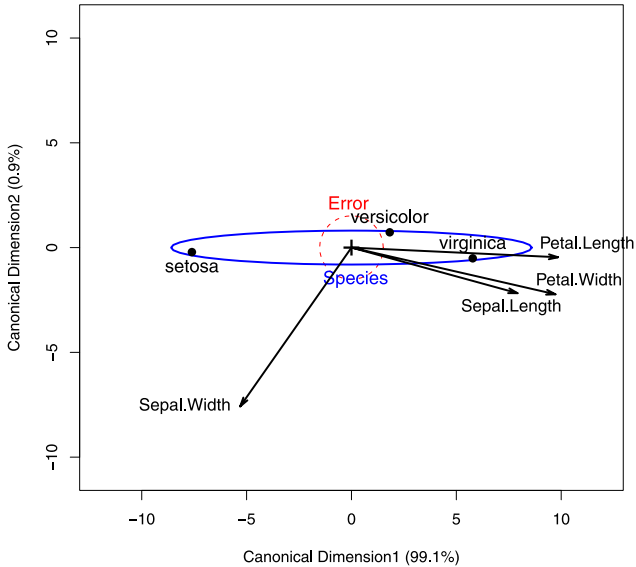}

\caption{Canonical HE plot for the Iris data.
In this plot, the $\mathbf{H}$ ellipse is shown using effect-size scaling
to preserve resolution,
and the variable vectors have been multiplied by a constant to
approximately fill the plot space.
The projections of the variable
vectors on the coordinate axes show the correlations of the variables
with the canonical dimensions.}\label{figHE-can-iris}
\end{figure}

Figure~\ref{figHE-can-iris} shows the canonical HE plot for the iris
data, the
view in canonical space corresponding to Figure~\ref
{figHE-contrasts-iris} in data space
for two of the variables (omitting the contrast vectors).
Note that for $g=3$ groups, $df_h=2$, so $s=2$ and the representation
in 2D is exact.
This provides a very simple interpretation: Nearly all (99.1\%)
of the variation in species means can be accounted for by the first
canonical dimension,
which is seen to be aligned with three of the four variables, most
strongly with
petal length. The second canonical dimension is mostly related to
variation in the
means on sepal width, and this variable is negatively correlated with the
other three.

%
%

Finally, imagine a 4D version of the HE plot of Figure~\ref
{figHE-contrasts-iris} in data space,
showing the four-dimension\-al ellipsoids for $\mathbf{H}$ and $\mathbf
{E}$. Add
to this plot
unit vectors corresponding to the coordinate axes, scaled to some
convenient constant length.
Some rotation would
show that the $\mathbf{H}$
ellipsoid is really only two-dimensional, while $\mathbf{E}$ is 4D.
Applying the transformation given by ${\mathbf{E}}^{-1}$ as in
Figure~\ref{figellipse-geneig}
and projecting into the 2D subspace of the nonzero dimensions of
$\mathbf{H}$
would give a view equivalent to the
canonical HE plot in Figure~\ref{figHE-can-iris}.
The variable vectors in this plot are just the shadows of the original
coordinate axes.

\section{Kissing Ellipsoids}\label{seckiss}

In this section we consider some circumstances in which there is a data
stratification factor or there are two (or more)
principles or procedures for deriving estimates of a parameter vector
$\bolds{\beta}$
of a linear model,
each with its associated estimated covariance matrix, for example,
$\widehat{\bolds{\beta}}{}^A$ with covariance matrix $\widehat{\Var
}(\hat{\bolds{\beta}}{}^A)$
and
$\widehat{\bolds{\beta}}{}^B$ with covariance matrix $\widehat{\Var
}(\hat{\bolds{\beta}}{}^B)$.
The simplest motivating example is two-group discriminant analysis
(Section~\ref{secdiscrim}).
In data space, solutions to this statistical problem can be described
geometrically
in terms of the property that the data ellipsoids around the group centroids
will just ``kiss''
(or \emph{osculate}) along a path between the two centroids. We call
this path
the \emph{locus of osculation}, whose properties are described in
Section~\ref{seclocus}.

Perhaps more interesting and more productive is that the same geometric ideas
apply equally well in parameter ($\beta$) space. Consider, for example,
method A to be OLS estimation and several alternatives
for method B, such as ridge regression (Section~\ref{secridge}) or
Bayesian estimation
(Section~\ref{secbayesian}). The remarkable fact is that the geometry of
such kissing
ellipsoids provides a clear visual interpretation of these cases and others,
whenever we consider a convex combination of information from two
sources. In all cases,
the locus of osculation is interpretable in
terms of the statistical goal to be achieved, taking precision into account.

\subsection{Locus of Osculation}\label{seclocus}
\begin{figure}

\includegraphics{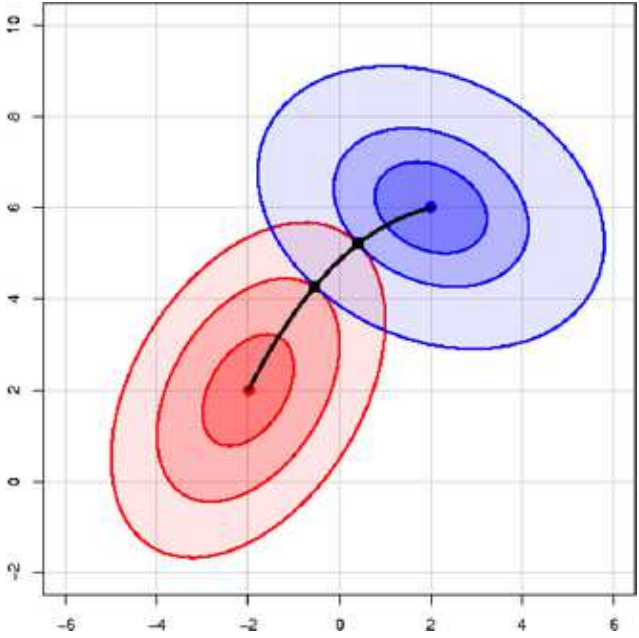}

\caption{Locus of osculation for two families of ellipsoidal level
curves, with centers at $\mathbf{m}_1 = (-2, 2)$ and $\mathbf{m}_2 =
(2, 6)$,
and shape matrices $\mathbf{A}_1$ and $\mathbf{A}_2$ given in
equation~(\protect\ref{eqkiss-demoA}).
The left ellipsoids (red) have $\mathrm{radii}=1, 2, 3$. The right ellipsoids
have $\mathrm{radii}=1, 1.74, 3.1$, where the last two values were
chosen to make
them kiss at the points marked with squares. The black curve is an
approximation to the path of osculation, using a
spline function connecting $\mathbf{m}_1$ to $\mathbf{m}_2$ via the marked
points of osculation.}\vspace*{-5pt}
\label{figkiss-demo}
\end{figure}

The problems mentioned above all have a similar and simple physical
interpretation: Imagine two stones dropped
into a pond at locations with coordinates $\mathbf{m}_1$ and $\mathbf
{m}_2$. The waves emanating from the centers
form concentric circles which osculate along the line from $\mathbf{m}_1$
to $\mathbf{m}_2$.
Now imagine a world with ellipse-generating stones, where instead of
circles, the waves form concentric ellipses determined by
the shape matrices $\mathbf{A}_1$ and $\mathbf{A}_2$.
The \emph{locus of osculation} of these ellipses will be the set of
points where the tangents
to the two ellipses are parallel (or, equivalently, that their normals
are parallel). An
example is shown in Figure~\ref{figkiss-demo}, using $\mathbf{m}_1 = (-2,
2)$, $\mathbf{m}_2 = (2, 6)$, and
%
%
\begin{equation}
\label{eqkiss-demoA}
 \qquad\mathbf{A}_1 = \pmatrix{
  1.0 & 0.5
\vspace*{2pt}\cr
0.5 & 1.5 } , \quad \mathbf{A}_2 =\pmatrix{
1.5 & -0.3
\vspace*{2pt}\cr
-0.3 & 1.0 },
\end{equation}
where we have found points of osculation by trial and error.

%
%
%

%
\begin{figure*}[b]

\includegraphics{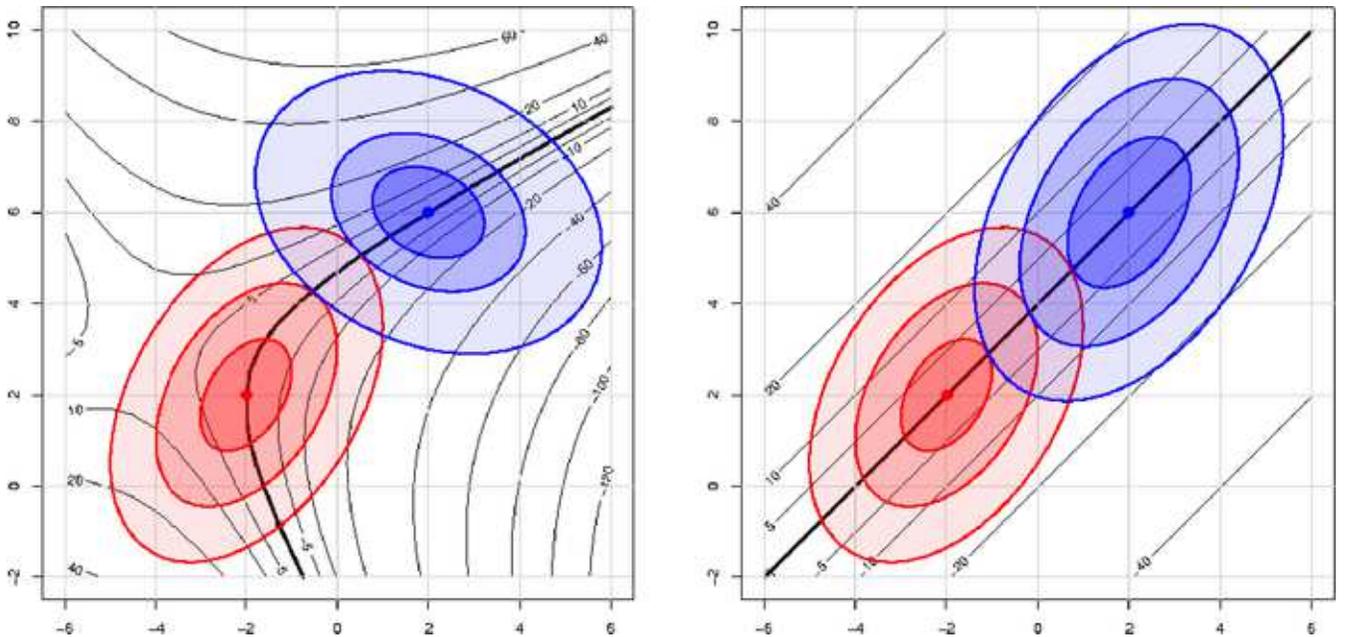}

\caption{Locus of osculation for two families of ellipsoidal level
curves, showing contour lines of the vector cross-product function
equation~(\protect\ref{eqlocus}).
The thick black curve shows the complete locus of osculation for these
two families of ellipses, where the cross-product function equals 0.
Left: with parameters as in Figure~\protect\ref{figkiss-demo} and
equation~(\protect\ref{eqkiss-demoA}). Right: with the same shape matrix
$\mathbf{A}_1$ for both
ellipsoids.}
\label{figkiss-demo2}
\end{figure*}

An exact general solution can be described as follows: Let the ellipses
for $i=1, 2$ be given by
%
\begin{eqnarray*}
f_i(\mathbf{x}) & = & (\mathbf{x}-\mathbf{m}_i)^{\mathsf{T}}
\mathbf{A}_i (\mathbf{x}-\mathbf{m}_i),
\\
\bigl\{ \mathbf{x}\dvtx f_i(\mathbf{x}) = c^2 \bigr\} &
= &\mathbf{m}_i \oplus\sqrt{\mathbf{A}_i}
\end{eqnarray*}
and denote their gradient-vector functions as
%
%
\begin{equation}
\nabla f(x_1, x_2) = \biggl(\frac{\partial f}{\partial x_1},
\frac
{\partial f}{\partial x_2} \biggr)
\end{equation}
so that
\begin{eqnarray*}
\nabla f_1(\mathbf{x}) & = & 2 \mathbf{A}_1 (
\mathbf{x}-\mathbf{m}_1),
\\
\nabla f_2(\mathbf{x}) & = & 2 \mathbf{A}_2 (\mathbf{x}-
\mathbf{m}_2) .
\end{eqnarray*}

Then, the points where $\nabla f_1$ and $\nabla f_2$ are parallel can
be expressed in terms of the
condition that their vector cross product
$\mathbf{u} \circledast\mathbf{v} = u_1 v_2 - u_2 v_1 = \mathbf
{v}^{\mathsf{T}}
\mathbf{C} \mathbf{u} = 0$, where $\mathbf{C}$ is the skew-symmetric matrix
\[
\mathbf{C} = \pmatrix{ 0 & 1
\vspace*{2pt}\cr
-1 & 0 }
\]
satisfying $\mathbf{C} = -\mathbf{C}^{\mathsf{T}}$.
Thus, the locus of osculation is the set $\mathcal{O}$, given by
$\mathcal{O} = \{\mathbf{x} \in\mathbb{R}^{2} \given\nabla
f_1(\mathbf{x})
\circledast\nabla f_2(\mathbf{x}) = 0 \}$,
which implies
%
%
\begin{equation}
(\mathbf{x}-\mathbf{m}_2)^{\mathsf{T}}\mathbf{A}_2^{\mathsf{T}}
\mathbf{C} \mathbf{A}_1 (\mathbf{x}-\mathbf{m}_1) = 0 .
\label{eqlocus}
\end{equation}
Equation~(\ref{eqlocus}) is a bilinear form in $\mathbf{x}$, with
central matrix
$\mathbf{A}_2^{\mathsf{T}}\mathbf{C} \mathbf{A}_1$,
implying that $\mathcal{O}$ is a conic section in the general case.
Note that when $\mathbf{x}=\mathbf{m}_1$ or $\mathbf{x}=\mathbf{m}_2$,
equation~(\ref{eqlocus}) is necessarily zero, so the locus of osculation
always passes through $\mathbf{m}_1$ and $\mathbf{m}_2$.

A visual demonstration of the theory above is\break
shown in Figure~\ref{figkiss-demo2} (left), which overlays the ellipses
in Figure~\ref{figkiss-demo} with contour lines
(hyperbolae, here)
of
the vector cross-product function contained in equation~(\ref{eqlocus}).
When the contours of $f_1$ and $f_2$ have the same shape ($\mathbf
{A}_1 =
c \mathbf{A}_2 $), as in the right panel of Figure~\ref{figkiss-demo2},
equation~(\ref{eqlocus})
reduces to a line, in accord with the stones-in-pond interpretation.
The above can be readily extended to ellipsoids in a higher dimension,
where the development is more easily understood
in terms of normals to the surfaces.


\subsection{Discriminant Analysis}\label{secdiscrim}

The right panel of Figure~\ref{figkiss-demo2}, considered in data space,
provides a
visual interpretation of the classical, normal theory two-group
discriminant analysis problem
under the assumption of equal population covariance matrices, $\bolds
{\Sigma}_1 = \bolds{\Sigma}_2$.
Here, we imagine that the plot shows the contours of data ellipsoids
for two groups,
with mean vectors $\mathbf{m}_1$ and $\mathbf{m}_2$, and common
covariance matrix
$\mathbf{A} = \mathbf{S}_{\mathrm{pooled}} = [ (n_1 -1)\mathbf{S}_1
+ (n_2
-1)\mathbf{S}_2 ] / (n_1 +n_2 -2) $.

The discriminant axis is the locus of osculation between the two
families of ellipsoids.
The goal in discriminant analysis, however, is to determine a
classification rule based on
a linear function, $\mathcal{D}(\mathbf{x}) = \mathbf{b}^{\mathsf{T}}
\mathbf{x}$,
such that
an observation $\mathbf{x}$ will be classified as belonging to Group 1 if
$\mathcal{D}(\mathbf{x}) \le d$, and to Group 2 otherwise. In linear
discriminant
analysis, the discriminant function coefficients are
given by
\[
\mathbf{b} = \mathbf{S}_{\mathrm{pooled}}^{-1} ( \mathbf{m}_1
- \mathbf{m}_2 ) .
\]

All boundaries
of the classification regions determined by $d$
will then be the tangent lines (planes) to the ellipsoids at points of
osculation.
The location of the classification region along the line from $\mathbf
{m}_1$ to $\mathbf{m}_2$
typically takes into account both the
prior probabilities of membership in Groups 1 and 2, and the costs of
misclassification.
Similarly, the left panel of Figure~\ref{figkiss-demo2} is a visual
representation of the
same problem when $\bolds{\Sigma}_1 \ne\bolds{\Sigma}_2$, giving rise
to quadratic classification
boundaries.

\subsection{Ridge Regression}\label{secridge}

In the univariate linear model, $\mathbf{y} = \mathbf{X} \bolds
{\beta} +
\bolds{\varepsilon}$,
high multiple correlations among the predictors in $\mathbf{X}$ lead
to problems
of \emph{collinearity}---unstable OLS
estimates of the parameters in $\bolds{\beta}$ with inflated standard errors
and coefficients that tend to be too large in absolute value.
Although collinearity is essentially a data problem
[\citet{Fox2008}],
one popular (if questionable) approach is ridge regression, which
shrinks the estimates toward
$\mathbf{0}$ (introducing bias) in an effort to reduce sampling variance.

Suppose the predictors and response have been centered at their means
and the unit vector is
omitted from $\mathbf{X}$. Further, rescale the columns of $\mathbf
{X}$ to unit
length, so that $\mathbf{X}^{\mathsf{T}}\mathbf{X}$ is a correlation matrix.
Then, the OLS estimates are given by
%
%
\begin{equation}
\widehat{\bolds{\beta}}{}^{\mathrm{OLS}} = \bigl(\mathbf{X}^{\mathsf
{T}}\mathbf
{X}\bigr)^{-1} \mathbf{X}^{\mathsf{T}}\mathbf{y} .
\end{equation}
Ridge regression replaces the standard residual sum of squares
criterion with a penalized
form,
%
%
\begin{eqnarray}\label{eqridgeRSS}
\operatorname{RSS}(k) = (\mathbf{y}-\mathbf{X} \bolds{\beta})^{\mathsf{T}}(
\mathbf{y}-\mathbf{X} \bolds{\beta}) + k \bolds{\beta}^{\mathsf
{T}}\bolds{\beta},
\nonumber
\\[-8pt]
\\[-8pt]
\eqntext{(k \ge0) ,}
\end{eqnarray}
whose solution is easily seen to be
%
%
\begin{eqnarray}\label{eqridge-beta}
\widehat{\bolds{\beta}}{}^{\mathrm{RR}}_k & = &\bigl(
\mathbf{X}^{\mathsf{T}} \mathbf{X} + k \mathbf{I}\bigr)^{-1}
\mathbf{X}^{\mathsf{T}}\mathbf{y}
\nonumber
\\[-8pt]
\\[-8pt]
\nonumber
& = & \mathbf{G} \widehat{\bolds{\beta}}{}^{\mathrm{OLS}},
\end{eqnarray}
where $\mathbf{G} = [\mathbf{I} + k (\mathbf{X}^{\mathsf{T}}\mathbf
{X})^{-1}
]^{-1}$.
Thus, as the ``ridge constant'' $k$ increases, $\mathbf{G}$ decreases,
driving $\widehat{\bolds{\beta}}{}^{\mathrm{RR}}_k$ toward~$\mathbf{0}$
[\citeauthor{HoerlKennard1970a} (\citeyear
{HoerlKennard1970a,HoerlKennard1970b})]. The addition of a
positive constant $k$ to the diagonal of $\mathbf{X}^{\mathsf
{T}}\mathbf{X}$
drives $\operatorname{det} (\mathbf{X}^{\mathsf{T}}\mathbf{X}+ k
\mathbf
{I})$ away from zero even
if $\operatorname{det} (\mathbf{X}^{\mathsf{T}}\mathbf{X})\approx0$.

%
%
\begin{figure}[b]
\vspace*{-3pt}
\includegraphics{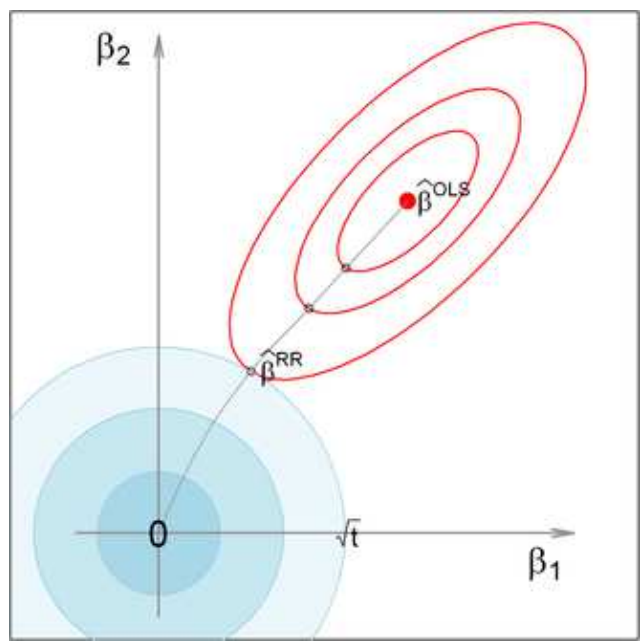}

\caption{Elliptical contours of the OLS residual sum of squares for
two parameters in a regression, together with
circular contours for the constraint function, $\beta_1^2 + \beta_2^2
\le t$. Ridge regression finds the point $\bolds{\beta}^{\mathrm{RR}}$ where
the OLS contours just kiss the constraint region.}%
\label{figridge-demo}
\end{figure}

The penalized Lagrangian formulation in equation~(\ref{eqridgeRSS})
has an
equivalent form as a constrained
minimization problem,
%
%
\begin{eqnarray}\label{eqridge}
\widehat{\bolds{\beta}}{}^{\mathrm{RR}} = \mathop{\argmin}_{\bolds{\beta}} (\mathbf{y}-
\mathbf{X} \bolds{\beta})^{\mathsf{T}}(\mathbf{y}-\mathbf{X}
\bolds{\beta})
\nonumber
\\[-8pt]
\\[-8pt]
\eqntext{\mbox{subject to } \bolds{\beta}^{\mathsf{T}}\bolds{\beta} \le t(k) ,}
\end{eqnarray}
which makes the size constraint on the parameters explicit, with $t(k)$
an inverse function of $k$. This form provides
a visual interpretation of ridge regression, as shown in Figure~\ref
{figridge-demo}.\vadjust{\goodbreak} Depicted in the figure are the
elliptical contours of the OLS regression sum of squares,
$\operatorname
{RSS}(0)$ around $\widehat{\bolds{\beta}}{}^{\mathrm{OLS}}$. Each
ellipsoid marks the point closest to the origin, that is, with $\min
\bolds{\beta}^{\mathsf{T}}\bolds{\beta}$.
It is easily seen that the ridge regression solution is the point where
the elliptical contours just kiss the
constraint contour.

Another insightful interpretation of ridge regression [\citet
{Marquardt1970}] sees the ridge estimator as equivalent to
an OLS estimator, when the actual data in $\mathbf{X}$ are
supplemented by
some number of fictitious observations, $n(k)$,
with uncorrelated predictors,
giving rise to
an orthogonal $\mathbf{X}_k^0$ matrix, and where $y=0$ for all
supplementary observations. The linear model then becomes
%
%
\begin{equation}
\label{eqridge-sup1}\pmatrix{
\mathbf{y}
\vspace*{2pt}\cr
\mathbf{0} } =\pmatrix{\mathbf{X}
\vspace*{2pt}\cr
\mathbf{X}_k^0}
\bolds{\beta}^{\mathrm{RR}} + \pmatrix{
\mathbf{e}
\vspace*{2pt}\cr
\mathbf{e}_k^0 },
\end{equation}
which gives rise to the solution
%
%
\begin{equation}
\label{eqridge-sup2} \widehat{\bolds{\beta}}{}^{\mathrm{RR}} = \bigl[
\mathbf{X}^{\mathsf{T}}\mathbf{X} + \bigl(\mathbf{X}_k^0
\bigr)^{\mathsf{T}} \mathbf{X}_k^0
\bigr]^{-1} \mathbf{X}^{\mathsf{T}} \mathbf{y} .
\end{equation}
But because $\mathbf{X}_k^0$ is orthogonal, $(\mathbf
{X}_k^0)^{\mathsf{T}}
\mathbf{X}_k^0$ is a scalar multiple of $\mathbf{I}$, so there
exists some value of $k$ making equation~(\ref{eqridge-sup2})
equivalent to
equation~(\ref{eqridge-beta}). As promised, the
ridge regression estimator then reflects a weighted average of the data
$[\mathbf{X}, \mathbf{y}]$ with $n(k)$ observations
$[\mathbf{X}_k^0, \mathbf{0}]$
biased toward $\bolds{\beta}=\mathbf{0}$. In Figure~\ref
{figridge-demo}, it
is easy to imagine that there is a direct translation between
the size of the constraint region, $t(k)$, and an equivalent
supplementary sample size, $n(k)$, in this interpretation.

This classic version of the ridge regression problem can be generalized
in a variety of ways, giving other geometric
insights. Rather than a constant multiplier $k$ of $\bolds{\beta
}^{\mathsf{T}}
\bolds{\beta}$ as the penalty term in equation~(\ref{eqridgeRSS}),
consider a penalty of the form $\bolds{\beta}^{\mathsf{T}}\mathbf{K}
\bolds{\beta}$ with
a positive definite matrix $\mathbf{K}$.
The choice $\mathbf{K} = \operatorname{diag} (k_1, k_2, \ldots)$
gives rise to a
version of
Figure~\ref{figridge-demo} in which the constraint contours are ellipses
aligned with the coordinate axes, with
axis lengths inversely proportional to $k_i$. These constants allow for
differential shrinkage of the OLS coefficients.
The visual solution to the obvious modification of equation~(\ref
{eqridge}) is
again the point where the elliptical
contours of $\operatorname{RSS}(0)$ kiss the contours of the (now
elliptical) constraint region.\looseness=-1

%
%
\begin{figure*}

\includegraphics{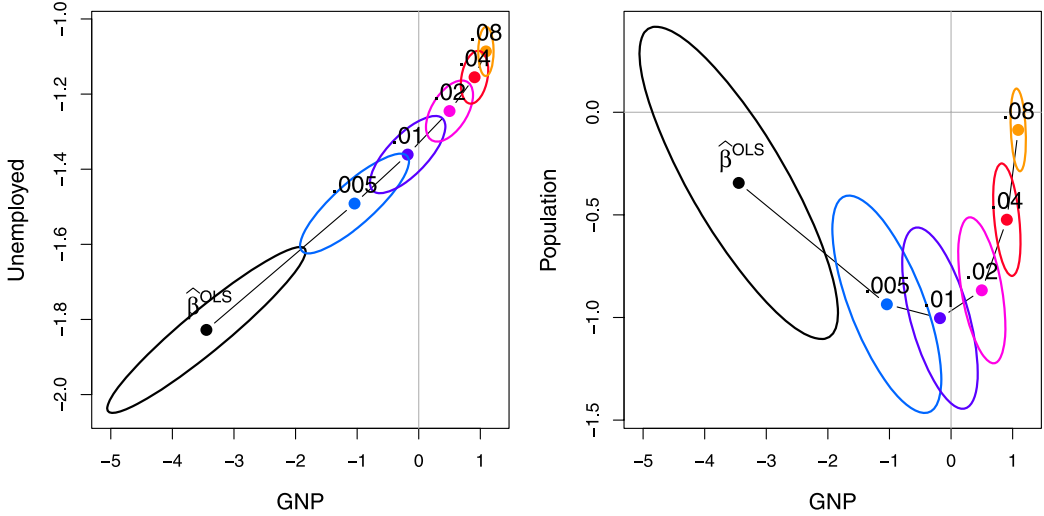}

\caption{Bivariate ridge trace plots for the coefficients of
Unemployed and Population
against the coefficient for GNP in Longley's data, with
$k = 0, 0.005, 0.01, 0.02, 0.04, 0.08$.
In both cases the coefficients are driven on average toward zero, but
the bivariate
plot also makes clear the reduction in variance.
To reduce overlap, all variance ellipses are shown with $1/2$ the
standard radius.}%
\label{figridge2}
\end{figure*}

\subsubsection{Bivariate ridge trace plots}\label{secridge2}
Ridge regression is touted (optimistically we think) as a method to
counter the effects
of collinearity by trading off a small amount of bias for an
advantageous decrease in variance. The results are often
visualized in a \emph{ridge trace plot}
[\citet{HoerlKennard1970b}],
showing the changes\vadjust{\goodbreak}
in individual coefficient estimates as a function of $k$.
A bivariate version of this plot, with confidence ellipses for
the parameters, is introduced here. This plot provides greater
insight into the effects of $k$ on coefficient variance.%
\footnote{Bias and mean-squared error are a different matter: Although
\citet{HoerlKennard1970a} demonstrate that there is a range of values
for the ridge constant $k$ for which the MSE of the ridge estimator is
smaller than that of the OLS estimator, to know where this range is
located requires knowledge of $\bolds{\beta}$. As we explain in the
following subsection, the constraint on $\hat{\bolds{\beta}}$
incorporated in the ridge estimator can be construed as a Bayesian
prior; the fly in the ointment of ridge regression, however, is that
there is no reason to suppose that the ridge-regression prior is in
general reasonable.}

Confidence ellipsoids for the OLS estimator are generated
from the estimated covariance matrix of the coefficients,
\[
\widehat{\Var} \bigl(\bolds{\beta}^{\mathrm{OLS}} \bigr) = \hat{
\sigma}^2_e \bigl(\mathbf{X}^{\mathsf{T}}\mathbf{X}
\bigr)^{-1} .
\]
For the ridge estimator, this becomes
[\citet{Marquardt1970}]
%
%
\begin{eqnarray}
\widehat{\Var} \bigl(\bolds{\beta}^{\mathrm{RR}} \bigr) &=& \hat{
\sigma}^2_e \bigl[\mathbf{X}^{\mathsf{T}}\mathbf{X} + k
\mathbf{I}\bigr]^{-1} \bigl(\mathbf{X}^{\mathsf{T}}\mathbf{X}\bigr)
\nonumber
\\[-8pt]
\\[-8pt]
\nonumber
&&{}\cdot
\bigl[\mathbf {X}^{\mathsf{T}} \mathbf{X} + k \mathbf{I}\bigr]^{-1} ,
\end{eqnarray}
which coincides with the OLS result when $k=0$.

Figure~\ref{figridge2} uses the classic \citet{Longley1967}
data to illustrate
bivariate ridge trace plots. The data consist of an economic time
series ($n=16$)
observed yearly from 1947 to 1962, with the number of people Employed
as the
response and the following predictors:
GNP, Unemployed, Armed.Forces, Population, Year, and GNP.deflator (using
1954 as 100).%
\footnote{\citet{Longley1967} used these data to demonstrate the
effects of
numerical instability and round-off error in least squares computations
based on direct computation of the cross-products matrix, $\mathbf
{X}^{\mathsf{T}}\mathbf{X}$.
Longley's paper sparked the development of a wide variety of
numerically stable least squares algorithms (QR, modified Gram-Schmidt, etc.)
now used is almost all statistical software.
Even ignoring numerical problems
(not to mention problems due to lack of independence), these
data would be anticipated to exhibit high collinearity because
a number of the predictors would be expected to have strong associations
with year and/or population, yet both of these are also included among the
predictors.}
For each value of $k$, the plot
shows the estimate $\widehat{\bolds{\beta}}$, together with the variance
ellipse. For the sake of this example, we assume that GNP is a primary
predictor of Employment, and we wish to know how other predictors
modify the
regression estimates and their variance when ridge regression is used.

For these data, it can be seen that even small values of $k$ have substantial
impact on the estimates~$\widehat{\bolds{\beta}}$. What is perhaps
more dramatic
(and unseen in univariate trace plots) is the impact on the size of the variance
ellipse. Moreover, shrinkage in variance is generally in a similar
direction to the
shrinkage in the coefficients. This new graphical method is developed
more fully in
\citet{Friendlygenridge2012},
including 2D and 3D plots, as well as more informative representations
of shrinkage by ellipsoids in
the transformed space of the SVD of the predictors.

\subsection{Bayesian Linear Models}\label{secbayesian}

In a Bayesian alternative to standard least squares estimation,
consider the case where our prior
information about $\bolds{\beta}$ can be encapsulated in a distribution
with a prior mean
$\bolds{\beta}^{\mathrm{prior}}$ and covariance matrix $\mathbf
{A}$. We
show that under reasonable conditions
the Bayesian posterior
estimate, $\widehat{\bolds{\beta}}{}^{\mathrm{posterior}}$, turns out
to be a weighted average of the prior coefficients $\bolds{\beta
}^{\mathrm{prior}}$ and the OLS solution $\widehat{\bolds{\beta
}}{}^{\mathrm{OLS}}$,
with weights proportional to the conditional prior precision, $\mathbf
{A}^{-1}$, and the data precision given by
$\mathbf{X}^{\mathsf{T}}\mathbf{X}$. Once again, this can be understood
geometrically as the locus of osculation of
ellipsoids that characterize the prior and the data.

Under Gaussian assumptions,
the conditional likelihood can be written as
\begin{eqnarray*}
&&\mathcal{L} \bigl(\mathbf{y} \given\mathbf{X}, \bolds{\beta},
\sigma^2 \bigr)
\nonumber
\\[-8pt]
\\[-8pt]
\nonumber
&&\quad\propto \bigl(\sigma^2
\bigr)^{-n/2} \exp \biggl[ -\frac{1}{2 \sigma^2} (\mathbf{y}-\mathbf{X}
\bolds {\beta})^{\mathsf{T}}( \mathbf{y}-\mathbf{X} \bolds{\beta}) \biggr] .
\end{eqnarray*}
To focus on alternative estimators, we can complete the square around
$\widehat{\bolds{\beta}} = \widehat{\bolds{\beta}}{}^{\mathrm{OLS}}$
to give
%
%
\begin{eqnarray}
\label{eqbayes-qform}\quad (\mathbf{y}- \mathbf{X} \bolds{\beta })^{\mathsf{T}}(
\mathbf{y}- \mathbf{X} \bolds{\beta}) &= &(\mathbf{y}- \mathbf{X} \widehat {\bolds{
\beta}})^{\mathsf{T}}(\mathbf{y}- \mathbf{X} \widehat{\bolds{\beta}})
\nonumber
\\[-4pt]
\\[-12pt]
\nonumber
&&{} + (\bolds{
\beta} - \widehat{\bolds{\beta}})^{\mathsf{T}}\bigl(\mathbf{X}^{\mathsf
{T}}
\mathbf{X}\bigr) (\bolds{\beta} - \widehat{\bolds{\beta}}) .
\end{eqnarray}
With a little manipulation, a conjugate prior, of the form $\Pr
(\bolds{\beta}, \sigma^2) = \Pr(\bolds{\beta} \given\sigma^2)
\times\Pr
(\sigma^2)$,
can be expressed with $\Pr(\sigma^2)$ an inverse gamma distribution
depending on the first term on the right-hand side of equation~(\ref
{eqbayes-qform})
and $\Pr(\bolds{\beta} \given\sigma^2)$ a normal distribution,
%
%
\begin{eqnarray}
&&\Pr \bigl(\bolds{\beta} \given\sigma^2 \bigr)\nonumber\\
&&\quad \propto \bigl(
\sigma^2 \bigr)^{-p} \\
&&\qquad{}\cdot\exp \biggl[ -\frac{1}{2 \sigma^2}
\bigl(\bolds{\beta} - \bolds{\beta}^{\mathrm{prior}} \bigr)^{\mathsf{T}}\mathbf{A}
\bigl( \bolds{\beta} - \bolds{\beta}^{\mathrm{prior}} \bigr) \biggr]
.\nonumber
\end{eqnarray}

The posterior distribution is then
$\Pr(\bolds{\beta}, \sigma^2 \given\mathbf{y}, \mathbf{X})
\propto\Pr
(\mathbf{y} \given\mathbf{X}, \bolds{\beta}, \sigma^2) \times
\Pr(\bolds{\beta} \given\sigma^2) \times\Pr(\sigma^2)$,
whence, after some simplification,
the posterior mean can be expressed as
%
%
\begin{equation}
\label{eqbayes-posterior} \widehat{\bolds{\beta}}{}^{\mathrm
{posterior}} = \bigl(
\mathbf{X}^{\mathsf{T}}\mathbf{X}+\mathbf{A}\bigr)^{-1} \bigl(
\mathbf{X}^{\mathsf
{T}}\mathbf{X} \widehat{\bolds{\beta}}{}^{\mathrm
{OLS}} +
\mathbf{A} \bolds{\beta}^{\mathrm{prior}} \bigr)\hspace*{-30pt}
\end{equation}
with covariance matrix $(\mathbf{X}^{\mathsf{T}}\mathbf{X}+\mathbf
{A})^{-1}$. The
posterior coefficients are
therefore a weighted average of
the prior coefficients and the OLS estimates, with weights given by the
conditional prior precision, $\mathbf{A}$,
and the data precision, $\mathbf{X}^{\mathsf{T}}\mathbf{X}$. Thus,
as we increase
the strength of our prior precision (decreasing
prior variance), we place greater weight on our prior beliefs relative
to the data.

In this context, ridge regression can be seen as the special case
where $\widehat{\bolds{\beta}}{}^{\mathrm{prior}} = \mathbf{0}$ and
$\mathbf{A} = k \mathbf{I}$,
and where Figure~\ref{figridge-demo} provides an elliptical
visualization. In equation~(\ref{eqridge-sup2}), the number of observations,
$n(k)$ corresponding to $\mathbf{X}_k^0$, can be seen as another way of
expressing the weight of the prior in relation to the data.


\subsection{Mixed Models: BLUEs and BLUPs}

In this section we make implicit use of the duality between data space
and $\beta$
space, where lines in one map into points in the other and ellipsoids
help to
visualize the precision of estimates in the context of the linear mixed
model for hierarchical data. We also show visually how the best linear unbiased
predictors (BLUPs) from the mixed model can be seen as a weighted average
of the best linear unbiased estimates (BLUEs) derived from OLS
regressions performed \emph{within} clusters of related data
and the overall mixed model GLS estimate.

The mixed model for hierarchical data provides a general framework for
dealing with dependence among observations in linear models,
such as occurs when students are sampled within schools, schools within
counties and so forth
[e.g., \citet{RaudenbushBryk2002}].
In these situations, the assumption of OLS that
the errors are conditionally independent is probably violated,
because, for example, students nested within the same school are
likely to have more similar outcomes than those from different schools.
Essentially the same model, with provision for serially correlated
errors, can be applied to longitudinal data [e.g.,
\citet{LairdWare1982}], although we will not pursue this application here.

The mixed model for the $n_i \times1$ response vector $\mathbf{y}_i$ in
cluster $i$ can be given as
%
%
\begin{eqnarray}
\label{eqmixed1} \mathbf{y}_i & = & \mathbf{X}_i
\bolds{\beta} + \mathbf{Z}_i \mathbf{u}_i + \bolds{
\varepsilon}_i ,\nonumber
\\
\mathbf{u}_i & \sim& \mathcal{N}_q (\mathbf{0},
\mathbf{G}_i),
\\
\bolds{\varepsilon}_i & \sim& \mathcal{N}_{n_i} (
\mathbf{0}, \mathbf{R}_i),
\nonumber
\end{eqnarray}
where
$\bolds{\beta}$ is a $p \times1$ vector of parameters corresponding to
the fixed effects
in the $n_i \times p$ model matrix $\mathbf{X}_i$;
$\mathbf{u}_i$ is a $q \times1$ vector of coefficients corresponding to
the random effects
in the $n_i \times q$ model matrix $\mathbf{Z}_i$;
$\mathbf{G}_i$ is the $q \times q$ covariance matrix of the random
effects in $\mathbf{u}_i$;
and $\mathbf{R}_i$ is the $n_i \times n_i$ covariance matrix of the
errors in $\bolds{\varepsilon}_i$.

Stacking the $\mathbf{y}_i$, $\mathbf{X}_i$, $\mathbf{Z}_i$ and so
forth in
the obvious way then gives
%
%
\begin{equation}
\label{eqmixed2} \mathbf{y} = \mathbf{X} \bolds{\beta} + \mathbf {Z} \mathbf{u}
+ \bolds{\varepsilon},
\end{equation}
where $\mathbf{u}$ and $\bolds{\varepsilon}$ are assumed to have normal
distributions with mean $\mathbf{0}$
and
%
%
\begin{equation}
\label{eqmixed3} \Var%
\pmatrix{ \mathbf{u} \vspace*{2pt}
\cr
\bolds{
\varepsilon} } %
= %
\lleft[\matrix{ \mathbf{G} & \mathbf{0}
\vspace*{2pt}
\cr
\mathbf{0} & \mathbf{R} } \rright] %
,
\end{equation}
where $ \mathbf{G} = \operatorname{diag} (\mathbf{G}_1,\ldots
,\mathbf
{G}_m)$, $\mathbf{R} =
\operatorname{diag} (\mathbf{R}_1,\ldots,\mathbf{R}_m)$ and $m$ is the
number of clusters.
The variance of $\mathbf{y}$ is therefore $\mathbf{V} = \mathbf{Z}
\mathbf{G}
\mathbf{Z}^{\mathsf{T}}+ \mathbf{R}$, and when
$\mathbf{Z} = \mathbf{0}$ and $\mathbf{R} = \sigma^2 \mathbf{I}$,
the mixed
model in equation~(\ref{eqmixed2}) reduces to the
standard linear model.


We now consider the case in which
$\mathbf{Z}_i = \mathbf{X}_i$ and we wish to predict $\bolds{\beta
}_i =
\bolds{\beta} + \mathbf{u}_i$, the vector of parameters for the
$i$th cluster.
At one extreme, we could simply ignore clusters and use the common
mixed-model generalized-least-square estimate,
%
%
\begin{equation}
\label{eqmixed4} \widehat{\bolds{\beta}}{}^{\mathrm{GLS}} = \bigl(
\mathbf{X}^{\mathsf{T}} \mathbf{V}^{-1} \mathbf{X}
\bigr)^{-1} \mathbf{X}^{\mathsf{T}} \mathbf{V}^{-1}
\mathbf{y},
\end{equation}
whose sampling variance is $\Var(\widehat{\bolds{\beta}}{}^{\mathrm
{GLS}}) =  \mathbf{(\mathbf{X}^{\mathsf{T}}\mathbf{V}^{-1} \mathbf
{X})}^{-1}$.
It is an unbiased predictor of $\bolds{\beta}_i$ since $E( \widehat
{\bolds{\beta}}{}^{\mathrm{GLS}} - \bolds{\beta}_i) = 0$.
With moderately large $m$, the sampling variance may be small relative
to $\mathbf{G}_i$ and
$\Var(\widehat{\bolds{\beta}}{}^{\mathrm{GLS}} - \bolds{\beta}_i)
\approx\mathbf{G}_i$.

At the other extreme, we ignore the fact that clusters come from a
common population and we calculate the separate
BLUE estimate within each cluster,
%
%
\begin{eqnarray}
\label{eqmixed5} \widehat{\bolds{\beta}}{}^{\mathrm{blue}}_i =
\mathbf{\bigl(\mathbf{X}_i^{\mathsf{T}}\mathbf{X}_i
\bigr)}^{-1} \mathbf{X}_i^{\mathsf{T}}
\mathbf{y}_i
\nonumber
\\[-8pt]
\\[-8pt]
\eqntext{\mbox{with } \Var \bigl( \widehat{\bolds{
\beta}}{}^{\mathrm{blue}}_i \given\bolds{ \beta}_i \bigr)
\equiv\mathbf{S}_i = \sigma^2 \mathbf{\bigl(
\mathbf{X}_i^{\mathsf{T}}\mathbf{X}_i
\bigr)}^{-1} .}
\end{eqnarray}

Both extremes have drawbacks: whereas the pooled overall GLS estimate
ignores variation between clusters,
the unpooled within-cluster BLUE ignores the common population and
makes clusters appear to differ more than they actually do.

%
\begin{figure*}

\includegraphics{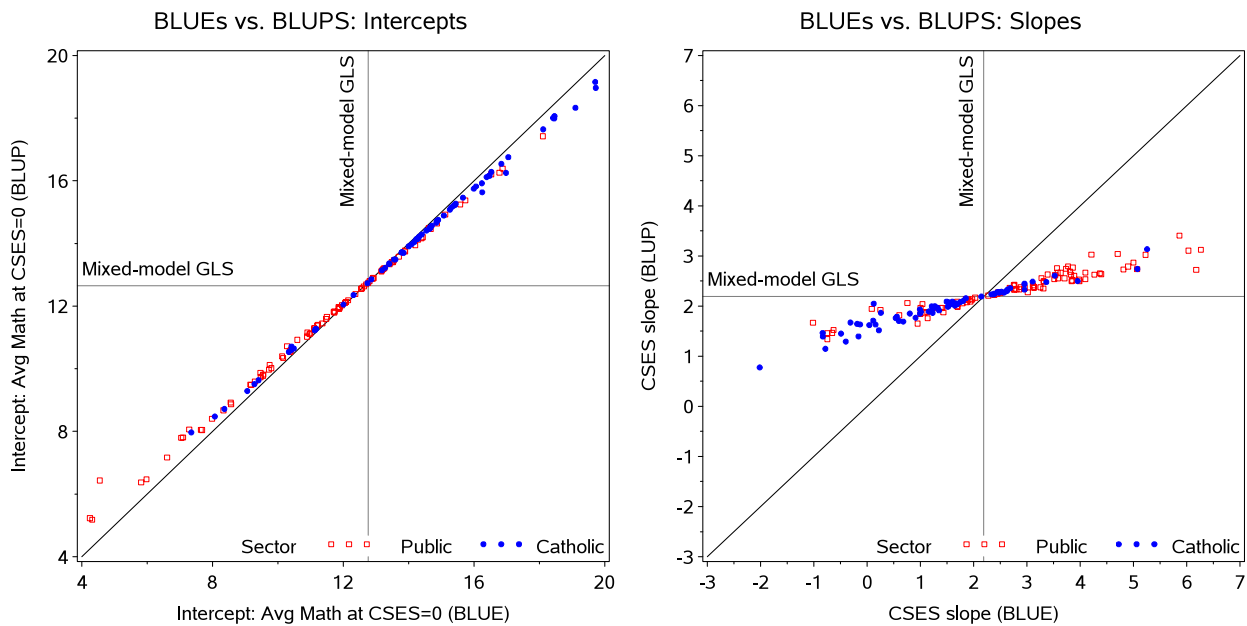}

\caption{Comparing BLUEs and BLUPs. Each panel plots the OLS estimates
from separate regressions for each school (BLUEs)
versus the mixed model estimates from the random intercepts and slopes
model (BLUPs).
Left: intercepts; Right: slopes for CSES. The shrinkage of the BLUPs
toward the GLS estimate is much greater for slopes than
intercepts.}
\label{fighsbmix4}
\end{figure*}

This dilemma led to the development of BLUPs (best linear unbiased
predictor) in models with random effects [\citet
{Henderson1975}, \citet{Robinson1991},\break \citet{Speed1991}].
In the case considered here, the BLUPs are an inverse-variance weighted
average of the mixed-model GLS estimates and of the BLUEs. The BLUP is then
%
%
\begin{eqnarray}
\label{eqmixed6} \widetilde{\bolds{\beta}}{}^{\mathrm{blup}}_i &=&
 \bigl( {\mathbf{S}_i}^{-1} + {\mathbf
{G}}_i^{-1} \bigr) ^{-1}
\nonumber
\\[-8pt]
\\[-8pt]
\nonumber
&&{}\cdot \bigl( {
\mathbf{S}}{}^{-1}_i \widehat{\bolds{\beta}}{}^{\mathrm
{blue}}_i
+ {\mathbf{G}}^{-1}_i \widehat{\bolds{
\beta}}{}^{\mathrm{GLS}}_i \bigr) .
\end{eqnarray}
This ``partial pooling'' optimally combines the information from cluster
$i$ with the information from all clusters,
shrinking $\widehat{\beta}_i^{\mathrm{blue}}$ toward $\widehat
{\beta}^{\mathrm{GLS}}$.
Shrinkage for a given parameter $ \beta_{ij} $ is greater when the
sample size $n_i$ is small or when the variance of the
corresponding random effect, $g_{ijj}$, is small.

Equation~(\ref{eqmixed6})
is of the same form as equation~(\ref{eqbayes-posterior}) and other convex
combinations of estimates considered
earlier in this section. So once again, we can understand these results
geometrically as the locus of
osculation of ellipsoids. Ellipsoids kiss for a reason: to provide an optimal
convex combination of information from two sources, taking precision
into account.

\subsubsection{Example: Math achievement and SES}

To illustrate, we use a classic data set from 
\citet{BrykRaudenbush1992} and \citet{RaudenbushBryk2002}
dealing with math achievement scores for a subsample of 7185 students
from 160 schools
in the 1982 High School \& Beyond survey of U.S. public and Catholic high
schools conducted by the National Center for Education Statistics (NCES).
The data set contains 90 public schools and 70 Catholic schools, with
sample sizes ranging from 14 to 67.

The response is a standardized measure of math achievement, while
student-level predictor variables include sex and student socioeconomic
status (SES), and
school-level predictors include sector (public or Catholic) and mean
SES for the
school (among other variables).
Following \citet{RaudenbushBryk2002}, student SES is considered the
main predictor and is
typically analyzed centered within schools,\vadjust{\goodbreak}
$\mathrm{CSES}_{ij} = \mathrm{SES}_{ij} - \mathrm{(mean\ SES)}_i$,
for ease of interpretation (making the within-school intercept for
school $i$
equal to the mean SES in that school).

For simplicity, we consider the case of CSES as
a single quantitative predictor in $\mathbf{X}$ in the example below. We
fit and compare the following models:
%
%
\begin{eqnarray}
\mathbf{y}_{i} & \sim& \mathcal{N} \bigl( \beta_0 +
x_{i} \beta_1 , \sigma^2 \bigr)\quad
\mbox{pooled OLS} ,
\\
\mathbf{y}_{i} & \sim& \mathcal{N} \bigl( \beta_{0i} +
x_{i} \beta_{1i} , \sigma^2_i
\bigr)\quad \mbox{unpooled BLUEs},
\\
\qquad\mathbf{y}_{i} & \sim& \mathcal{N} \bigl( \beta_{0} +
x_{i} \beta_{1} + u_{0i} + x_{i}
u_{1i} , \sigma^2_i \bigr)
\nonumber
\\[-8pt]
\\[-8pt]
\eqntext{\mbox{BLUPs:
random intercepts and slopes,} }
\end{eqnarray}
and also include a fixed effect of sector, common to all models; for
compactness, the sector effect is elided in the notation above.

In expositions of mixed-effects models, such models are often compared
visually by plotting predicted values in data space, where each school appears
as a fitted line under one of the models above (sometimes called
``spaghetti plots'').
Our geometric approach leads us to consider the equivalent but simpler
plots in the dual $\beta$ space,
where each school appears as a point.

Figure~\ref{fighsbmix4} plots the unpooled BLUE estimates\break against the
BLUPs from the random effects model, with separate panels for intercepts
and slopes to illustrate the shrinkage of different parameters. In
these data, the variance in intercepts\vadjust{\goodbreak} (average math achievement
for students at $\mathrm{CSES}=0$), $g_{00}$,
among schools in each sector is large, so the mixed-effects estimates
have small weight and there is little
shrinkage. On the other hand, the variance component for slopes,
$g_{11}$, is relatively small, so there is greater
shrinkage toward the GLS estimate.

%
%
\begin{figure}

\includegraphics{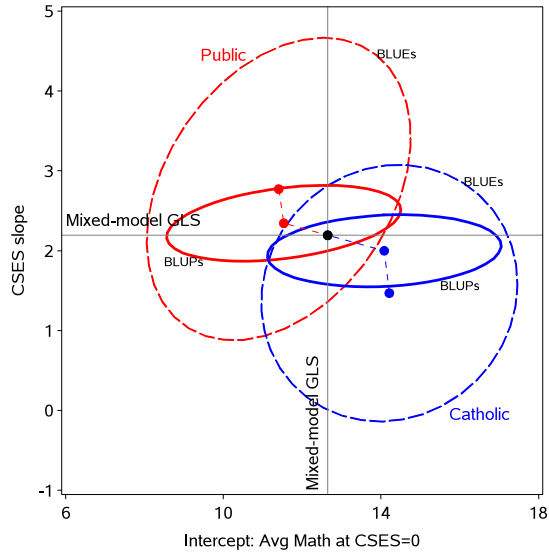}

\caption{Comparing BLUEs and BLUPs. The plot shows ellipses of 50\%
coverage for the estimates of intercepts and slopes
from OLS regressions (BLUEs) and the mixed model (BLUPs), separately
for each sector.
The centers of the ellipses illustrate how the BLUPS can be considered
a weighted average of the BLUEs and the
mixed-model GLS estimate, ignoring sector. The relative sizes of the
ellipses reflect the smaller variance for the
BLUPs compared to the BLUEs, particularly for slope estimates.}%
\label{fighsbmix43}
\end{figure}

For the present purposes, a more useful visual representation of these
model comparisons can be shown together in
the space of $(\beta_0, \beta_1)$, as in Figure~\ref{fighsbmix43}.
Estimates for individual schools are not shown,
but rather these are summarized by the ellipses of 50\% coverage for
the BLUEs and BLUPs within each sector.
The centers of the ellipsoids indicate the relatively greater shrinkage
of slopes compared to intercepts.
The sizes of ellipsoids show directly the greater precision of the
BLUPs, particularly for slopes.

\subsection{Multivariate Meta-Analysis}
A related situation arises in random effects multivariate
meta-analysis
[\citet{Berkey-etal1998}, \citet{Nam-etal2003}],
where several outcome measures are observed in a series of
similar research studies and it is desired to synthesize
those studies to provide an overall (pooled) summary of the outcomes,
together with meta-analytic inferences and
measures of heterogeneity across studies.

%
%
\begin{figure*}

\includegraphics{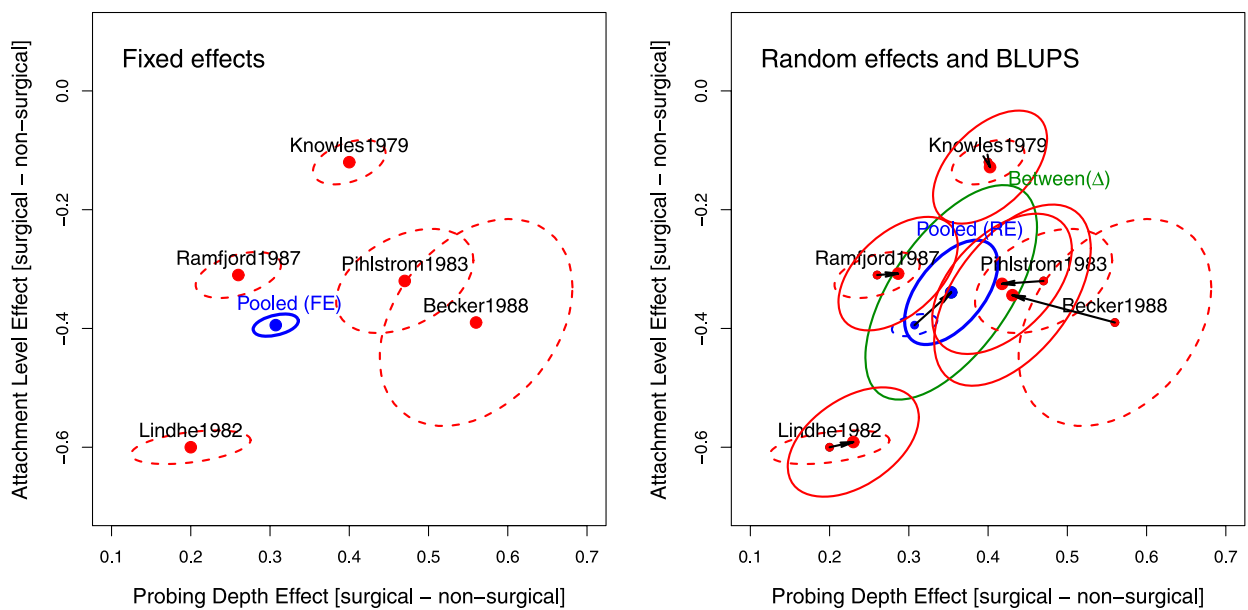}

\caption{Multivariate meta-analysis visualizations for five
periodontal treatment studies with outcome measures PD and AL.
Left: Individual study estimates $\mathbf{y}_i$ and 40\% standard
ellipses for $\mathbf{S}_i$ (dashed, red) together with the pooled,
fixed effects estimate and its associated covariance ellipse (blue).
Right: BLUPs from the random-effects multivariate meta-analysis model
and their associated covariance ellipses
(red, solid), together with the pooled, population-averaged estimate
and its covariance ellipse (blue), and the estimate
of the between-study covariance matrix, $\Delta$ (green). Arrows show
the differences between the FE and the RE models.}
\label{figmvmeta2}
\end{figure*}

The application of mixed model ideas in this context differs
from the standard situation in that individual data are
usually unavailable and use is made instead of
summary data (estimated treatment effects and their covariances)
from the published literature. The multivariate extension of standard
univariate methods of meta-analysis allows the correlations among
outcome effects to be taken into account and estimated, and
regression versions can incorporate study-specific covariates to
account for some inter-study heterogeneity. More importantly,
we illustrate a graphical method based (of course) on ellipsoids
that serves to illustrate bias, heterogeneity and shrinkage in BLUPs,
and the optimism of fixed-effect estimates when
study heterogeneity is ignored.\looseness=-1

The general mixed-effects multivariate meta-anal\-ysis model can be
written as
%
%
\begin{equation}
\label{eqmvmeta1} \mathbf{y}_i = \mathbf{X}_i \bolds{
\beta} + \bolds{\delta}_i + \mathbf{e}_i ,
\end{equation}
where $\mathbf{y}_i$ is a vector of $p$ outcomes (means or treatment effects)
for study $i$; $\mathbf{X}_i$ is the matrix of study-level predictors
for study $i$ or a unit vector when no covariates are available;
$\bolds{\beta}$ is the population-averaged vector of regression parameters
or effects (intercepts, means) when there are no covariates;
$\bolds{\delta}_i$ is the $p$-vector of random effects associated with
study $i$,
whose $p\times p$ covariance matrix $\bolds{\Delta}$ represents the
be-\break tween-study
heterogeneity unaccounted for by $\mathbf{X}_i \bolds{\beta}$; and,
finally, $\mathbf{e}_i$ is the $p$-vector of random sampling errors
(independent of $\bolds{\delta}_i$)
within study $i$, having the $p\times p$ covariance matrix $\mathbf{S}_i$.

With suitable distributional assumptions, the\break mixed-effects model in
equation~(\ref{eqmvmeta1}) implies that
%
%
\begin{equation}
\label{eqmvmeta2} \mathbf{y}_i \sim\mathcal{N}_p (
\mathbf{X}_i \bolds{\beta}, \bolds{\Delta} + \mathbf{S}_i)
\end{equation}
with $\Var(\mathbf{y}_i) = \bolds{\Delta} + \mathbf{S}_i$. When
all the
$\bolds{\delta}_i = \mathbf{0}$, and thus $\bolds{\Delta}=\mathbf{0}$,
equation~(\ref{eqmvmeta1}) reduces to a fixed-effects model, $\mathbf
{y}_i =
\mathbf{X}_i \bolds{\beta} + \mathbf{e}_i$,
which can be estimated by GLS to give
%
%
\begin{eqnarray}
\widehat{\bolds{\beta}}{}^{\mathrm{GLS}} & = & \bigl(\mathbf{X}^{\mathsf
{T}}
\mathbf{S} \mathbf{X}\bigr)^{-1} \mathbf{X}^{\mathsf{T}}
\mathbf{S}^{-1} \mathbf{y} ,\label{eqmvmeta3}
\\
\widehat{\Var} \bigl(\widehat{\bolds{\beta}}{}^{\mathrm{GLS}} \bigr) & = & \bigl(
\mathbf{X}^{\mathsf{T}}\mathbf{S} \mathbf{X}\bigr)^{-1}, \label
{eqmvmeta4}
\end{eqnarray}
where $\mathbf{y}$ and $\mathbf{X}$ are the stacked $\mathbf{y}_i$
and $\mathbf{X}_i$, and
$\mathbf{S}$ is the block-diagonal matrix containing the $\mathbf{S}_i$.
The fixed-effects model ignores unmodeled heterogeneity among the
studies, however,
and consequently the estimated effects in
equation~(\ref{eqmvmeta3})
may be
biased and
the estimated uncertainty of these effects in equation~(\ref
{eqmvmeta4}) may
be too small.

The example we use here concerns the comparison of surgical (S) and
nonsurgical (NS) procedures for the treatment of moderate periodontal
disease in five randomized split-mouth design clinical trials
[\citet{Antczak-Bouckoms-etal1993}, \citet{Berkey-etal1998}].
The two outcome measures for each patient were pre- to post-treatment
changes after one year
in probing depth (PD) and attachment level (AL), in mm, where
successful treatment
should decrease probing depth and increase attachment level.
Each study was summarized by the mean difference, $\mathbf{y}_i =
(\mathbf{y}_i^S - \mathbf{y}_i^{NS})$,
between S and NS treated teeth, together with the covariance matrix
$\mathbf{S}_i$ for each study.
Sample sizes ranged from 14 to 89 across studies.\looseness=-1

The left panel of Figure~\ref{figmvmeta2} shows the individual study
estimates of PD and AL together with their covariances ellipses
in a generic form that we propose as a more useful visualization of
multivariate meta-analysis results than standard
tabular displays: individual estimates plus model-based summary, all
with associated covariance ellipsoids.%
\footnote{The analyses described here were carried out using the
\pkg{mvmeta} package for R [\citet{mvmeta}].}

It can be seen that all studies show that surgical treatment yields
better probing depth (estimates are positive), while
nonsurgical treatment results in better attachment level (all estimates
are negative). As well, within each study, there is a consistently
positive correlation between the two outcome effects: patients with a
greater surgical vs. nonsurgical difference on one
measure tend to have a greater such difference on the other, and
greater within-study variation on PD than on AL.%
\footnote{Both PD and AL are measured on the same scale (mm), and the
plots have been scaled to have unit aspect ratio,
justifying this comparison.}
The overall sizes of the ellipses largely reflect (inversely) the
sample sizes in the various studies.
As far as we know, these results were not noted in previous analyses of
these data.
Finally, the fixed-effect estimate, $\widehat{\bolds{\beta}}{}^{\mathrm{GLS}} = (0.307, -0.394)$,
and its covariance ellipse suggest that these effects are precisely estimated.

The random-effects model is more complex because $\bolds{\beta}$ and
$\bolds{\Delta}$ must be estimated jointly.
A variety of methods have been proposed (full maximum likelihood,
restricted maximum likelihood, method of
moments, Bayesian methods, etc.), whose details [for which see \citet
{Jackson-etal2011}] are not relevant to\vadjust{\goodbreak} the present discussion.
Given an estimate $\widehat{\Delta}$, however, the pooled,
population-averaged point estimate of effects under the
random-effects model can be expressed as
%
%
\begin{equation}
\label{eqmvmeta5}\qquad \widehat{\bolds{\beta}}{}^{\mathrm{RE}} = \biggl(\sum
_i \mathbf{X}_i^{\mathsf{T}}\bolds{
\Sigma}_i^{-1} \mathbf{X}_i
\biggr)^{-1} \biggl(\sum_i
\mathbf{X}_i^{\mathsf{T}}\bolds{\Sigma}_i^{-1}
\mathbf{y}_i \biggr) ,
\end{equation}
where $\bolds{\Sigma}_i = \mathbf{S}_i + \widehat{\bolds{\Delta}}$.
The first term in equation~(\ref{eqmvmeta5}) gives the estimated covariance
matrix $\mathbf{V} \equiv\widehat{\Var}(\widehat{\bolds{\beta
}}{}^{\mathrm{RE}})$
of the random-effect pooled estimates.
For the present example, this is shown in the right panel of
Figure~\ref{figmvmeta2} as the blue ellipse. The green ellipse shows the
estimate of the between-study covariance, $\widehat{\bolds{\Delta}}$,
whose shape indicates that studies with a larger estimate
of PD also tend to have a larger estimate of AL (with correlation = 0.61).
It is readily seen that, relative to
the fixed-effects estimate $\widehat{\bolds{\beta}}{}^{\mathrm{GLS}}$,
the unbiased estimate $\widehat{\bolds{\beta}}{}^{\mathrm{RE}} =
(0.353, -0.339)$
under the random-effects model
has been shifted toward the centroid of the individual study estimates
and that its covariance ellipse is now considerably larger, reflecting
between-study heterogeneity.
In contrast to the fixed-effect estimates, inferences on $H_0 \dvtx
\widehat{\bolds{\beta}}{}^{\mathrm{RE}} = \mathbf{0}$
pertain to the entire population of potential studies of these effects.

Figure~\ref{figmvmeta2} (right) also shows the best linear unbiased
predictions of individual study estimates
and their associated covariance ellipses, superposed (pure\-ly for
didactic purposes)
on the fixed-effects estimates to allow direct comparison.
For random-effects models, the BLUPs have the form
%
%
\begin{equation}
\label{eqmvmeta6} \widehat{\bolds{\beta}}_i^{\mathrm{BLUP}} =
\widehat{\bolds{\beta}}{}^{\mathrm{RE}} + \widehat{\bolds{\Delta}} \bolds{
\Sigma}_i^{-1} \bigl(\mathbf{y}_i - \widehat{
\bolds{\beta}}{}^{\mathrm{RE}} \bigr) ,
\end{equation}
with covariance matrices
%
%
\begin{equation}
\label{eqmvmeta7} \widehat{\Var} \bigl(\widehat{\bolds{\beta
}}_i^{\mathrm{BLUP}} \bigr) = \mathbf{V} + \bigl(\widehat{\bolds{
\Delta}} - \widehat {\bolds{\Delta}} \bolds{\Sigma}_i^{-1}
\widehat{\bolds{\Delta}} \bigr) .
\end{equation}

Algebraically, the BLUP outcome estimates in\break equation~(\ref{eqmvmeta6})
are thus a weighted average of the~po\-pulation-averaged estimates and
the study-specific estimates,
with weights depending on the relative sizes of the within- and
between-study covariance~ma\-trices $\mathbf{S}_i$ and $\bolds{\Delta}$.
The point BLUPs borrow strength from the assumption of an underlying
multivariate distribution of study parameters with covariance matrix
$\bolds{\Delta}$, shrinking toward the mean inversely proportional to
the within-study covariance.
Geometrically, these estimates may be described as occurring along the
locus of osculation of the ellipses
$\mathcal{E}(\mathbf{y}_i, \mathbf{S}_i)$ and $\mathcal{E}(\widehat
{\bolds{\beta}}, \widehat{\bolds{\Delta}})$.

Finally, the right panel of Figure~\ref{figmvmeta2} also shows the
covariance ellipses of the BLUPs
from equation~(\ref{eqmvmeta6}). It is clear that their orientation is a
blending of the correlations in\vadjust{\goodbreak}
$\mathbf{V}$ and $\mathbf{S}_i$, and their size reflects the error in the
average point estimates $\mathbf{V}$
and the error in the random deviation predicted for each study.

\section{Discussion and Conclusions}\label{secdiscussion}

\begin{quote}
I know of scarcely anything so apt to impress the imagination as the
wonderful form of cosmic order expressed by the ``[Elliptical] Law of
Frequency of
Error.''
The law would have been personified by the Greeks and deified, if they
had known of it$.\ldots$
It is the supreme law of Unreason.
Whenever a large sample of chaotic elements are taken in hand
and marshaled in the order of their magnitude,
an unsuspected and most beautiful form of regularity proves to have been
latent all along.
\flushright{Sir Francis Galton, \emph{Natural Inheritance}, London: Macmillan, 1889
(``[Elliptical]'' added).
}
\end{quote}

We have taken the liberty to add the word ``Elliptical'' to this famous
quotation from Galton (\citeyear{Ga89}).
His ``supreme law of Unreason'' referred to univariate distributions of
observations tending to the
Normal distribution in large samples. We believe he would not take us
remiss, and might perhaps
welcome us for extending this view to
two and more dimensions, where ellipsoids often provide
an ``unsuspected and most beautiful form of regularity.''

In statistical data, theory and graphical methods, one fundamental organizing
distinction can be made depending
on the dimensionality of the problem. A~coarse but useful scale
considers the essential defining
distinctions to be among:
\begin{itemize}
\item ONE (univariate),
\item TWO (bivariate),
\item MANY (multivariate).
\end{itemize}
This scale%
\footnote{This idea, as a unifying classification principle for data
analysis and
graphics,
was first suggested to the first author in
seminars by John Hartigan at Princeton, c.~1968.}
at least implicitly organizes much of current statistical teaching,
practice and software.
But within this classification, the data, theory and graphical methods
are often treated separately (1D, 2D, $n$D),
without regard to geometric ideas and visualizations that help to unify them.

This paper starts from the premise that one geometric form---the
ellipsoid---provides a unifying\break framework for many statistical
phenomena, with simple representations in
1D (a line) and 2D (an ellipse)\vadjust{\goodbreak} that extend naturally to higher
dimensions (an ellipsoid). The intellectual leap
in statistical thinking from ONE to TWO in \citet{Galton1886} was enormous.
Galton's visual insights derived from the ellipse quickly led to an
understanding of the
ellipse as a contour of a bivariate normal surface. From here, the step
from TWO to MANY
would take another 20--30 years, but it is hard to escape the
conclusion that
geometric insight from the ellipse to the general ellipsoid in $n$D
played an important role in the development of multivariate statistical methods.

In this paper, we have tried to show how ellipsoids can be useful tools for
visual statistical thinking, data analysis and pedagogy in a variety of
contexts often
treated separately and from a univariate perspective. Even in bivariate
and multivariate problems, first-moment summaries (a 1D regression line
or 2${}+{}$D regression surface) show only part of the story---that of the
expectation of a response $\mathbf{y}$ given predictors $\mathbf{X}$.
In many cases, the more interesting part of the story concerns the
\emph{precision} of various methods of estimation, which we've shown
to be easily revealed through data ellipsoids and
elliptical confidence regions for parameters.

The general relationships among statistical methods, matrix algebra and
geometry are
not new here. To our knowledge, \citet{Dempster69} was the first to
exploit these relationships
in a systematic fashion, establishing the connections among abstract
vector spaces,
algebraic coordinate systems, matrix operations and properties, the dualities
between observation space and variable space,
and the geometry
of ellipses and projections. %
The roots of these connections go back much further---to
\citet{Cramer1946} (the idea of the concentration ellipsoid), to
\citet{Pearson1901} and \citet{Hotelling1933} (principal components),
and, we maintain, ultimately to \citet{Galton1886}.\break
Throughout this development, elliptical geometry has played
a fundamental role, leading to important visual insights.

The separate and joint roles of statistical computation and
computational graphics should not be underestimated
in appreciation of these developments. Dempster's analysis of the
connections among geometry, algebra and
statistical methods was fueled by the development and software
implementation of algorithms
[Gram-Schmidt orthogonalization, Choles\-ky decomposition, sweep and
multistandardize operators from
\citet{Beaton64}]
that allowed him to show precisely the translation of theoretical relations
from abstract algebra to numbers and from there to graphs and diagrams.

\citet{Monette90}
took these ideas several steps further, developing interactive 3D
graphics focused on linear
models, geometry and ellipsoids, and demonstrating how many statistical
properties
and results could be understood through the geometry of ellipsoids.
Yet, even at this
later date, the graphical facilities of readily available statistical
software were still
rather primitive, and 3D graphics was available only on high-end workstations.

Several features of the current discussion may help to present these
ideas in a
new light. First, the examples we have presented rely heavily on
software for
statistical graphics developed separately and jointly by all three authors.
These have allowed us to create what we hope are compelling illustrations,
all statistically and geometrically exact, of the principles and ideas that
form the body of the paper. Moreover, these are now general methods, implemented
in a variety of R packages, for example,
\citet{FoxWeisberg2011}, \citet{Friendly07manova},
and a large collection of SAS macros (\url{http://datavis.ca/sasmac}),
so we hope this paper will contribute to turning the theory we describe
into practice.

Second, we have illustrated, in a wide variety of contexts,
comprising all classical (Gaussian) linear models, multivariate linear models
and several extensions,
how ellipsoids can contribute substantially to the understanding of statistical
relationships, both in data analysis and in pedagogy. One graphical
theme underlying
a number of our examples is how the simple addition of ellipses to
standard 2D graphical
displays provides an efficient \emph{visual} summary of important bivariate
statistical quantities (means, variances, correlation, regression
slopes, etc.).
While first-moment visual summaries are now common adjuncts to
graphical displays
in standard software, often by default, we believe that the
second-moment visual summaries
of ellipses (and ellipsoids in 3D)
now deserve a similar place in statistical practice.

Finally, we have illustrated several recent or entirely new
visualizations of
statistical methods and results, all based on elliptical geometry.
HE plots for MANOVA designs [\citet{Friendly07manova}]
and their projections into canonical space (Section~\ref{secmlm})
provide one class of examples where ellipsoids provide simple visual
summaries of
otherwise complex statistical results.
Our analysis of the geometry of added variable-plots suggested the idea
of superposing
mar\-ginal and conditional plots, as in Figure~\ref{figcoffee-avplot-B},
leading to
direct visualization of the difference between marginal and conditional
relationships
in linear models.
The bivariate ridge trace plots described in\vadjust{\goodbreak} Section~\ref{secridge2}
are a
direct outgrowth
of the geometric approach taken here, emphasizing the duality between
views in data
space and in parameter ($\bolds{\beta}$) space.
We believe these all embody von Humboldt's (\citeyear{Humboldt1811a}) dictum, quoted in the
\hyperref[sec1]{Introduction}.


%
%
\begin{appendix}\label{app}
\section*{Appendix: Geometrical and Statistical Ellipsoids}\label{secAppendixA}
This appendix outlines useful results and properties concerning the
representation of geometric and statistical ellipsoids.
A number of these can be traced to or have more general descriptions
within the abstract formulation of
\citet{Dempster69},
but casting them in terms of ellipsoids provides a simpler and more
easily visualized framework.
\subsection{Taxonomy and Representation of Generalized
Ellipsoids}\label{sectaxonomy}
Section~\ref{secgeometric} defined a \emph{proper} (origin-centered)
ellipsoid in $\mathbb{R}^{p}$ by
$\mathcal{E} := \{ \mathbf{x}\dvtx\mathbf{x}^{\mathsf{T}}\mathbf
{C} \mathbf
{x} \le1 \}
$ that is bounded with nonempty interior (call these ``fat'' ellipsoids).
For more general purposes, particularly for statistical applications,
it is useful to give ellipsoids a wider definition.
To provide a complete taxonomy, this wider definition should also
include ellipsoids that may be unbounded in some directions in $\mathbb{R}^{p}$
(an infinite cylinder of ellipsoidal cross-section) and
degenerate (singular) ellipsoids that are ``flat'' in $\mathbb{R}^{p}$ with
empty interior, such as when a 3D ellipsoid has no extent in
one dimension (collapsing to an ellipse), or in two dimensions
(collapsing to a line). These ideas are made precise below
with a definition of the \emph{signature}, $\mathcal{G}(\mathbf{C})$,
of a generalized ellipsoid.

The motivation for this more general representation is to allow a
notation for a class of generalized ellipsoids to be
algebraically closed under operations (a) image and preimage under a
linear transformation, and (b) inversion.
The goal is to be able to think about, visualize and \emph{compute} a
linear transformation of an ellipsoid with central matrix
$\mathbf{C}$ or its inverse transformation via an analog of $\mathbf
{C}^{-1}$, which applies equally to unbounded
and/or degenerate ellipsoids.
Algebraically, the vector space of $\mathbf{C}$ is the \emph{dual} of
that of $\mathbf{C}^{-1}$
[\citet{Dempster69}, Chapter 6]
and vice-versa.
Geometrical applications can show how points, lines and hyperplanes
in $\mathbb{R}^{p}$ are all special cases of ellipsoids.
Statistical applications concern the relationship between a predictor
data ellipsoid and the
corresponding $\bolds{\beta}$ confidence ellipsoid (Section~\ref
{secbetaspace}): The $\bolds{\beta}$ ellipsoid will be unbounded (some
linear combinations
will have infinite confidence intervals) \emph{iff} the corresponding
data ellipsoid is flat, as when $p>n$ or some predictors are collinear.

Defining ellipsoids with $\{ \mathbf{x}\dvtx\mathbf{x}^{\mathsf
{T}}\mathbf{C}
\mathbf{x}
\le1 \}$ produces proper ellipsoids for $\mathbf{C}$
positive definite and unbound\-ed, fat ellipsoids for $\mathbf{C}$ positive
semi-definite.
But it does not produce degenerate (i.e., flat) ellipsoids. On the
other hand,
the representation in equation~(\ref{eqellisoidsph}),
$\mathcal{E} := \mathbf{A} \mathcal{S}$, with $\mathcal{S}$ the
unit sphere,
produces proper ellipsoids when $\mathbf{C} = ( \mathbf{A}^{\mathsf{T}}
\mathbf{A} )^{-1}$
where $\mathbf{A}$ is a nonsingular $p \times p$ matrix and degenerate
ellipsoids when $\mathbf{A}$ is a singular, but does not
produce unbounded ellipsoids.

%
%
\setcounter{figure}{0}
\begin{figure*}

\includegraphics{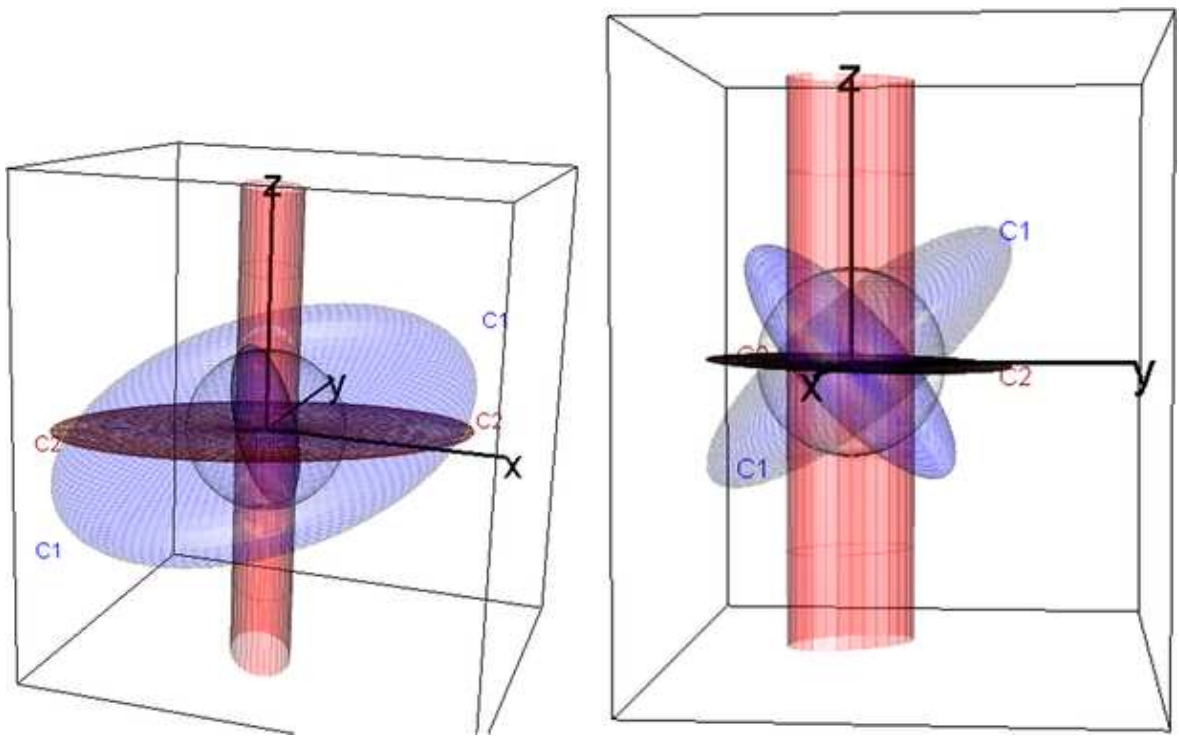}

\caption{Two views of an example of generalized ellipsoids. $\mathbf
{C}_1$ (blue) determines a proper, fat ellipsoid;
its inverse $\mathbf{C}_1^{-1}$ also generates a proper ellipsoid.
$\mathbf{C}_2$ (red) determines an improper, flat ellipsoid,
whose inverse $\mathbf{C}_2^{-1}$ is an unbounded cylinder of elliptical
cross section. The scale of these images is defined
by a unit sphere (gray). The left panel shows that $\mathbf{C}_2$ is a
projection of $\mathbf{C}_1$ onto the plane where
$z=0$.
The right panel shows a view illustrating the orthogonality of each
$\mathbf{C}$ and its dual, $\mathbf{C}^{-1}$.}
\label{figgell3d}
\end{figure*}

One representation that works for all ellipsoids---fat or flat \emph
{and} bounded or unbounded---can be based on
a singular value decomposition (SVD) representation
$\mathbf{A} = \mathbf{U} \bolds{\Delta} \mathbf{V}^{\mathsf{T}}$, with
%
%
\setcounter{equation}{0}
\begin{equation}
\mathcal{E} := \mathbf{U} (\bolds{\Delta} \mathcal{S}) ,
\end{equation}
where $\mathbf{U}$ is orthogonal and $\bolds{\Delta}$ is diagonal with
nonnegative reals or infinity.%
\footnote{Note that the parentheses in this notation are obligatory:
$\bolds
{\Delta}$ as defined transforms the
unit sphere, which is then transformed by $\mathbf{U}$. $\mathbf
{V}^{\mathsf{T}}
$, also orthogonal, plays no role
in this representation, because an orthogonal transformation of
$\mathcal{S}$ is still a unit sphere.}
The ``inverse'' of an ellipsoid $\mathcal{E}$ is then simply $\mathbf{U}
(\bolds{\Delta}^{-1} \mathcal{S})$.
The connection with traditional representations is that,
if $\bolds{\Delta}$ is finite,
$\mathbf{A} = \mathbf{U} \bolds{\Delta} \mathbf{V}^{\mathsf{T}}$,
where $\mathbf{V}$ can be any orthogonal matrix, and
if $\bolds{\Delta}^{-1}$ is finite, $\mathbf{C} = \mathbf{U} \bolds
{\Delta}^{-2} \mathbf{U}^{\mathsf{T}}$.

The $\mathbf{U} (\bolds{\Delta} \mathcal{S})$ representation also allows
us to characterize any such
generalized ellipsoid in $\mathbb{R}^{p}$ by its \emph{signature},
%
%
\begin{eqnarray}
\mathcal{G}(\mathbf{C}) = \#[ \delta_i >0, \delta_i
=0, \delta_i =\infty]
\nonumber
\\[-8pt]
\\[-8pt]
\eqntext{\mbox{with } \sum\mathcal{G}(
\mathbf{C}) = p,}
\end{eqnarray}
a 3-vector containing the number ($\#$) of positive, zero and infinite
singular values. For example, in $\mathbb{R}^{3}$, any proper
ellipsoid has
the signature $\mathcal{G}(\mathbf{C})=(3, 0, 0)$; a flat, 2D ellipsoid
has $\mathcal{G}(\mathbf{C})=(2, 1, 0)$; a flat, 1D ellipsoid (a line)
has $\mathcal{G}(\mathbf{C})=(1, 2, 0)$. Unbounded examples include an
infinite flat plane, with $\mathcal{G}(\mathbf{C})=(0, 1, 2)$,
and an infinite cylinder of elliptical cross-section, with $\mathcal
{G}(\mathbf{C})=(2, 0, 1)$.

Figure~\ref{figgell3d} illustrates these ideas, using two generating
matrices, $\mathbf{C}_1$ and $\mathbf{C}_2$, in this more
general representation,
\[
\mathbf{C}_1 = \left[ \matrix{
6 & 2 & 1
\vspace*{2pt}\cr
2 & 3 & 2
\vspace*{2pt}\cr
1 & 2 & 2 }
\right] ,\quad \mathbf{C}_2 = \left[
\matrix{ 6 & 2 & 0
\vspace*{2pt}\cr
2 & 3 & 0
\vspace*{2pt}\cr
0 & 0 & 0 }\right],
\]
where $\mathbf{C}_1$ generates a proper ellipsoid and $\mathbf{C}_2$
generates an improper, flat ellipsoid.
$\mathbf{C}_1$ and its dual, $\mathbf{C}_1^{-1}$, both have
signatures $(3,
0, 0)$.
$\mathbf{C}_2$ has the signature $(2, 1, 0)$, while its inverse (dual)
has the signature $(2, 0, 1)$.
These varieties of ellipsoids are more easily understood in the 3D
movies included in the online supplements.

\subsection{Properties of Geometric Ellipsoids}\label{secproperties}

\begin{itemize}
\item Translation: An ellipsoid centered at $\mathbf{x}_0$ has the
definition $\mathcal{E} := \{ \mathbf{x}\dvtx(\mathbf{x}-\mathbf
{x}_0)^{\mathsf{T}}
\mathbf{C} (\mathbf{x}-\mathbf{x}_0) =1
\}$ or $\mathcal{E} := \mathbf{x}_0 \oplus\mathbf{A} \mathcal{S}$
in the
notation of Section~\ref{secstatistical}.

\item Orthogonality: If $\mathbf{C}$ is diagonal, then the
origin-centered ellipsoid has its axes aligned with the coordinate axes and
has the equation
%
%
\begin{equation}
\label{eqellisoid2}\qquad \mathbf{x}^{\mathsf{T}}\mathbf{C} \mathbf{x} =
c_{11} x_1^2 + c_{22}
x_2^2 + \cdots+ c_{pp} x_p^2
=1 ,
\end{equation}
where $1/\sqrt{c_{ii}} = c_{ii}^{-1/2}$ are the radii (semi-diameter
lengths) along the coordinate axes.

\item Area and volume: In two dimensions, the area of the axis-aligned
ellipse is $\pi(c_{11} c_{22})^{-1/2}$.
For $p=3$, the volume is $\frac{4}{3}\pi(c_{11} c_{22} c_{33})^{-1/2}$.
In the general case, the hypervolume of the ellipsoid is pro\-portional
to $|\mathbf{C}|^{-1/2}=\Vert\mathbf{A}\Vert$
and is given by\break $\pi^{p/2} \operatorname{det} (\mathbf{C})^{-1/2} /
[ \Gamma
(\frac{p}{2}+1 ) ]$
[\citet{Dempster69}, Section 3.5],
where the first two factors are familiar as the normalizing constant of
the multivariate normal
density function.
%
%
\begin{figure}

\includegraphics{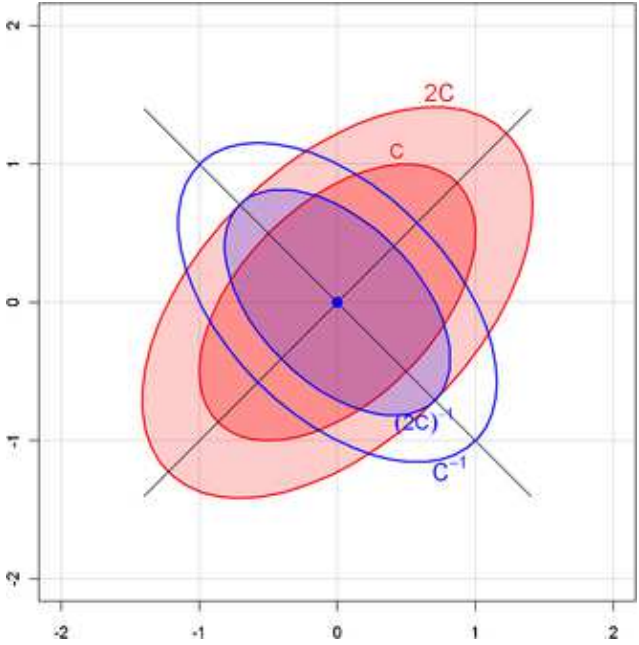}

\caption{Some properties of geometric ellipsoids, shown in 2D.
Principal axes of an ellipsoid are given by the eigenvectors of
$\mathbf{C}$, with radii $1/\sqrt{\lambda_i}$. For an ellipsoid defined
by equation~(\protect\ref{eqellisoid1}),
the comparable ellipsoid for $2\mathbf{C}$ has radii multiplied by
$1/\sqrt{2}$.
The ellipsoid for $\mathbf{C}^{-1}$ has the same principal axes, but with
radii $\sqrt{\lambda_i}$, making it
small in the directions where $\mathbf{C}$ is large and vice-versa.
} \label{figinverse}
\end{figure}

\item Principal axes: In general, the eigenvectors, $\mathbf{v}_i,
i=1,\ldots,p$,
of $\mathbf{C}$ define the principal axes of the ellipsoid and
the inverse of the square roots of the ordered
eigenvalues, $\lambda_1 > \lambda_2 ,\ldots, \lambda_p$, are the
principal radii.
Eigenvectors belonging to eigenvalues that are 0 are directions in
which the ellipsoid is unbounded.
With $\mathcal{E} = \mathbf{A} \mathcal{S}$, we consider the
singu\-lar-value decomposition
$ \mathbf{A}= \mathbf{U} \mathbf{D} \mathbf{V}^{\mathsf{T}}$,
with $\mathbf{U}$ and $\mathbf{V}$ orthogonal matrices and $\mathbf
{D}$ a
diagonal nonnegative matrix
with the same dimension as $\mathbf{A}$.
The column vectors of $\mathbf{U}$, called the left singular vectors,
correspond to the eigenvectors of $\mathbf{C}$ in the case of a proper
ellipsoid.
The positive diagonal elements of $\mathbf{D}$, $d_1 > d_2 > \cdots> d_p>0$,
are the principal radii of the ellipsoid with $d_i = 1/\sqrt{\lambda_i}$.
In the singular case, the left singular vectors form a set of principal
axes for the flattened ellipsoid.%
\footnote{Corresponding left singular vectors and eigenvectors are not
necessarily equal, but sets that belong to the same eigenvalue/singular
value span the same space.}

\item Inverse: When $\mathbf{C}$ is positive definite, the eigenvectors
of $\mathbf{C}$ and $\mathbf{C}^{-1}$ are identical, while
the
eigenvalues of $\mathbf{C}^{-1}$ are $1/\lambda_i$. It follows that the
ellipsoid for
$\mathbf{C}^{-1}$ has the same axes as that of $\mathbf{C}$, but with
inversely proportional radii.
In $\mathbb{R}^{2}$, the ellipsoid for $\mathbf{C}^{-1}$
is, with rescaling, a ${90}^{\circ}$ rotation of the ellipsoid for
$\mathbf{C}$,
as illustrated in Figure~\ref{figinverse}.

\item Generalized inverse: A definition for an inverse ellipsoid that
is equivalent in the case of proper ellipsoids,
%
%
\begin{equation}
\label{eqellipseginv} \mathcal{E}^{-1} := \bigl\{ \mathbf{y} \dvtx\bigl|
\mathbf{x}^{\mathsf{T}}\mathbf{y}\bigr| \le1, \forall\mathbf{x} \in \mathcal{E} \bigr
\} ,
\end{equation}
generalizes to all ellipsoids. The inverse of a singular ellipsoid is
an improper ellipsoid and vice versa.

\item Dimensionality: The ellipsoid is bounded if $\mathbf{C}$ is
positive definite (all $\lambda_i > 0$).
Each $\lambda_i = 0$ increases the dimension of the space along which
the ellipsoid is unbounded by one.
For example, with $p=3$, $\lambda_3=0$ gives a
cylinder with an elliptical cross-section in 3-space, and $\lambda_2 =
\lambda_3=0$ gives an infinite slab with thickness $2 \sqrt{\lambda_1}$. With $\mathcal{E} = \mathbf{A} \mathcal{S}$, the dimension of the
ellipsoid is equal to the number of positive singular values of
$\mathbf{A}$.
\item Projections: The projection of a $p$-dimensional ellipsoid into
any subspace
is $\mathbf{P} \mathcal{E}$, where
$\mathbf{P}$ is an idempotent $p \times p$ (projection) matrix, that is,
$\mathbf{P} \mathbf{P}= \mathbf{P}^2 = \mathbf{P}$.
For example, in $\mathbb{R}^{2}$ and $\mathbb{R}^{3}$,
the matrices
\[
\mathbf{P}_2 = \left[ \matrix{ 1
& 1
\vspace*{2pt}\cr
0 & 0}
\right] ,\quad \mathbf{P}_3 = \left[
\matrix{ 1 & 0 & 0
\vspace*{2pt}\cr
0 & 1 & 0
\vspace*{2pt}\cr
0 & 0 & 0}
\right]
\]
project, respectively, an ellipse onto the line $x_1 = x_2$ and an
ellipsoid into the ($x_1, x_2$) plane. If $\mathbf{P}$ is symmetric, then
$\mathbf{P}$ is the matrix of an orthogonal projection, and it is easy to
visualize $\mathbf{P} \mathcal{E}$ as the shadow of $\mathcal{E}$ cast
perpendicularly onto $\operatorname{span}(\mathbf{P})$. Generally,
$\mathbf{P}
\mathcal{E}$ is the shadow of $\mathcal{E}$ onto $\operatorname
{span}(\mathbf{P})$ along the null space of $\mathbf{P}$.

\item Linear transformations: A linear transformation of an ellipsoid
is an ellipsoid, and the pre-image of an ellipsoid under a linear
transformation is an ellipsoid.
A nonsingular linear transformation maps a proper ellipsoid into a
proper ellipsoid in the form shown in Section~\ref{secstatistical},
equation~(\ref{eqLimage}).

%
%
\begin{figure*}

\includegraphics{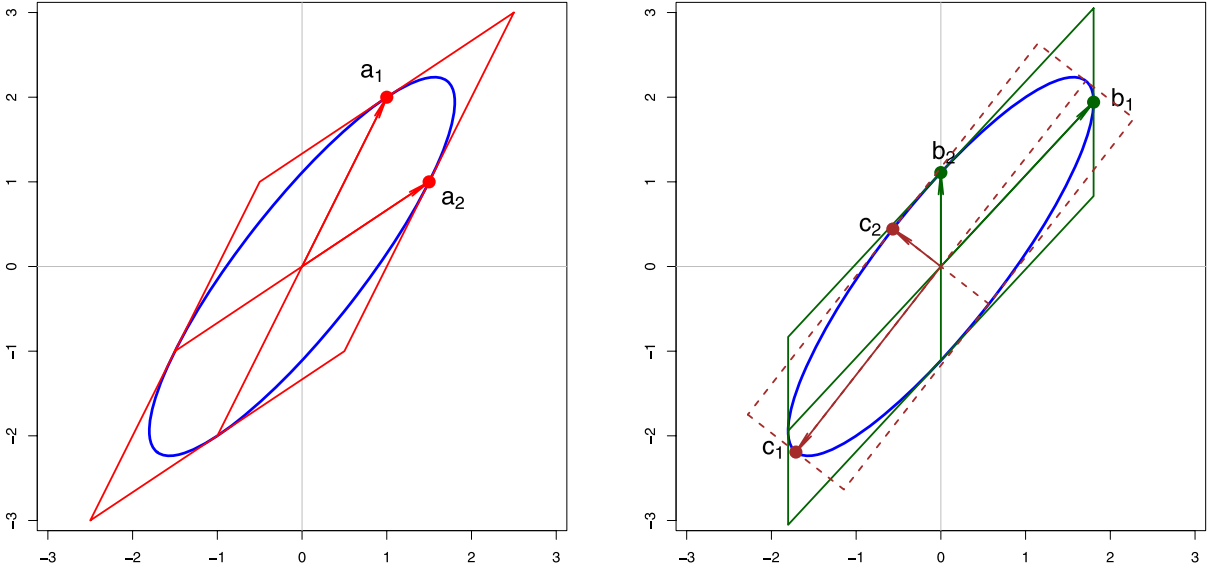}

\caption{Conjugate axes of an ellipse with various factorizations of
$\mathbf{W}$ and corresponding
basis vectors.
The conjugate vectors lie on the ellipse, and their tangents can be
extended to form a parallelogram framing it.
Left: for an arbitrary factorization, given in equation~\textup{(\protect\ref
{eqfac1})}.
Right: for the Choleski factorization (solid, green, $b_1, b_2$) and
the principal component factorization (dashed, brown, $c_1, c_2$).}
\label{figconjugate}
\end{figure*}

\item Slopes and tangents: The slopes of the ellipsoidal surface in the
directions of the coordinate
axes are given by $\partial/ \partial\mathbf{x} (\mathbf
{x}^{\mathsf{T}}
\mathbf{C} \mathbf{x}) = 2 \mathbf{C} \mathbf{x}$.
From this, it follows that the tangent hyperplane to the unit
ellipsoidal surface at the point
$\mathbf{x}_\alpha$, where
$\mathbf{x}_\alpha^{\mathsf{T}}\partial/\break
\partial\mathbf{x} (\mathbf{x}^{\mathsf{T}}\mathbf{C} \mathbf{x}) = 0$,
has the equation $\mathbf{x}_\alpha^{\mathsf{T}}\mathbf{C} \mathbf
{x} = 1$.
\end{itemize}

\subsection{Conjugate Axes and Inner-Product Spaces}\label{secconjugate}

For any nonsingular $\mathbf{A}$ in equation~(\ref{eqellAS}) that
generates an
ellipsoid, 
the columns of
$\mathbf{A}= [
\mathbf{a}_{1}, \mathbf{a}_{2}, \ldots, \mathbf{a}_{p} ]
$
form a set of ``conjugate axes'' of the ellipsoid. (Two diameters are
conjugate \emph{iff}
the tangent line at the endpoint of one diameter is parallel to the
other diameter.)
Each vector
$\mathbf{a}_{i}$
lies on the ellipsoid, and the tangent hyperplane at that point is
parallel to the span of all the other column vectors of
$\mathbf{A}$.
For
$p=2$
this result is illustrated in Figure~\ref{figconjugate} (left)
in which
%
%
\begin{eqnarray}
\label{eqfac1} \mathbf{A}&=&\left[ %
\matrix{ \mathbf{a}_{1}
& \mathbf{a}_{2} } %
\right]=\left[
\matrix{ 1 & 1.5 \vspace*{2pt}
\cr
2 & 1 } %
\right]
\quad\Rightarrow
\nonumber
\\[-8pt]
\\[-8pt]
\nonumber
 \mathbf{W}&=&\mathbf{A A^{\mathsf{T}}}= \left [
\matrix{ 3.25 & 3.5 \vspace*{2pt}
\cr
3.5 & 5 } %
\right].
\end{eqnarray}

Consider the inner-product space with inner product matrix
%
\begin{eqnarray}
\mathbf{W}^{-1}=\left[ %
\matrix{ 1.25 & -0.875
\vspace*{2pt}
\cr
-0.875 & 0.8125 } %
\right]\nonumber\\
\eqntext{\mbox{and inner product }
\langle\mathbf{x},\mathbf{y} \rangle=\mathbf
{{x}'W}^{-1}\mathbf{y} .}
\end{eqnarray}
Because
$\mathbf{A}^{\mathsf{T}}\mathbf{W}^{-1}\mathbf{A}
=\mathbf{A}^{\mathsf{T}}( \mathbf{A}\mathbf{A}^{\mathsf{T}}
)^{-1}\mathbf{A}
=\mathbf{A}^{\mathsf{T}}(\mathbf{A}^{\mathsf{T}})^{-1}\cdot \mathbf
{A}^{-1}\mathbf{A}
=\mathbf{I}
$,
we see that
$\mathbf{a}_{1}$
and
$\mathbf{a}_{2}$
are orthogonal unit vectors (in fact, an orthonormal basis) in this
inner product:

\begin{eqnarray*}
\langle\mathbf{a}_{i},\mathbf{a}_{i} \rangle& = &
\mathbf{{a}^{\mathsf{T}}}_{i}\mathbf{W}^{-1}
\mathbf{a}_{i}=1 ,
\\
\langle\mathbf{a}_{1},\mathbf{a}_{2} \rangle&
= &\mathbf{{a}'}_{1}\mathbf{W}^{-1}
\mathbf{a}_{2}=0 .
\end{eqnarray*}

Now, if $\mathbf{W}=\mathbf{B{B}^{\mathsf{T}}}$ is any other
factorization of
$\mathbf{W}$,
then the columns of
$\mathbf{B}$
have the same properties as the columns of
$\mathbf{A}$.
Particular factorizations yield interesting and statistically useful
sets of conjugate axes.
The illustration in Figure~\ref{figconjugate} (right) shows two such
cases with special properties:
In the Choleski factorization (shown solid in green), where
$\mathbf{B}$ is lower triangular, the last conjugate axis, $\mathbf{b}_2$,
is aligned with the coordinate
axis $\mathbf{x}_2$. Each previous axis ($\mathbf{b}_1$, here) is the
orthogonal complement to
all later axes in the inner-product space of
$\mathbf{W}^{-1}$.
The Choleski factorization is unique in this respect, subject to a
permutation of the rows and columns of $\mathbf{W}$.
The subspace $\{ c_1 \mathbf{b}_1 + \cdots+ c_{p-1} \mathbf{b}_{p-1}
, c_i
\in\mathbb{R}\}$ is the plane of the regression of the last variable
on the others,
a fact that generalizes naturally to
ellipsoids that are not necessarily centered at the origin. 

%
%
\begin{figure*}

\includegraphics{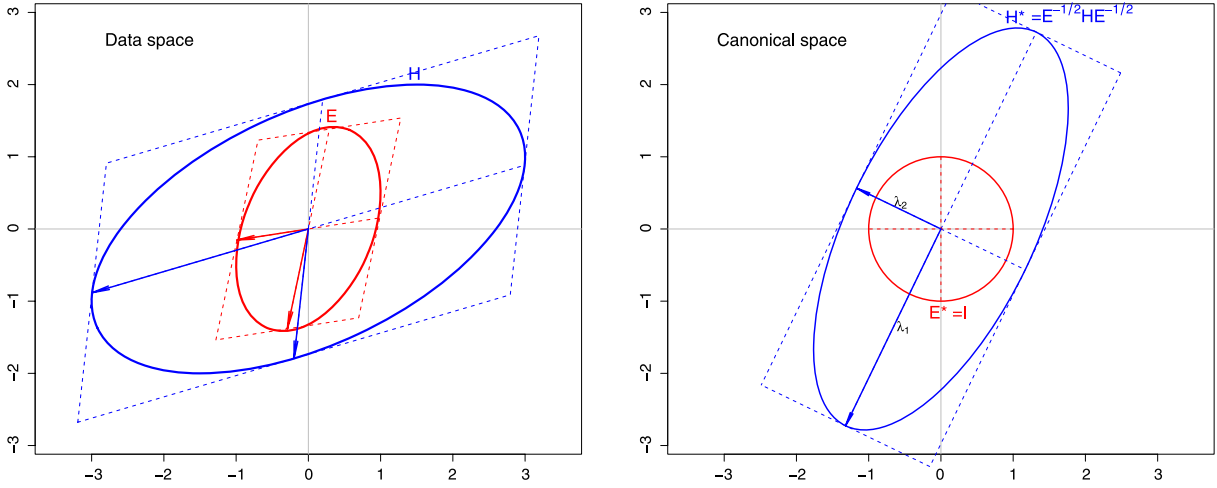}

\caption{Left: Ellipses for $\mathbf{H}$ and $\mathbf{E}$ in
Euclidean ``data space.''
Right: Ellipses for $\mathbf{H}^\star$ and $\mathbf{E}^\star$ in the
transformed ``canonical space,''
with the eigenvectors of $\mathbf{H}$ relative to $\mathbf{E}$ shown
as blue
arrows, whose radii
are the corresponding eigenvalues, $\lambda_1, \lambda_2$.}%
\label{figellipse-geneig}
\end{figure*}

In the principal-component (PC) factorization\break (shown dashed in brown in
Figure~\ref{figconjugate}, right) $\mathbf{W}=\mathbf{C} \mathbf
{C}^{\mathsf{T}}$, where
$\mathbf{C}=\bolds{\Gamma\Lambda}^{1/2}$
and, hence,
$\mathbf{W}=\bolds{\Gamma\Lambda{\Gamma}'}$
is the spectral decomposition of
$\mathbf{W}$. Here, the ellipse axes are orthogonal in the space of
the ellipse
(so the bounding tangent parallelogram is a rectangle) \emph{as well
as} in the inner-product space of
$\mathbf{W}^{-1}$. The PC factorization is unique in this respect (up to
reflections of the axis vectors).

As illustrated in Figure~\ref{figconjugate}, each pair of conjugate axes
has a corresponding bounding tangent
parallelogram. It can be shown that all such parallelograms have the
same area
and equal sums of squares of the lengths of their diameters.

\subsection{Ellipsoids in a Generalized Metric Space}


In Appendix~\ref{secconjugate}, we considered the positive semi-definite
matrix $\mathbf{W}$ and
corresponding ellipsoid to be
referred to a Euclidean space, perhaps with different basis vectors.
We showed that various measures of the ``size'' of the ellipsoid could
be defined
in terms of functions of the eigenvalues $\lambda_i$ of $\mathbf{W}$.

We now consider the generalized
case of an analogous $p \times p$ positive semi-definite symmetric
matrix $\mathbf{H}$, but where measures of
length, distance and angles are referred to a metric defined by a
positive-definite symmetric
matrix $\mathbf{E}$. As is well known, the generalized eigenvalue problem
is to find the scalars
$\lambda_i$ and vectors $\mathbf{v}_i, i=1, 2, \ldots, p$,
such that $\mathbf{H} \mathbf{v} = \lambda\mathbf{E} \mathbf{v}$,
that is, the
roots of
$\operatorname{det} (\mathbf{H} - \lambda\mathbf{E})=0$.

%

For such $\mathbf{H}$ and $\mathbf{E}$, we can always find a factor
$\mathbf{A}$ of
$\mathbf{E}$, so that
$\mathbf{E} = \mathbf{A} \mathbf{A}^{\mathsf{T}}$, whose columns
will be conjugate
directions for $\mathbf{E}$
and whose rows will also be conjugate directions for $\mathbf{H}$, in that
$\mathbf{H} = \mathbf{A}^{\mathsf{T}}\mathbf{D} \mathbf{A}$,
where $\mathbf{D}$ is diagonal. Geometrically, this means that there
exists a unique pair of
bounding parallelograms for the $\mathbf{H}$ and $\mathbf{E}$
ellipsoids whose
corresponding sides are parallel. A~linear transformation of $\mathbf{E}$
and $\mathbf{H}$
that transforms the parallelogram
for $\mathbf{E}$ to a square (or cuboid), and hence $\mathbf{E}$ to a circle
(or spheroid), generates an
equivalent view in what we describe below as canonical space.

In statistical applications (e.g., MANOVA, canonical correlation), the
generalized
eigenvalue problem is transformed to an ordinary eigenvalue problem by
considering
the following equivalent forms with the same $\lambda_i$, $\mathbf{v}_i$:
\begin{eqnarray*}
(\mathbf{H} - \lambda\mathbf{E}) \mathbf{v} & = & \mathbf{0},
\\
\Rightarrow\quad \bigl(\mathbf{H} {\mathbf{E}}^{-1} - \lambda\mathbf{I}
\bigr) \mathbf{v} & = & \mathbf{0} ,
\\
\Rightarrow \quad \bigl({\mathbf{E}}^{-1/2} \mathbf{H} {\mathbf
{E}}^{-1/2} - \lambda\mathbf{I} \bigr) \mathbf{v} & = & \mathbf{0} ,
\end{eqnarray*}
where the last form gives a symmetric matrix, $\mathbf{H}^\star=
\mathbf{E}^{-1/2} \mathbf{H} \mathbf{E}^{-1/2}$.
Using the square root of $\mathbf{E}$ defined by the
principal-component factorization $\mathbf{E}^{1/2} = \bolds{\Gamma}
\bolds{\Lambda}^{1/2}$ produces
the ellipsoid $\mathbf{H}^\star$, the
orthogonal axes of which correspond to the $\mathbf{v}_i$, whose squared
radii are the corresponding 
eigenvalues $\lambda_i$. This can be seen geometrically as a rotation
of ``data space''
to an orientation defined by the principal axes of $\mathbf{E}$, followed
by a re-scaling, so
that the $\mathbf{E}$ ellipsoid becomes the unit spheroid. In this
transformed space
(``canonical space''), functions of the squared radii $\lambda_i$ of
the axes of $\mathbf{H}^\star$ give direct measures of 
the ``size'' of $\mathbf{H}$ relative to $\mathbf{E}$. The
orientation of the
eigenvectors
$\mathbf{v}_i$ can be related to the (orthogonal) linear combinations
of the
data variables that are successively largest in the metric of $\mathbf{E}$.

To illustrate, Figure~\ref{figellipse-geneig} (left) shows
the ellipses generated by
\[
\mathbf{H} = \left[ %
\matrix{9 & 3
\vspace*{2pt}\cr
3 & 4 }
\right]\quad \mbox{and}\quad \mathbf{E} = \left[
\matrix{ 1 & 0.5
\vspace*{2pt}\cr
0.5 & 2 }
\right]
\]
together with their conjugate axes. For $\mathbf{E}$, the conjugate axes
are defined by the columns of the right factor,
$\mathbf{A}^{\mathsf{T}}$,
in $\mathbf{E} = \mathbf{A} \mathbf{A}^{\mathsf{T}}$; for $\mathbf{H}$,
the conjugate
axes are defined by the columns of $\mathbf{A}$.
The transformation to $\mathbf{H}^\star= \mathbf{E}^{-1/2} \mathbf
{H} \mathbf{E}^{-1/2}$ is shown in the right panel
of Figure~\ref{figellipse-geneig}. In this ``canonical space,'' angles
and lengths have the ordinary interpretation
of Euclidean space, so the size of $\mathbf{H}^\star$ can be interpreted
directly in terms of functions of
the radii $\sqrt{\lambda_1}$ and $\sqrt{\lambda_2}$.
\end{appendix}

\section*{Acknowledgments}
We acknowledge first the inspiration we found in \citet{Dempster69}
for this geometric approach to statistical thinking.
This work was supported by Grant OGP0138748 from the National Sciences and Engineering Research
Council of Canada to Michael Friendly,
and by grants from the Social Sciences and Humanities Research Council of Canada to John Fox.
We are grateful to Antonio Gasparini for helpful discussion regarding
multivariate meta-analysis models, to Duncan Murdoch for
advice on 3D graphics and comments on an earlier draft, and to the
reviewers and Associate Editor for many helpful suggestions.

\begin{supplement}[id=suppA]
\stitle{Supplementary materials for Elliptical insights:
Understanding statistical methods through elliptical geometry}
\slink[doi]{10.1214/12-STS402SUPP} 
\sdatatype{.pdf}
\sfilename{sts402\_supp.pdf}
\sdescription{The supplementary materials include SAS and R scripts to
generate all
of the figures for this article.
Several 3D movies are also included to show phenomena better than can
be rendered in static print images.
A new R package, \pkg{gellipsoid}, provides computational support for
the theory described in Appendix~\ref{sectaxonomy}.
These are also available at
\texttt{\href{http://datavis.ca/papers/ellipses}{http://}
\href{http://datavis.ca/papers/ellipses}{datavis.ca/papers/ellipses}} and described
in\break
\url{http://datavis.ca/papers/ellipses/supp.pdf}\break [\citet{FriendlyMonetteFox2012ellipses-supp}].}
\end{supplement}

%

%

\end{document}